\documentclass[12pt,preprint]{aastex}

\usepackage{graphicx}
\newcommand{\nel}{n_e(\gamma,\Omega_e)}
\newcommand{\nph}{n_{ph}(\epsilon,\Omega_{ph})}
\newcommand{\nelg}{n_e(\gamma)}
\newcommand{\nphm}{n_{ph}(\epsilon,\mu_{ph})}
\newcommand{\nsca}{\dot n_(\epsilon_s,\Omega_s)}

\slugcomment{Submitted to {\it The Astrophysical Journal}}

\shorttitle{HE emission from colliding winds of massive stars}
\shortauthors{Reimer et al.}

\begin{document}

\title{Non-thermal high-energy emission from colliding winds of massive stars}

\author{A. Reimer\altaffilmark{1}, M. Pohl\altaffilmark{2} \and O. Reimer\altaffilmark{1}}

\altaffiltext{1}{W.W. Hansen Experimental Physics Laboratory, Stanford University,
445 Via Palou, Stanford, CA 94305, USA; afr@stanford.edu, olr@stanford.edu}
\altaffiltext{2}{Department of Physics and Astronomy,
	  Iowa State University,
	  Ames, Iowa 50011, USA; mkp@iastate.edu}

\begin{abstract}
Colliding winds of massive star binary systems are considered as potential 
sites of non-thermal high-energy photon production. This is motivated merely
by the detection of synchrotron radio emission from the expected colliding wind location.
Here we investigate the properties of high-energy photon production 
in colliding winds of long-period WR+OB-systems. We found that in 
the dominating leptonic radiation process anisotropy and Klein-Nishina 
effects will likely yield spectral and variability signatures in the 
$\gamma$-ray domain at or above the sensitivity of current or upcoming 
$\gamma$-ray telescopes. 
Individual colliding wind binary (CWB) systems are therefore regarded as candidate 
$\gamma$-ray sources for the satellite-based $\gamma$-ray missions AGILE and 
GLAST, and ground-based imaging atmospheric Cherenkov telescopes (MAGIC) or 
telescope arrays (CANGAROO, H.E.S.S., and VERITAS).

Analytical formulae for the steady-state proton- and electron particle spectra are 
derived assuming diffusive particle acceleration out of a pool of thermal wind particles, 
and taking into account adiabatic and all relevant radiative losses  (Inverse Compton scattering, 
synchrotron radiation, non-thermal bremsstrahlung, Coulomb losses and inelastic proton-proton collisions).
For the first time we include their advection/convection in the wind collision zone, 
and distinguish two regions within this extended region: the acceleration region
where spatial diffusion is superior to convective/advective motion, and the convection
region defined by the convection time shorter than the diffusion time scale.
We show that electron losses my well extend into the Klein-Nishina transition
regime, and find analytical approximations for the loss rate in this regime.

The calculation of the Inverse Compton radiation uses the full Klein-Nishina
cross section, and takes into account the anisotropic nature of the scattering process.
This leads to orbital flux variations by up to several orders of magnitude
which may, however, be blurred by the geometry of the system (eccentricity, inclination).
The maximum inverse Compton radiation occurs when the WR-star is
located behind the OB-star. Both, anisotropy and Klein-Nishina effects will likely yield 
characteristic spectral and variability signatures in the $\gamma$-ray domain.
Non-thermal bremsstrahlung emission is found to be mostly of minor relevance.
Propagation effects lead to a deficit of low-energy particles in the convection
zone, which -- if its size is sufficiently large -- may leave visible
imprints in the radiation spectra.
If protons are accelerated to at least several GeV, $\pi^0$-decay $\gamma$-rays contribute
to the high energy SED, and, depending on the injected electron-to-proton ratio,
might be visible with upcoming $\gamma$-ray instruments.
We show that pair production from photon-photon collisions can not be neglected 
in these systems in general, and might affect the emitted spectrum above $\sim 50$GeV
depending on orbital phase and system inclination.

The calculations are applied to the typical WR+OB-systems \object{WR 140} and \object{WR 147} to 
yield predictions of their expected spectral and temporal characteristica and to evaluate chances
to detect high-energy emission with the current and upcoming $\gamma$-ray experiments.
\end{abstract}


\section{Introduction}

Early type stars (O-, early B-, Wolf-Rayet (WR) stars) are hot stars
($T_{\rm eff} > 10000$~K) with masses $>20~M_\sun$.
They are known to possess some of the strongest sustained
winds among galactic objects. As a class Wolf-Rayet stars have the highest known mass
loss $\dot M \sim 10^{-4...-5}~M_\sun/$yr of any stellar type.
The supersonic winds of massive stars
may reach terminal velocities of $v_\infty >1000-5000$~km/s \citep{Cassinelli},
their kinetic energy $L_w = \dot M v_\infty^2/2$, however, rarely exceed 1\% of the
bolometric radiative energy output of typically $\sim 10^{38}$~erg/s.

In recent years massive stars have been connected to several high energy phenomena:
they are suspected to be the progenitor of some type of $\gamma$-ray bursts \citep{Wooseley,Paczynski}, 
and are known to be an interacting medium for the blast wave expelled from supernova explosions.
Triggered by the detections of several $\gamma$-ray sources by EGRET that are
not unambiguously identified yet but found in positional coincident with massive binary systems
\citep{Kau97,corr_analysis}, and motivated by the detection of synchrotron radiation from 
the collision region in some massive binaries, these systems have also been proposed as potential
sites of non-thermal high-energy photon production \citep{first,Eichler93,corr_analysis}.

Unlike in a wind of a single massive star where particles has been proposed to be accelerated
by either multiple weak shocks from line-driven instabilities \citep{Lucy,White85} to compensate for
the expansion and radiative losses even close to the stellar photosphere, or at the terminal shocks created
by the interaction with the swept up interstellar medium \citep{Casse,voelki},
the collision of supersonic winds
produces strong shocks where both electrons and protons can be efficiently accelerated to
high energies through first order Fermi acceleration \citep{Eichler93}.

In this latter scenario the shock region is exposed to both, a strong radiation field in the UV range
from the participating hot stars, and their magnetic fields. The detection of
synchrotron radiation from such collision regions, implied by the flat or negative spectral indices
and brightness temperatures of $10^{6...7}$~K,  proofs the existence of magnetic fields
as well as relativistic electrons in the collision region of some massive binary systems
\citep{Abbott86}.
In addition, observations of radio emission reveal an extended region of the synchrotron radiation \citep{Dougherty2000}.
In fact, in some binaries (e.g. \object{WR 147}, \object{WR 146}, \object{OB2~No.5}) the collision region 
between the main sequence
stars has been resolved in the radio band showing an extended, slightly elongated non-thermal feature
on VLA and MERLIN images \citep{Dougherty1996,Dougherty2000,Contreras97}
in addition to the free-free emission from the spherical wind of the stars.
Recently, the extended wind-wind collision region from WR~147 has even been detected in the X-ray band by
Chandra \citep{Pittard2002}.

Those electrons with energy $\gamma_e m_e c^2$
will inevitably also be responsible for a non-thermal high energy component at
$\sim \gamma_e^2 \epsilon_{\rm T}$ produced through inverse Compton scattering of the
dense stellar radiation field with characteristic energy $\epsilon_{\rm T}$. Interestingly, 
recent XMM and simultanous VLA observations
of the WR-star 9~Sgr might already lurk the hint of a non-thermal X-ray component. While
the hard X-ray component of 9~Sgr could be equally well fitted by either a hot multi-temperature
thermal model with $kT\geq 1.5$~keV, or a steep power-law with power-law index $\geq 2.9$, 
suggesting
a compression ratio $\leq 1.8$, the VLA spectrum at the time of the XMM-observations was
clearly non-thermal, thus indicating a similar compression ratio $\sim 1.7$ \citep{Rauw2002}.

The relativistic electrons will also loose some fraction of their energy due to non-thermal
bremsstrahlung in the field of ions that are embedded in the stellar winds.
It has been estimated, however, that the resulting non-thermal bremsstrahlung component
in the $\gamma$-ray domain at $\sim (\gamma_e/2) m_e c^2$ from electrons
of energy $\gamma_e m_e c^2$ will be a minor contribution to the overall emission
because of the ambient gas densities involved \citep{bene2003}.

If electron acceleration is also accompagnied with acceleration of protons out of the thermal pool
of wind material, then proton interactions with the ambient ions will produce $\gamma$-rays through
the hadronic neutral pion decay channel $p + p \longrightarrow \pi^0 + X$, $\pi^0 \longrightarrow \gamma + \gamma$.
Additionally, radiation from the secondary pairs, generated through the decay of
charged mesons that are produced by the hadronic $pp$-collisions, is expected to contribute
also to the overall broad band spectrum.
The $\pi^0$-decay radiative channel has already been considered in the past in the context of
winds from massive stars
\cite[\cite{ChenWhite91,White92};][for the case of single O-stars]{White85},
however, no propagation effects inside the extended collision region nor competing loss mechanisms
(such as expansion losses in the wind) has been taken into account here.

The goal of this work is to extend the model of non-thermal emission in the high-energy domain
from the colliding wind region of binary systems of massive stars to include all
relevant energy losses and simultanously also consider the propagation effects that affect
a relativistic particle distibution in such an environment.
After a brief discription of the geometrical model considered in this work (Sect.~2),
we evaluate the evolution of both, proton and primary electron spectra, on the basis
of a simplified diffusion-loss equation. In Sect.~3 we calculate the expected
photon emission due to the inverse Compton process (for the first time including Klein-Nishina and 
anisotropy effects),
relativistic bremsstrahlung and the $\gamma$-rays from the decay of $\pi^0$ that is produced
in hadronic proton-proton collisions. The total expected gamma-ray spectrum
is corrected for photon absorption in the UV radiation field of the massive stars.
We apply our model to the famous long-period binary system WR~140 and the above mentioned WR~147
in Sect.4. Our conclusions summarize our results and provide an outlook on the detectability
of colliding wind binary systems with upcoming instruments like GLAST, and contemporary Imaging Atmospheric Cherenkov 
Telescopes (IACTs).


\section{The geometric model of a colliding wind region}

The typical radial velocity profile $V(x)$ of winds from hot massive stars obeys
the relation
$$
V(x) = v_\infty (1-r_s/x)^\beta
$$
\citep{Windbuch}
with $\beta \simeq 0.8$, $r_s$ the radius of the star and $v_\infty$ the terminal velocity.
Recent observations indicate the existence of clumping in the wind \citep[e.g.][]{Tony96,Schild2004}.
For our schematic picture here we shall postpone the effects of clumping to later work and will consider
a homogeneous uniform wind. We also neglect any small-scale shocks within the
wind. The winds from binary systems flow nearly radially until they collide
to form a discontinuity at the location of ram pressure balance, and forward and reverse shocks. 
In the shock region the gas is heated
to temperatures of $10^{6...8}$~K \citep{Luo90,Stevens92,Usov92} which causes strong thermal X-ray
emission. Behind the shock the gas expands sideways from the wind collision region along the contact
surface out to larger $r$. It is therefore suggestive to distinguish two regions of the extended
emission site (see Fig.~1): In the acceleration zone $r \leq r_0$ (first-order) diffusive acceleration
provides high-energy particles out of a thermal pool. While the stellar winds prohibit the escape of energetic
particles on the upstream side of the shocks, we anticipate that in the downstream region
spatial diffusion is more efficient than convective/advective motion (which we call "convection" in the following)
in transporting particles to the boundary of the acceleration zone at $r=r_0$. Subsequent to their diffusive
escape from the acceleration zone, the energetic particles enter
the convection dominated zone $r>r_0$. The characteristic radius $r=r_0$ is defined by 
equality of the convection time scale
$t_{\rm conv}$ and the diffusion time scale $t_{\rm diff}$. 

\begin{figure}[t]
\resizebox{\hsize}{!}{\includegraphics{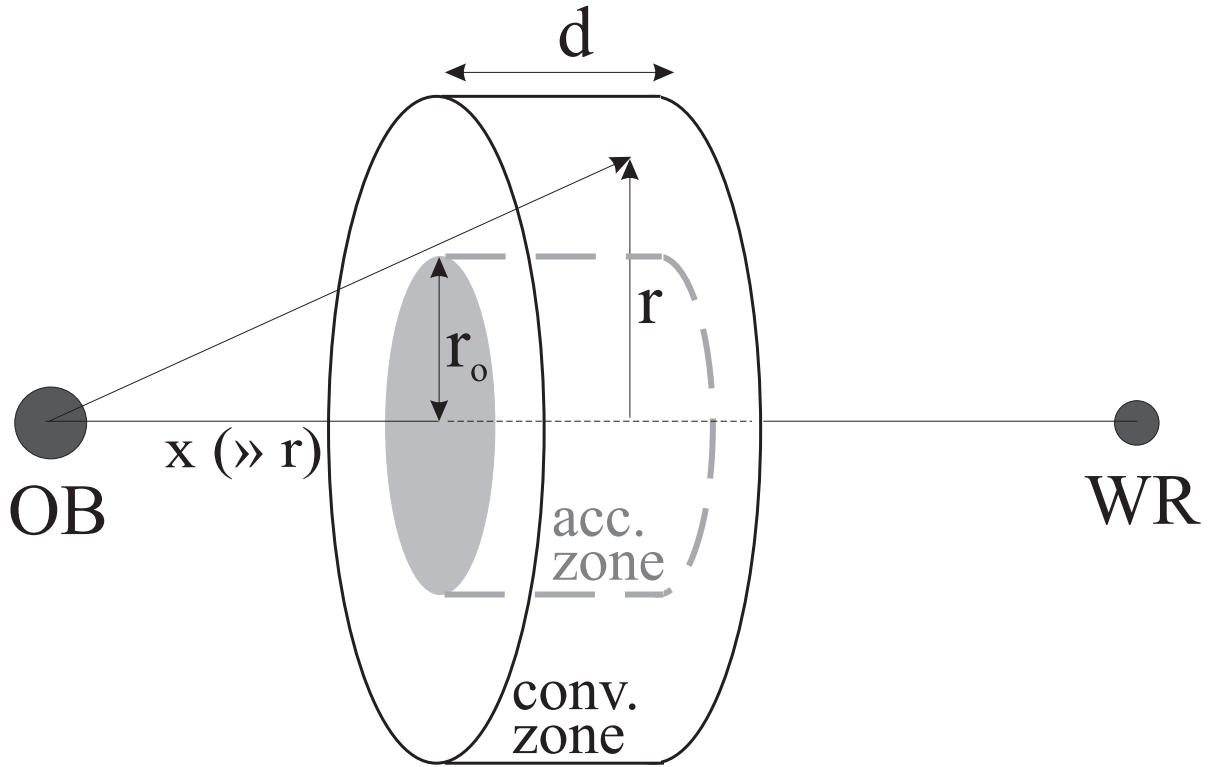}}
\caption{Model geometry for the collision of two stellar winds of a WR+OB binary system.
A strong shock is created in the wind collision region which is situated at a distance
$x=x_{\rm OB}$ from the OB-star. This region of thickness $d$ consists of an 
acceleration zone of size $r_0$ 
where suprathermal particles from the stellar wind are accelerated,
and an adjacent convection zone where particle streaming along the wind contact surface 
dominates over diffusion. For details see text.}
\label{Schema}
\end{figure}

Furthermore, we demand that
the distance of the emission region from the low-momentum-wind source is large compared
to the longitudinal extension of the emission region. While in reality the form of the
contact surface is bent towards the star that shows the lower wind momentum, observed as arcs of 
emission on radio images \citep[e.g.][]{Dough05},
we consider here the simplified geometry of a cylinder disk (see Fig.~\ref{Schema}). 
This may be justified by the rapid
particle energy loss rates that do not allow the transport of energetic particles to large distances
from their acceleration site. In other words, most particle energy losses are expected to
appear close to the acceleration zone where a cylinder geometry appears a reasonable approximation.
The thickness $d$ of the cylinder-like emission region is governed by the velocity of
the hitting winds. We shall further neglect here the interaction of the
stellar radiation fields on the wind structure \citep{Gayley,Stevens} which is justified for long-period binaries.

In the case of a collision between the spherical wind of a primary (e.g. a WR-star)
and a secondary companion (e.g. a OB-star) which has reached their
terminal velocities ($v_{\rm OB}$, $v_{\rm WR}$), the location of the shock is determined by the
balance of the ram pressure of both winds
\begin{equation}
\rho_{\rm OB} v^2_{\rm OB} = \rho_{\rm WR} v^2_{\rm WR}
\label{1}
\end{equation}
where $\rho_{\rm OB}$, $\rho_{\rm WR}$ are the densities of the gas ahead of and near the shock
of the stellar winds of the OB- and WR-stars.
In this case the distance of the shock front from the WR-star, $x_{\rm WR}$, and
from the OB-star, $x_{\rm OB}$, is given by
\begin{equation}
x_{\rm WR} = \frac{1}{1+\sqrt{\eta}}D,\hspace*{.8cm} x_{\rm OB} = \frac{\sqrt{\eta}}{1+\sqrt{\eta}}D
\end{equation}
where
\begin{equation}
\eta =  \frac{\dot M_{\rm OB} v_{\rm OB}}{ \dot M_{\rm WR} v_{\rm WR}}
\end{equation}
and $D=x_{\rm WR}+x_{\rm OB}$ is the separation of the binaries and $\dot M_{\rm OB}$, $\dot M_{\rm WR}$ are the mass loss
rates of the WR- and OB-star, respectively. Since $\dot M_{\rm OB} < \dot M_{\rm WR}$
and $v_{\rm OB} \approx v_{\rm WR}$, the shock location is rather close to the
OB-star, i.e. $D\gg x=x_{\rm OB}$. This appears to be in excellent agreement with
the observed locations of the collision region in the radio domain of e.g. \object{WR 147} 
\citep{Dougherty2000}, and this also implies that the shocked gas of the winds of both the
OB- and the WR-star have a comparable mass density.

If the wind collision occurs at a distance smaller than the Alfv\'en radius
$$
r_A \approx r_s \times (1+\xi) \hspace*{.5cm} \mbox{for} \hspace*{.3cm} \xi \ll 1,
$$
$$
r_A \approx r_s \times \xi^{1/4} \hspace*{.5cm} \mbox{for} \hspace*{.3cm} \xi \gg 1
$$
(where $\xi = B_s^2 r_s^2/(2 \dot M v_\infty)$, $B_s$ is the star's surface magnetic field)
from the OB-star, significant deceleration of
the WR-wind in the OB-radiation field is expected \citep{Eichler93}. In this case, the
colliding winds do not reach their terminal velocities.
Typically, $\xi\sim 0.01\ldots 0.1$ leading to
$r_A\sim 3\ldots 5 r_s$ for OB-stars \citep{Barlow82}.
For the present work we limit ourselves to situations, where the binary winds reach
their terminal velocities at the shock location.

The external magnetic field $B$ of a star changes in an outflow from the classical dipol field
$$
B \approx B_s \times (r_s/x)^3
$$
to a radially dominated one at the Alfv\'en radius
$$
B \approx B_s \times r_s^3/r_A x^2
$$
and finally to a toroidally dominated one for $x > r_s v_\infty/v_{\rm rot}$
$$
B \approx B_s \times v_{\rm rot} r_s^2/ (v_\infty r_A x)
$$
where $v_{\rm rot}$ is the surface rotation velocity with typical values of order $\sim 0.1 v_\infty$
for early type stars \citep{Conti,Penny}.
We use these equations to determine the magnetic field at the location of the collision front.
This value is held constant throughout the emission zone.

The surface magnetic field of massive stars are not well known.
\cite{Donati} report the detection of a 1.1~kG dipolar magnetic field of presumably fossil origin
at the surface of the young O star $\theta_1$ Ori~C.
\cite{Ignace98} argues for surface magnetic field strengths of order $10^4$~G in WR-stars.
On the other hand from the non-detection of the Zeeman effect for many O- and B-stars only upper limits of order 
a few $100$~G exist
\citep{Barker86,Mathys1999}.
For this work we shall fix the surface magnetic field at a reasonable value of $B_s = 100$~G unless
stated otherwise.

The plane of the binary system is inclined by an angle $i$ with respect
to the observer ($i=\pi/2$ corresponds to an observer in the plane of the stars), and $\phi_B$, the angle
between the projected sight line and the line connecting the stars
is a measure of the orbital phase of the system. In the following, periastron passage is defined
by the orbital phase $\Phi=0$, and $\phi_B=0$ for the WR-star being in front of the OB-star along the sight line.

In the co-rotating system centered on the OB-star the location of the emission site is 
defined in polar coordinates
by the azimuthal angle $\phi$ and the polar angle $\theta=\arcsin(r/x) \approx r/x$ for $r\ll x$
(see Fig.~1). In the same star-centered frame the line-of-sight to the observer is described 
by the angles $\phi_L$ with $\tan\phi_L=\sin\phi_B\,\cot i$, and
$\theta_L$ with $\cos{\theta_L}=\cos{\phi_B}\,\sin i$.

The radiation yield of inverse Compton scattering depends on the angle 
$\theta_{\rm ph}=\theta_{\rm sc}$ between the directions of
the incoming (from the OB-star) and outgoing photons, which obviously depends on the location of the 
scattering electron as well as the orbital phase. We find
\begin{displaymath}
\mu_{\rm ph} =\cos\theta_{\rm ph}= 
\cos{\theta_L}\,\cos{\theta} + \sin{\theta_L}\, \sin{\theta}\, \cos(\phi-\phi_L)
\end{displaymath}

The corresponding azimuthal angle $\phi_{\rm ph}$ is irrelevant for the scattering process.


\section{Particle spectra}

We expect two standing shocks and a discontinuity between them. For typical
massive stars the wind velocities are of comparable value and in addition the ram-pressure balance
(Eq.~\ref{1})
ensures a similar upstream gas density for both the OB- and the WR-wind shocks.
It thus appears justified as assume the two shocks as well as the corresponding
acceleration and escape rates to be identical.

As motivated above we distinguish two zones of the extended emission region.
In the acceleration zone suprathermal particles of energy $E_0$ from the stellar winds are energized
through diffusive shock acceleration
at a rate $\dot E = a E$ where $a=V_s^2(c_r-1)/(3 c_r\kappa_a)$ \citep{Schlickibuch} with
$V_s=v_{\rm OB}=v_{\rm WR}$ is the upstream velocity of the standing shock, $c_r$ the compression ratio
and $\kappa_a$ the (assumed) energy-independent diffusion coefficient
perpendicular to the wind contact surface. In this region diffusion dominates over
the convective flow along the wind contact surface. The acceleration region can in good
approximation be treated as a leaky box with a free escape boundary at $r_0$ (see Fig.~1).
Introducing an energy-independent escape time $T_0=r_0^2/(4\kappa_d)$, where
$\kappa_d$ is the diffusion coefficient along the wind contact surface, 
the continuity equation
for this zone then reads:
\begin{equation}
\frac{\partial}{\partial E}(\dot E\, N(E)) + \frac{N(E)}{T_0} = Q_0\, \delta(E-E_0)
\label{PDE1}
\end{equation}
where $\dot E = aE-\dot E_{\rm loss}$ includes energy gain through diffusive acceleration as well
as continuous energy losses. 
The solution $N(E)$ consists of a power law that is modified
at the high-energy end of the spectrum: $N(E) \propto E^{-s} f_c(E)$ where $f_c(E)$ depends on the radiative
energy losses employed, and $s=(a T_0)^{-1}+1$ where energy losses
are negligible.

Adjacent to the acceleration zone is the so-called convective zone, where convection along the
wind contact surface is a
faster transport process than is diffusion. For the convective flow we
assume for simplicity a constant velocity, $V = p\,v_{\rm OB}$ with 
$p={\rm const}\leq 1$. The continuity equation for this zone
also includes adiabatic losses, $\propto E\vec\nabla\,\vec V$, and is given by
\begin{equation}
\vec\nabla(\vec V\,\,N(E,r))+\frac{\partial}{\partial E}
\big[(\dot E - \frac{E}{3}\vec\nabla\,\vec V)\,\,N(E,r)\big]=0
\label{PDE2}
\end{equation}
where $\dot E=-\dot E_{\rm loss}$ represents the continuous energy losses in this region.
The boundary conditions $N(r\rightarrow\infty)=0$ and $N(r=r_0, E)=N_{\rm acc}(E)$ apply,
where $N_{\rm acc}(E)$ is the
homogeneous particle density in the acceleration region.

At the location $r=r_0$ the diffusive escape time scale $T_0 = r_0^2/4\kappa_d$ 
equals the convection time scale
$t_{\rm conv} = r_0/V$, and this allows to determine $r_0=4\kappa_d/V$. For the power-law index
$s$ one then finds 
\begin{equation}
s = 1+ p^2\, {{3\,c_r}\over {4\,(c_r-1)}}\,{{\kappa_a}\over {\kappa_d}}\,. 
\end{equation}
It is remarkable that for isotropic diffusion,
$\kappa_a = \kappa_d$, $s$ is fully determined by the compression ratio
$c_r$ of the shock and the ratio of the convection velocity to the shock velocity. 
For strong shocks, $c_r=4$, hard particle spectra with $s = 1+p^2$
are then expected in the acceleration zone while the generic $s=2$-spectrum requires
non-isotropic and/or energy-dependent diffusion, or an extreme value
for the convection velocity $V=v_{\rm OB}$.
The smallest possible size $r_0$ of the acceleration region corresponds to 
the diffusion coefficient $\kappa_d$ approaching the Bohm-limit.

\subsection{Electron spectra}

The general solution of Eq.~\ref{PDE1} is given by
\begin{equation}
N(E) = \frac{1}{\dot E} \int^E dE' Q(E') \exp\left[-\frac{1}{T_0}\int_{E'}^{E}\frac{dE''}{\dot E(E'')}\right] 
\label{acc_solution}
\end{equation}
It has been shown that inverse Compton scattering, if treated in the Thomson regime, in most 
cases determines the maximum electron energy
and is the most important radiative loss channel for ultrarelativistic electrons in
colliding winds of massive stars \citep{Eichler93,Muecke}, since typically $u_{\rm ph,T}/u_{\rm B} \geq 1$
($u_{\rm ph,T}$ is the energy density of the stellar radiation field, $u_{\rm B}$ the magnetic field
energy density in the emission region). For typical system parameters one finds
$u_{\rm ph,T}/u_{\rm B} \approx 67 L_{\rm bol,38}/B_G^2 x_{\rm OB,13}^2$ where $L_{\rm bol,38}$ is the bolometric
luminosity of the OB-star in $10^{38}$~erg/s, $B_G$ is the magnetic field in the collision region in Gauss
and $x_{\rm OB,13} = x_{\rm OB}/(10^{13}$~cm). At lower energies bremsstrahlung and Coulomb
scattering determine
the shape of the electron spectrum. Radiative losses included in our calculations for the
electron spectra are synchrotron losses, inverse Compton losses on the stellar radiation field of
differential photon density $n_{\rm ph,T}(\epsilon)=n_0\delta(\epsilon-\epsilon_T)$ (mono-chromatic approximation), 
electron-ion bremsstrahlung and Coulomb losses.
The total radiative energy loss rate is then given by
\begin{equation}
\dot E = -(b_{\rm syn}+b_{\rm IC,TL})E^2+(a-b_{\rm br})E-b_{\rm coul}
\label{elec_radlosses}
\end{equation}
with 
$$
b_{\rm IC,TL} = \frac{4}{3 m_e^2 c^3}\,\sigma_T\, u_{\rm ph}
$$
for the Thomson regime with $c$ the velocity of light and $\sigma_T=6.65\cdot 10^{-25}$cm$^2$ the
Thomson cross section,
$$
b_{\rm syn} = \frac{4}{3 m_e^2 c^3}\,\sigma_T\, u_B,
$$
$$
b_{\rm br} = \frac{2}{\pi}\,\alpha\, \sigma_T\,c\, N_H
$$
in the weak-shielding limit where we have neglected the logarithmic term, and where $\alpha$ is the fine structure constant,
$m_e$ the electron mass and $N_H$ the thermal ion density (in cm$^{-3}$), and \citep{Schlickibuch}
$$
b_{\rm coul} = 55.725\, c\,\sigma_T\, N_H\, m_e\, c^2.
$$

\begin{figure}[t]
\includegraphics[height=13cm]{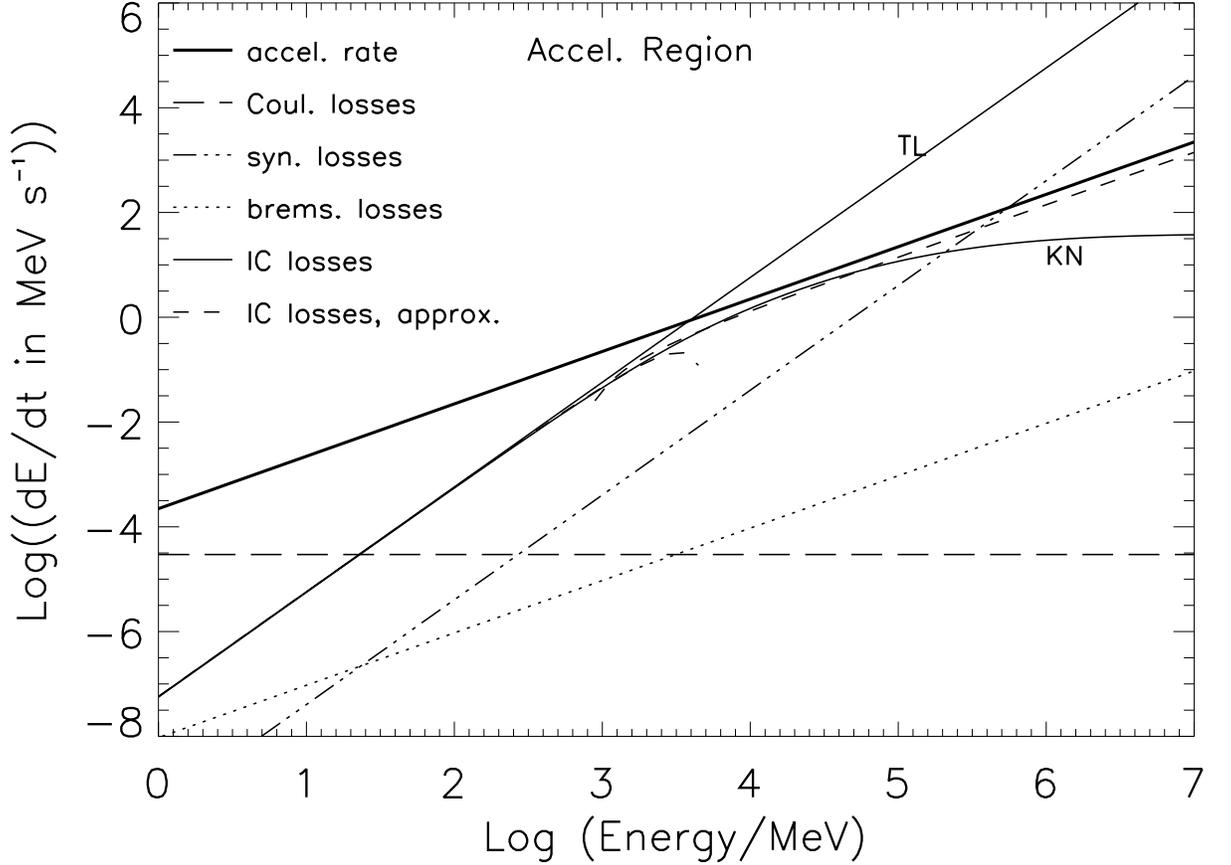}
\caption{Energy loss rates due to inverse Compton scattering (thin solid lines) in the Thomson regime (TL) and
Klein-Nishina regime (KN), synchrotron radiation (dashed-triple-dotted line), relativistic electron-ion
bremsstrahlung (dotted line) and Coulomb interactions (dashed line) in comparison to the acceleration rate
(thick solid line) in the acceleration zone. Parameters are: $L_{\rm bol,OB}=10^5 L_{\sun}$,
$\epsilon_T=10$~eV, $\dot M_{\rm OB}=10^{-6} M_{\sun}$yr$^{-1}$, $\dot M_{\rm WR}=10\dot M_{\rm OB}$,
$v_{\rm OB}=3768$km/s,
$D= 10^{14}$cm, $B_s = 100$G, $x_{\rm OB}\approx 0.24D$, $B\approx 0.5$G,
$n_{\rm ph,T}\approx 10^{11}$cm$^{-3}$, $N_H\approx 5\cdot 10^7$cm$^{-3}$,
$\kappa_a = 1.6\cdot 10^{20}$cm$^2$s$^{-1}$, $V\approx 1884$~km/s, $T_0\approx 1127$~s,
$r_0\approx 8.5\cdot 10^{11}$cm.
}
\label{acc_loss}
\end{figure}

Fig.~\ref{acc_loss} shows the energy loss rates in comparison to the acceleration rate for
a set of parameters typically found for colliding wind binaries. Obviously Coulomb losses
dominate the low energy end of the electron spectrum, and simultanously provide a lower limit
to the acceleration rate $a$.
At energies $E\ga m_e^2 c^4/\epsilon_T$ Klein-Nishina effects significantly 
modify the energy loss rate due to inverse Compton scattering. 
Fig.~\ref{acc_loss} demonstrates that firstly, Klein-Nishina
effects can in general not be neglected in the environment of colliding massive star winds. Secondly
the curvature of the loss rate
due to the Klein-Nishina decline of the inverse Compton cross section actually starts already at much
lower electron energies.
Consequently, in some cases the electron spectrum might be rather limited by synchrotron losses if the wind parameters
are favourable. 
In the following, we take into account Klein-Nishina effects already above $E>E_{\rm TL}$ where $E_{\rm TL}$
is considered the energy below which the Klein-Nishina energy loss deviates not more than 20-25\% from the
Thomson limit approximation. Typically, for the systems considered in this work $E_{\rm TL}\approx 10^{-2}\mbox{MeV}/\epsilon_{\rm T,MeV}$ with $\epsilon_{T,MeV}$ the target photon energy in MeV.

In the extreme Klein-Nishina regime the total electron energy losses in the acceleration zone 
at high energies are
effectively only provided by synchrotron radiation. In the transition range between the Thomson 
and Klein-Nishina limit, however, inverse Compton scattering may still dominate over synchrotron emission, thus mandating a proper treatment of the former.
To allow for analytical solutions of Eq.~\ref{PDE1} we
approximate the Klein-Nishina decline of the loss rate in the transition range by
\begin{eqnarray}
\dot E_{\rm IC,KN} & = & - b_{\rm IC,TL} E^2 (1-E\epsilon_{\rm T,MeV}/5\cdot 10^{-2}\mbox{MeV})
\label{KN1}
\end{eqnarray}
for all energies $E<E_s=10^{-1.7\ldots -1.6}$MeV$/\epsilon_{\rm T,MeV}$, and
\begin{eqnarray}
\dot E_{\rm IC,KN} \simeq - (0.27\sigma_T c n_{\rm ph,T} \epsilon_{\rm T,MeV}^{1/8} E+7\epsilon_{\rm T,eV}^{-1.846}*D_{14}^{-2}+2.1\cdot 10^{-4})\, \mbox{MeV/s}\\
                    = -(q_a E+q_b)
\label{KN2}
\end{eqnarray}
for $E\geq E_{s}$ and with $n_{\rm ph,T}$ the integrated target photon density and $D_{14}$ the binary 
separation in $10^{14}$cm.
These approximations are tested for $\epsilon_{\rm T}=1\ldots 100$eV, and are suitable 
for $0.01 u_{\rm ph,T} \leq u_{\rm B} \leq u_{\rm ph,T}$ (while they deviate by more than an order of magnitude
at energy $q_a/(2b_{\rm syn})\,[(1+\sqrt{(1-4b_{\rm syn}q_b/q_a^2)}]$ if $u_{\rm B}$ decreases to $\leq 0.003 u_{\rm ph,T}$).

This approximation is indicated in Fig.~\ref{acc_loss} by the long-dashed line and is suitable for
cases where relativistic bremsstrahlung losses are much smaller than synchrotron or inverse Compton losses.
While this approach takes reasonably into account the early deviation of the Klein-Nishina cross section
from the Thomson cross section,
it overestimates the energy loss rate in the extreme Klein-Nishina regime, which
is mainly responsible for the steepness of the declining electron spectrum.
Consequently, our derived electron and photon spectra shall be regarded with caution in their declining part.

\begin{figure}[t]
\includegraphics[height=13cm]{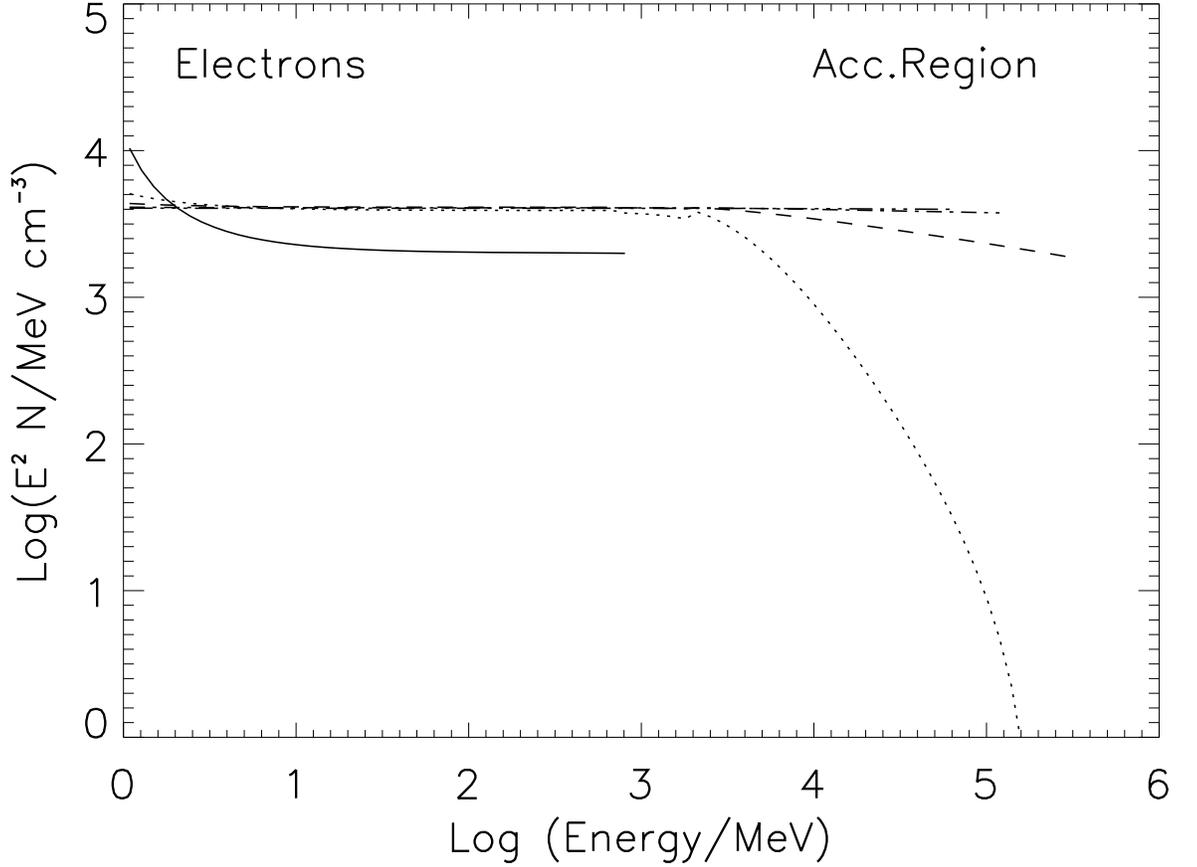}
\caption{Steady-state electron spectrum in the acceleration region for $D=5\cdot 10^{13}$cm (solid line),
$10^{14}$cm (dotted line), $2\cdot 10^{14}$cm (dashed line), $5\cdot 10^{14}$cm (dashed-dotted line) 
and $10^{15}$cm (dashed-tripple dotted line). Parameters are: $Q_0=1\mbox{cm}^{-3} \mbox{s}^{-1}$,
$B\approx 1.3, 0.5, 0.3 0.1 \mbox{ and } 0.05$~G,
$n_{\rm ph,T}\approx (44, 11, 2.8, 0.4 \mbox{ and } 0.1)\times 10^{10}$cm$^{-3}$, $N_H\approx (21, 5.2, 1.3, 
0.2 \mbox{ and } 0.05)\times 10^7$cm$^{-3}$,
$v_{\rm OB}=3537, 3768, 3884, 3954 \mbox{ and } 3977$km/s, $V\approx 1768, 1884, 1942, 1977 \mbox{ and } 1988$~km/s,
$r_0\approx (9, 8.5, 8.2, 8.1 \mbox{ and } 8\times 10^{11}$cm.
All other parameters are the same as used in Fig.~\ref{acc_loss}.
}
\label{elec_spec}
\end{figure}
\begin{figure}
\includegraphics[height=13cm]{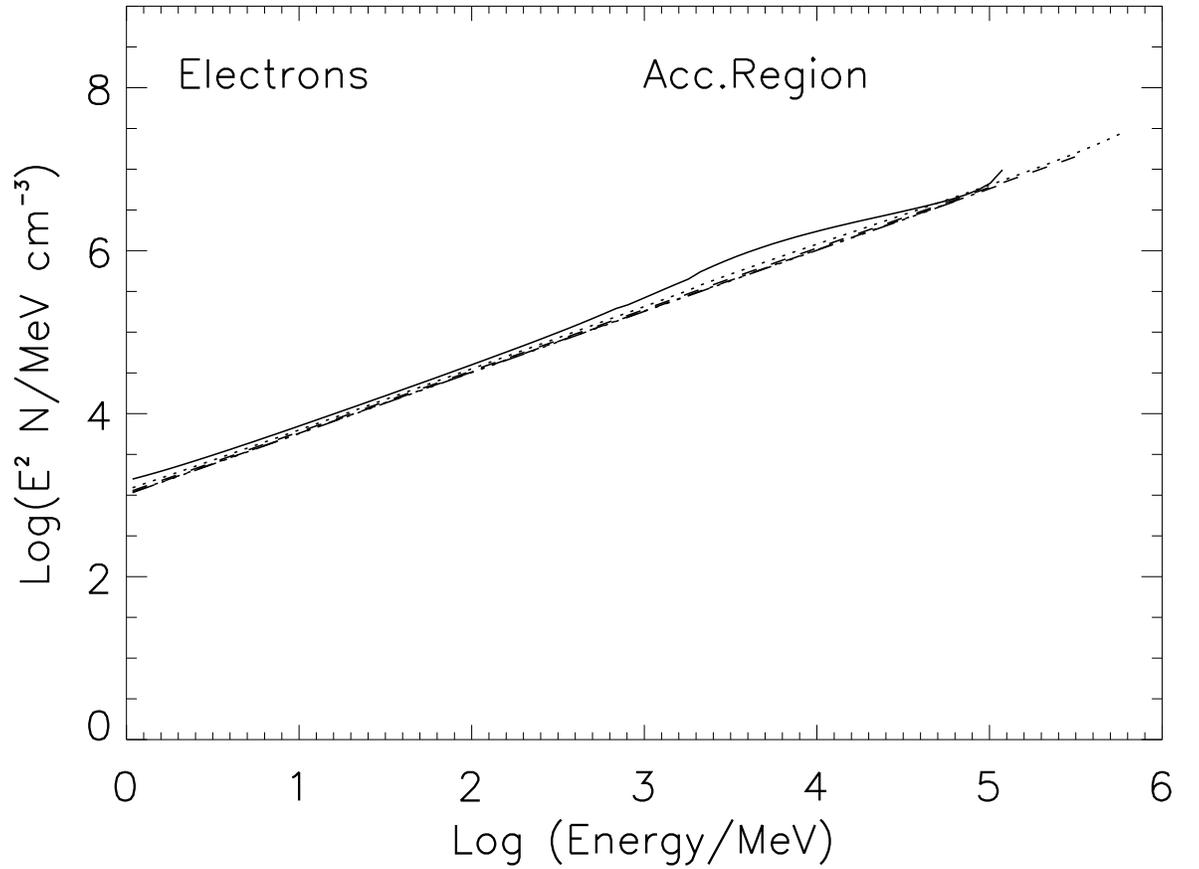}
\caption{Steady-state electron spectrum in the acceleration region assuming 
$\kappa_a = \kappa_d=4\cdot 10^{19}$cm$^2$s$^{-1}$
and for $D=5\cdot 10^{13}$cm (solid line),
$10^{14}$cm (dotted line), $2\cdot 10^{14}$cm (dashed line), $5\cdot 10^{14}$cm (dashed-dotted line) 
and $10^{15}$cm (dashed-tripple dotted line). 
All other parameters are the same as used in Fig.~\ref{elec_spec}.}
\label{elec_spec_iso}
\end{figure}

The analytical solutions of Eq.~\ref{PDE1} for electrons in the acceleration region are detailed in
App.~\ref{app_e}.
Fig.~\ref{elec_spec} shows examples of the resulting electron spectra for various
values of the binary separation,
while all other parameters are the same as used in Fig.~\ref{acc_loss}.
The general shape is determined by the interplay between acceleration gains and losses. At low energies
Coulomb losses dominate and lead to the well-known upturn towards low energies due to
the stronger increase of the acceleration rate with energy $\dot E_{\rm acc} \propto E$ when comparing
to the competing radiative losses $\dot E_{\rm loss}\propto const$. For a reasonable convection
velocity $V=1/2 v_{\rm OB}$ \citep[e.g.][]{Luo90} and assuming strong shocks $c_r=4$, a $\propto E^{-2}$ particle
spectrum develops in the acceleration region if $\kappa_a/\kappa_d=4$, which we shall use in the following
for demonstrative purposes if not noted otherwise.
Thus in Fig.~\ref{elec_spec} a $E^{-2}$-spectrum develops until
inverse Compton losses cause a steepening. The shape of this decline reflects the approximation
employed to simulate the losses in the Klein-Nishina regime. 
The kink at $\sim 10^{3.5}$~MeV corresponds to the transition from the Thomson regime to the
Klein-Nishina loss rate approximation. Finally, the cutoff at energy $E_c$ can be either due to
radiative losses dominating over the acceleration rate, or the diffusion coefficient $\kappa_d$
approaching the Bohm limit $\kappa_{\rm Bohm}\approx 1/3 r_L c$ where $r_L$ is the Larmor radius.
In the latter case the escape time scale $T_0$ decreases with energy. 
For simplicity (and in order to keep the solutions analytical)
we set $T_0=0$ at $\kappa_d \leq \kappa_{\rm Bohm}$ which then causes a sharp cutoff.
More sophisticated calculations may smooth this decline towards an approximately exponential shape 
\citep{cutoff}.
In the former case the cutoff
is determined by the synchrotron loss rate that dominates regarding the flattening loss curve at high energies.
In Fig.~\ref{elec_spec} the particle spectra cut off due to radiative losses for $D\leq 10^{14}$cm
while at larger binary separations the Bohm limit causes the cutoff. 
Here both, synchrotron and inverse Compton losses, cease to be able to significantly affect the
acceleration spectrum. 
For comparison we have also calculated electron spectra assuming isotropic diffusion.
Fig.~\ref{elec_spec_iso} shows the resulting spectra where all parameters are unchanged
with respect to Fig.~\ref{elec_spec} except for the ratio of the diffusion coefficients $\kappa_a/\kappa_d$. 
As expected hard particle spectra with spectral index $s=1+p^2=1.25$ develop, and radiative losses
alter the spectral shape depending on the binary separation. For $E\rightarrow E_{c}$
the spectra diverge for $((a-b_{\rm br})*T_0)^{-1}<1$ (see Eq.~\ref{elec_solution1}, \ref{elec_solution2}), 
which is fullfilled here.

\begin{figure}[t]
\includegraphics[height=13cm]{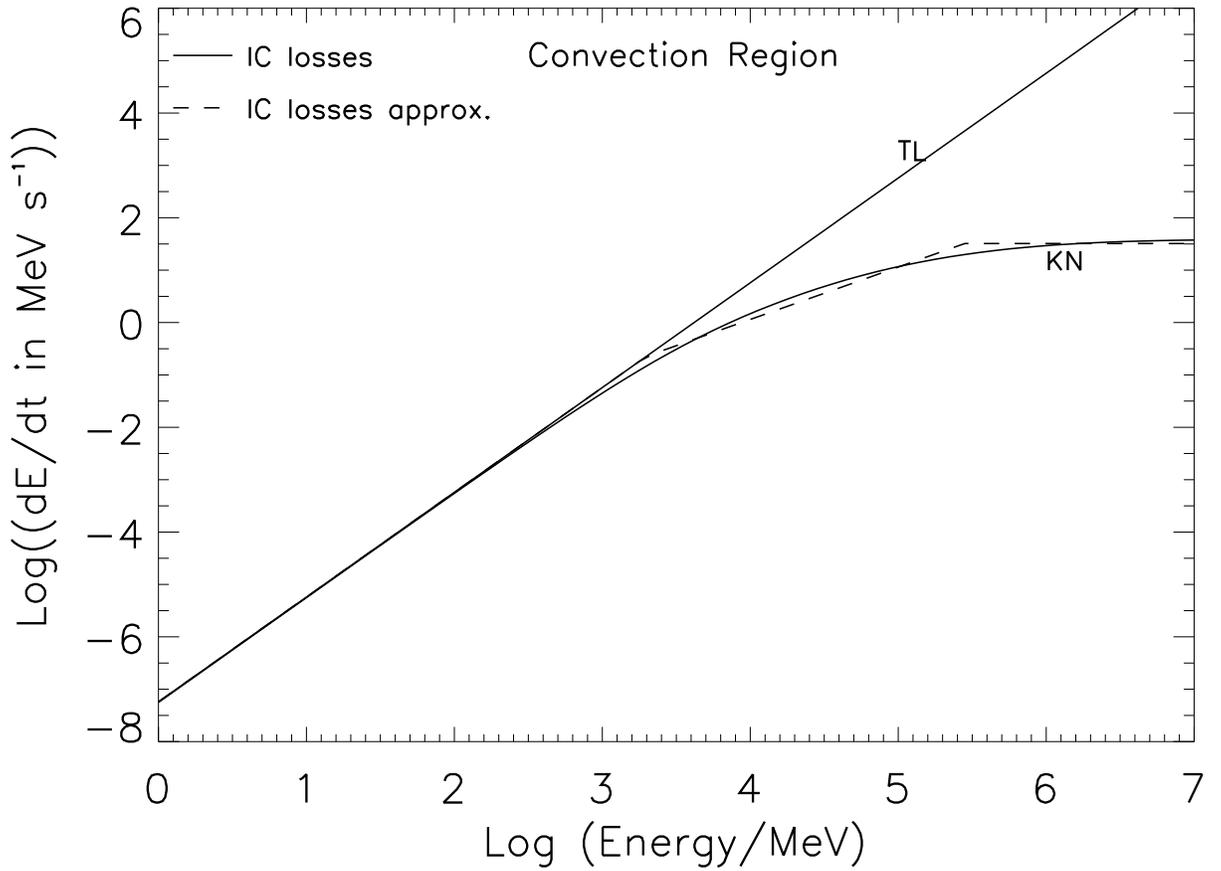}
\caption{
Energy loss rates in the convection zone due to inverse Compton scattering in the Thomson regime (TL,
thin solid lines) and the Klein-Nishina regime (KN). The parameters are the same as in Fig.~\ref{acc_loss}.
}
\label{conv_loss}
\end{figure}

Fig.~\ref{conv_loss} shows the energy loss rates in the convection zone for the same parameters as used for
Fig.~\ref{acc_loss}. On account of expansion in the cylindrical convection flow
the magnetic field strength and the gas density fall off quickly with distance from the acceleration
region. Coulomb interactions, bremsstrahlung, and synchrotron radiation will therefore loose
importance as energy loss mechanisms and the inverse Compton scattering effectively provides the bulk of
the electron energy losses.

Using a constant convection velocity the steady-state particle spectrum in the convection 
region can be found by solving
Eq.~\ref{PDE2}. 
With the boundary conditions $N(E,r)\rightarrow 0$ for $r\rightarrow \infty$ 
and $N(E,r)\rightarrow N_{\rm acc}(E)$ 
for $r\rightarrow r_0$ where $N_{\rm acc}(E)$ is the steady-state particle spectrum in the 
acceleration region we derive the following analytical solutions using the standard method of characteristics:
For $E<E_{\rm TL}$ the radiative losses are $\dot E = - b_{\rm syn\& IC}\,E^2$ and
\begin{equation}
N(E) = \left(\frac{r_0}{r}\right)^{2/3} 
\left[1+\frac{3b_{\rm syn\& IC}}{2V}Er\left(\left(\frac{r_0}{r}\right)^{2/3}-1\right)\right]^
{-2} 
N_{\rm acc}(\tilde E)
\end{equation}
with 
\begin{equation}
\tilde E = E \left(\frac{r}{r_0}\right)^{1/3}\left[1+\frac{3b_{\rm syn\& IC}}{2V}Er
\left(\left(\frac{r_0}{r}\right)^
{2/3}-1\right)\right]^{-1} 
\end{equation}
For $E\ge E_{\rm TL}$ an analytical solution can not be found using Eq.~\ref{KN1}.
In the convection region the approximation for the Inverse Compton energy losses
in the Klein-Nishina regime, $E > E_{TL}$,
\begin{equation}
\dot E_{\rm IC,KN} = -q_c E
\label{KN3}
\end{equation}
for $E_{\rm TL} < E < E_{\rm extr}$ with $q_c \approx b_{\rm IC}E_{\rm TL}$, and
\begin{equation}
\dot E_{\rm IC,KN} = -q_d
\label{KN4}
\end{equation}
for $E \geq E_{\rm extr}=q_d/q_c$ with $q_d \approx 9/16 c \sigma_T m_e^2 c^4 n_{\rm ph,T} \epsilon_T^{-1}$,
turns out reasonable if $E_{\rm TL}$ is increased to 
$E_{\rm TL}\approx 10^{-1.7}\mbox{MeV}/\epsilon_{\rm T,MeV}$
(see Fig.~\ref{conv_loss}).
In this case the solution of Eq.~\ref{PDE2} is
\begin{equation}
N(E) = \exp[\frac{q_c}{V}(r-r_0)] \left(\frac{r_0}{r}\right)^{2/3} N_{\rm acc}(\tilde E)
\end{equation}
with 
\begin{equation}
\tilde E = E \exp[\frac{q_c}{V}(r-r_0)] \left(\frac{r}{r_0}\right)^{1/3}
\end{equation}
for $E_{\rm TL} < E < E_{\rm ext}$, and
\begin{equation}
N(E) = \left(\frac{r_0}{r}\right)^{2/3} N_{\rm acc}(\tilde E)
\end{equation}
with 
\begin{equation}
\tilde E = E  \left(\frac{r}{r_0}\right)^{1/3} + \frac{3q_d}{4V}\left[\left(\frac{r}{r_0}\right)^{1/3}r-r_0\right]
\end{equation}
for $E \geq E_{\rm ext}$.

The total solution is then combined such to assure continuity for all functions $E(r)$, 
$N(E,r,r_0,N_{\rm acc}(\tilde E))$.

\begin{figure}[t]
\includegraphics[height=13cm]{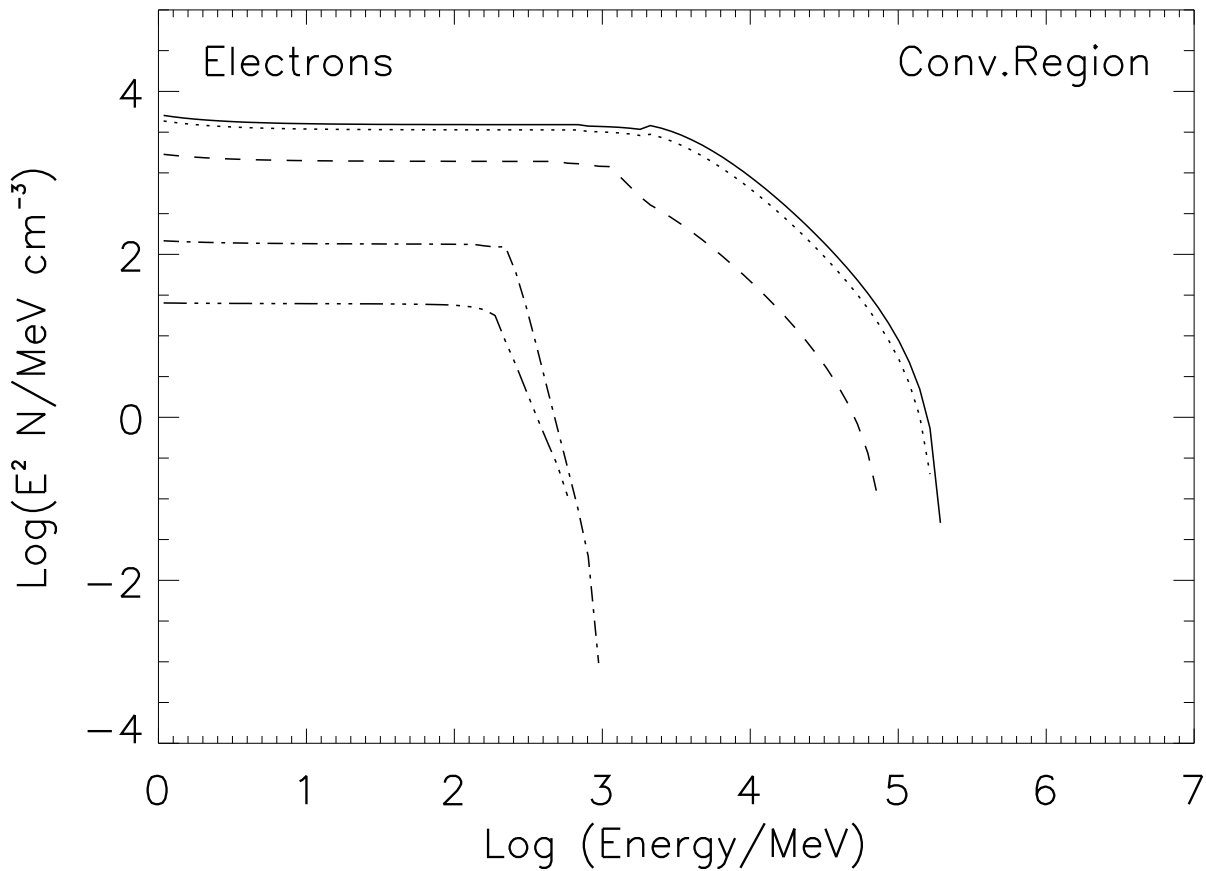}
\caption{Steady-state electron spectra in the convection region at distance 
$\Delta r=10^{11}$cm (dotted line), $10^{12}$cm (dashed line), $10^{13}$cm (dashed-dotted line) and 
$5\cdot 10^{13}$cm (dashed-triple dotted line) from the transition point $r_0$ as compared 
to the electron spectrum in the acceleration region (solid line)
and for a binary separation of $D=10^{14}$cm. All other parameters are the same
as in Fig.~\ref{acc_loss}.
}
\label{elec_specconv1}
\end{figure}

\begin{figure}[t]
\includegraphics[height=13cm]{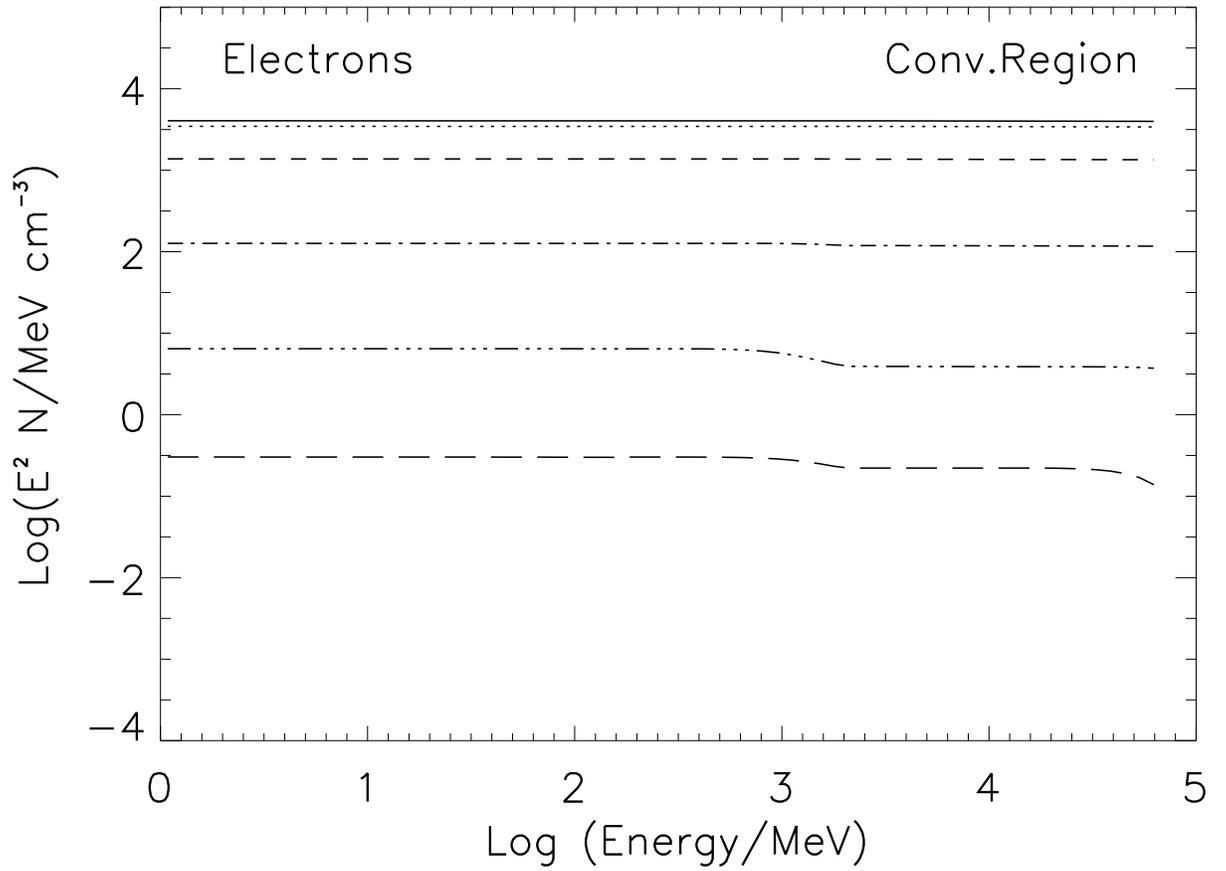}
\caption{Same as in  Fig.~\ref{elec_specconv1} but for a binary separation of $D=10^{15}$cm
and $\Delta r=10^{11}$cm (dotted line), $10^{12}$cm (short dashed line), $10^{13}$cm (dashed-dotted line), 
$10^{14}$cm (dashed-triple dotted line) and $10^{15}$cm (long dashed line).}
\label{elec_specconv2}
\end{figure}

Fig.~\ref{elec_specconv1} shows the resulting electron spectrum in the convection region 
at $r=r_0\ldots 10^{14}$cm with a binary separation
of $D=10^{14}$cm and using the parameters as described in Fig.~\ref{acc_loss} for the 
acceleration region. As the particles
convect along $r$ radiative losses alter the high energy end of the particle spectrum 
leading to a decreasing cutoff energy with increasing $r$ 
while adiabatic losses lower the overall particle density. In Fig.~\ref{elec_specconv2} the 
binary separation has been enlarged to
$D=10^{15}$cm. As a result radiative losses show a significant impact on the spectral 
shape only at large $r$. 

In summary, taking into account convection in the extended emission region alters the energy cutoff 
in the integrated particle spectrum if radiative losses prove to be important, and 
simultanously lowers the total non-thermal particle density.
This behaviour is reflected in the corresponding photon spectra (see Sect.~4).

\subsection{Proton spectra}

If protons are accelerated together with the electrons, hadronic nucleon-nucleon interaction 
take place with an approximated energy loss rate of
\begin{equation}
\dot E = -b_{\rm pp} E
\end{equation}
above the kinematic threshold for pion production $E>E_{\rm thr} \simeq 0.28$~GeV \citep{Mannheim94},
\begin{equation}
b_{\rm pp} = 1.3\times 3cN_H \sigma_{\rm pp}m_\pi/m_p \,
\label{bpp}
\end{equation}
where $\sigma_{\rm pp}=3\cdot 10^{-26}$cm$^2$ is the corresponding hadronic cross section,
and the factor 1.3 takes into account the here assumed metallicity (90\% H, 10\% He).
Note that this linear relation for the energy loss rate is exact only
above a nucleon kinetic energy of a few GeV, whereas at lower energies it shall be considered as
an approximation. The corresponding error drops below $\sim 30\%$ above $\sim 10$~GeV.  
In addition,
Coulomb-losses in the dense wind of the massive companion star of the WR-star will alter the injection spectrum.
The Coulomb-losses are \citep{Mannheim94}
\begin{equation}
\dot E = -{{3\,c\,\sigma_T\,m_e c^2\,Z^2\,\ln \lambda}\over 2}\,N_H\,
{{\beta^2}\over {x_m^3 + \beta^3}}
\end{equation}
with
$$
\beta={\sqrt{E\,(E+2m_p c^2)} \over {E+m_p c^2}}
$$
and
$$x_m = 0.2\,\sqrt{{T_e}\over {10^8\ {\rm K}}}\,.$$ 
The Coulomb-barrier occurs at $E_m \simeq 20\mbox{MeV}(T_e/10^8 {\rm K})$ where
$T_e\approx 10^8$~K is the electron temperature. We approximate the Coulomb-losses by
\begin{equation}
\dot E = -b_{\rm bel} E_{\rm MeV}
\end{equation}
below the Coulomb-barrier with
$b_{\rm bel} = 6.6\cdot 10^{-16} Z^2 N_H/x_m^3$~MeV/s, $E_{\rm MeV}$ is the particle kinetic energy in MeV,
and 
\begin{equation}
\dot E = \frac{-b_{\rm ab}}{\sqrt{E_{\rm MeV}}}
\end{equation}
above the Coulomb-barrier with 
$b_{\rm ab} = 6.7\cdot 10^{-12} Z^2 N_H$~MeV/s.
This approximation deviates at most a factor $\sim 2$ (by approaching the Coulomb-barrier) from the exact loss formula. 
In the relativistic regime, $\beta\geq 0.5$, we use
\begin{equation}
\dot E = -b_{\rm rel}
\end{equation}
with $b_{\rm rel} = 3.1\cdot 10^{-13} Z^2 N_H$~MeV/s.

Eq.~\ref{PDE1} describes the behaviour of the steady-state spectrum in the 
acceleration region. The solution, Eq.~\ref{acc_solution},
can again be derived analytically (see App.~\ref{app_p}). 

\begin{figure}[t]
\includegraphics[height=13cm]{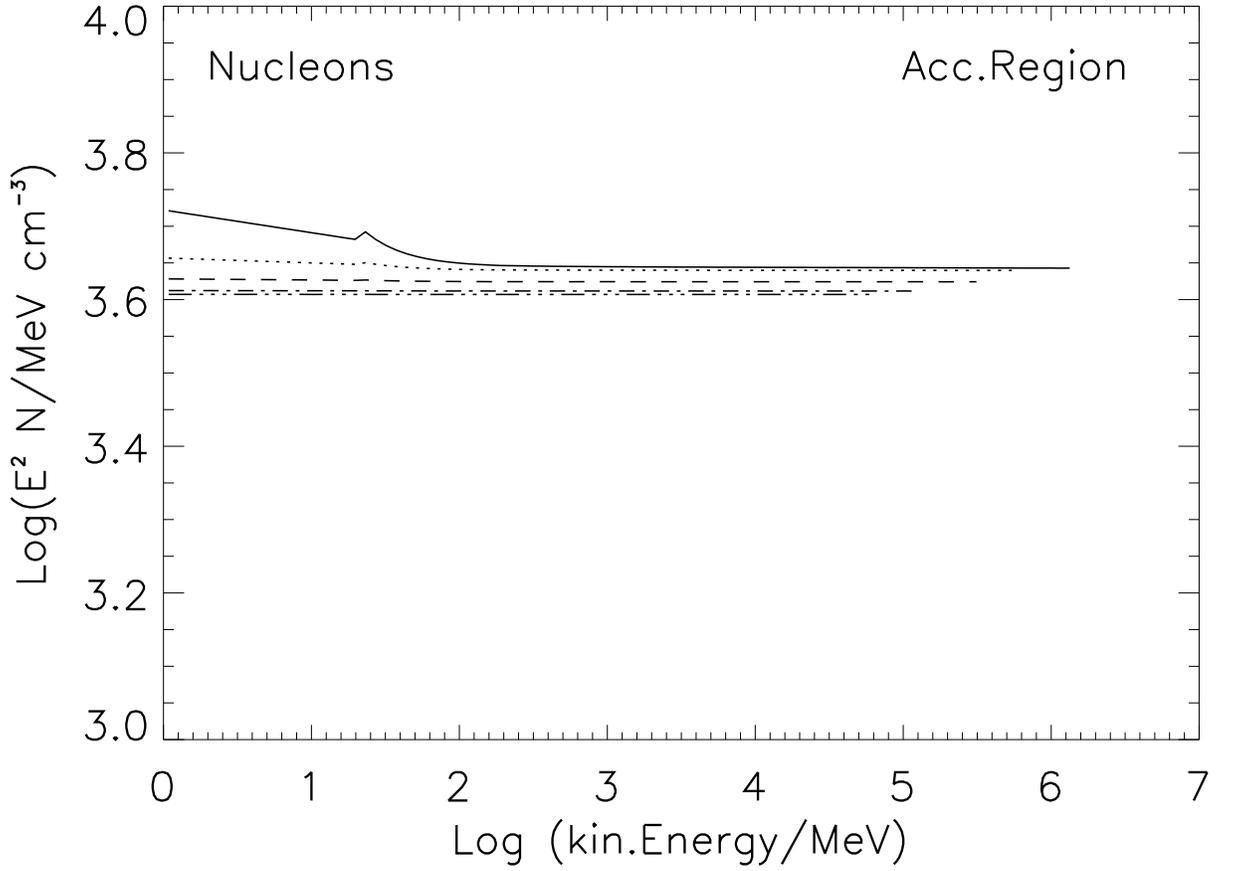}
\caption{Steady-state proton spectra in the acceleration region for binary separations
$D=5\cdot 10^{13}$cm (solid line),
$10^{14}$cm (dotted line), $2\cdot 10^{14}$cm (dashed line), $5\cdot 10^{14}$cm 
(dashed-dotted line) 
and $10^{15}$cm (dashed-tripple dotted line). $Q_0$ is set to unity, $T_e=10^8$K. All other parameters
are the same as in Fig.~\ref{elec_spec}.
}
\label{nucl_specacc}
\end{figure}

Fig.~\ref{nucl_specacc} shows examples of steady-state nucleon spectra with varying distance of the 
binary stars to each other.
Close binaries show an upturn at low energies due to a high rate of Coulomb losses, 
and a spectral shape at larger energies
that repeats the acceleration spectrum due to the same energy dependence of losses from
hadronic $pp$-interactions and energy gain. For large binary separations Coulomb losses are unimportant, 
and the loss corrected particle spectrum has the same spectral shape as the steady-state acceleration spectrum.
Due to the low hadronic cross section radiative losses hardly cause any cutoff in the particle spectrum. 
Instead, faster particle escape when approaching the Bohm diffusion limit leads to a steepening
of the emitting particle distribution. For simplicity, we chose to treat this effect analog to the electron
acceleration in Sect.~3.1, i.e. we set $T_0=0$ for $\kappa_d\leq\kappa_{\rm Bohm}$.

Applying mass conservation the continuity equation implies $N_H\propto r^{-1}$ for the target ion density $N_H$. 
With $r>r_0$, where $r_0$ indicates the transition from the acceleration to convection region with
typically $r_0\sim 10^{12}$cm in the star systems considered here, hadronic $pp$-interactions
and Coulomb-losses can readily be neglected.
Eq.~\ref{PDE2} can be solved to give 
the analytical solution for a constant convection velocity $V$:
\begin{equation}
N(E) = \left(\frac{r_0}{r}\right)^{2/3} N_{\rm acc}(\tilde E)
\end{equation}
with 
\begin{equation}
\tilde E = E \left(\frac{r}{r_0}\right)^{1/3}
\end{equation}

\begin{figure}[t]
\includegraphics[height=13cm]{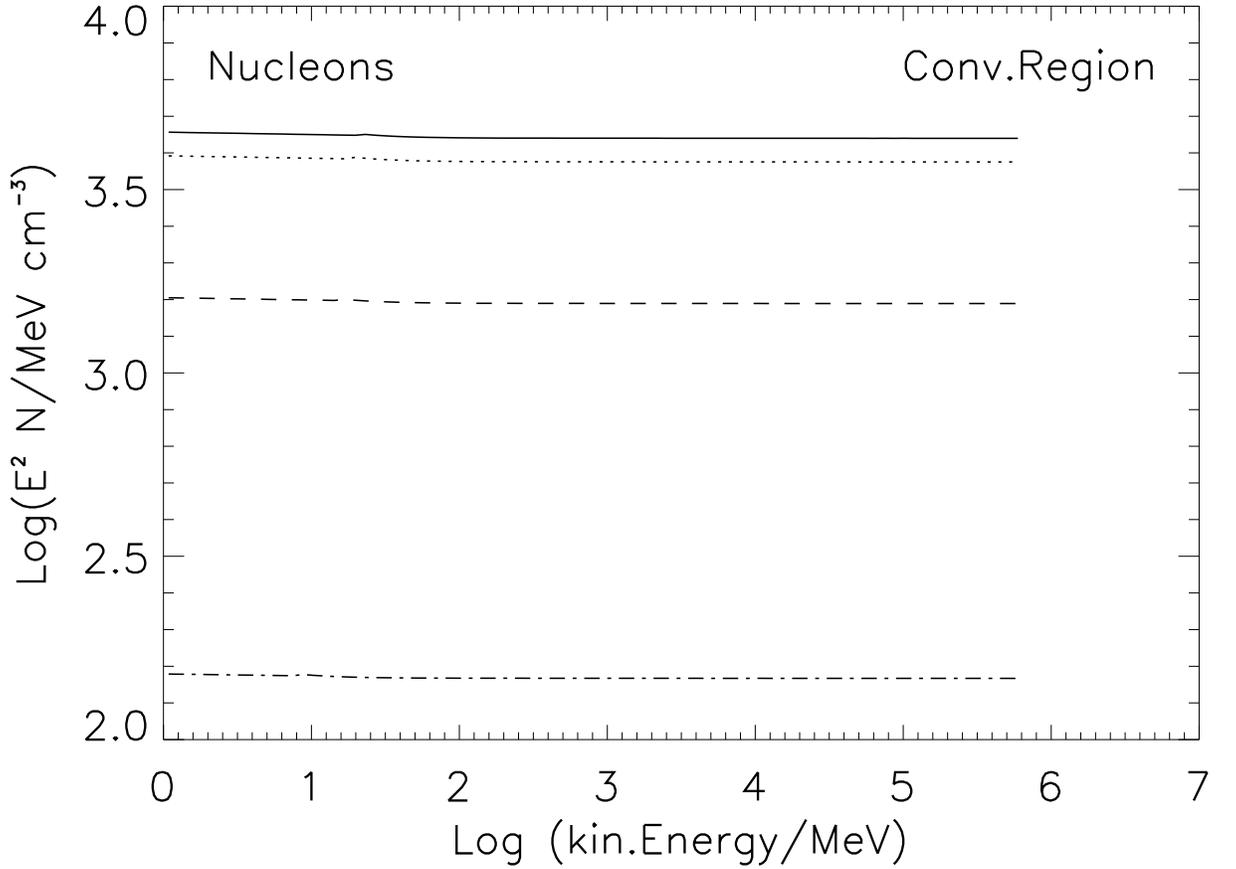}
\caption{Steady-state proton spectra in the convection region at distance 
$\Delta r=10^{11}$cm (dotted line), $10^{12}$cm (dashed line) and $10^{13}$cm (dashed-dotted line)
from the transition point $r_0$ as compared to the proton spectrum in the acceleration region (solid line)
and for a binary separation of $D=10^{14}$cm. All other parameters are the same
as in Fig.~\ref{nucl_specacc}.
}
\label{nucl_specconv}
\end{figure}

Fig.~\ref{nucl_specconv} shows the resulting nucleon spectrum for the same parameter set as used in 
Fig.~\ref{elec_specconv1}. 
Obviously the losses due to $pp$-interactions are of minor importance as compared to the adiabatic 
losses the nucleons suffer in the convection region. 

\subsection{Particle spectra normalization}

For applications to massive binary systems the normalization of the relativistic particle component is 
limited by several constraints: Firstly, the inverse Compton, bremsstrahlung and $\pi^0$-decay $\gamma$-rays
must not overproduce any observational limits imposed by $\gamma$-ray observations
(e.g. EGRET, ...). Secondly, since the particles are accelerated out of the pool of thermal
particles, particle number conservation dictates that the
relativistic particle flux injected into the system must be smaller than the wind particle
flux entering the acceleration zone, i.e. $Q_0 \leq \dot M/(m_p\,x^2 4\pi\, d)$
with $d$ the thickness of the acceleration site. 
Furthermore, due to energy conservation the total injected particle energy can not be larger than the
total kinetic wind energy of the binary system which rarely exceeds 1\% of the total radiative 
energy output of the stars, (typically $L_{\rm w}\leq 10^{37}$erg/s). 
The energy density of accelerated particles is given by $U_{\rm acc}\approx Q_0 E_0/a$
assuming an $E^{-2}$-spectrum ($aT_0=1$). In equilibirium the total acceleration
power equals the loss power due to escape, leading to $L_{\rm acc}=U_{\rm acc} V_{\rm acc}/T_0$
where $V_{\rm acc}=r_0^2\pi d$ is the acceleration volume.
By noting that $(r_0^2/4x^2) L_{\rm w}$ is the kinetic power available to the acceleration region,
$Q_0$ can not exceed $Q_0 \sim (r_0^2/4x^2) L_{\rm w}/(V_{\rm acc} E_0)=L_{\rm w}/(4x^2 \pi d E_0)$. 
These latter two arguments pose to date a stronger 
constraint on the normalization than the $\gamma$-ray limits from EGRET.
In the following we use this maximum possible injection power unless stated otherwise.

\section{Photon spectra}

This work is devoted to photon emission at high energies with emphasis on the $\geq 1$~MeV regime. Important
non-thermal continuum emission processes here are inverse Compton scattering of
the dense stellar radiation field to high energies, relativistic bremsstrahlung
of the electrons in the field of the ions in the wind and the decay of 
$\pi^0\longrightarrow \gamma+\gamma$ that are produced in hadronic
nucleon-nucleon collisions. For the calculations of the photon spectra we assume the
particle distributions to be isotropic in the wind-wind collision zone.

\subsection{Inverse Compton scattering}

\begin{figure}[t]
\includegraphics[height=10cm]{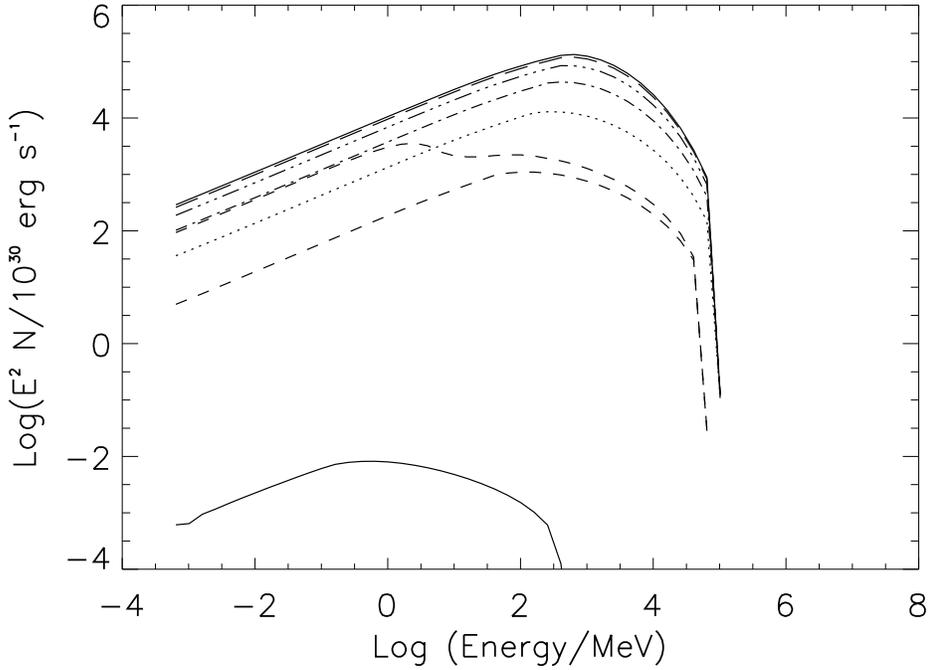}
\caption{IC spectra from the acceleration region for $D=10^{14}$cm and 
$\theta_{\rm L}=0\degr$ (lower solid line), $30\degr$ (dashed line),
$60\degr$ (dotted line), $90\degr$ (dashed-dotted line), $120\degr$ (dashed-triple dotted line) $150\degr$ 
(long dashed line)
and $180\degr$ (upper solid line). 
The normalization corresponds to a injected power of particles with
energy $E_0=1$~MeV prior to acceleration of $\sim 0.031\%$ of the total OB-wind luminosity, 
the emitting volume is $\sim 5\cdot 10^{37}$cm$^3$.
For $\theta_{\rm L}=30\degr$ the total volume-integrated (i.e. acceleration plus 
convection zone) IC spectrum is also shown.
All parameters except for $D$ are the same as used in Fig.~\ref{elec_spec}. The thickness $d$ of the emission region
is estimated by $d\approx x=x_{\rm OB}$ \citep{Eichler93}.
}
\label{ICincline}
\end{figure}

\begin{figure}[t]
\includegraphics[height=13cm]{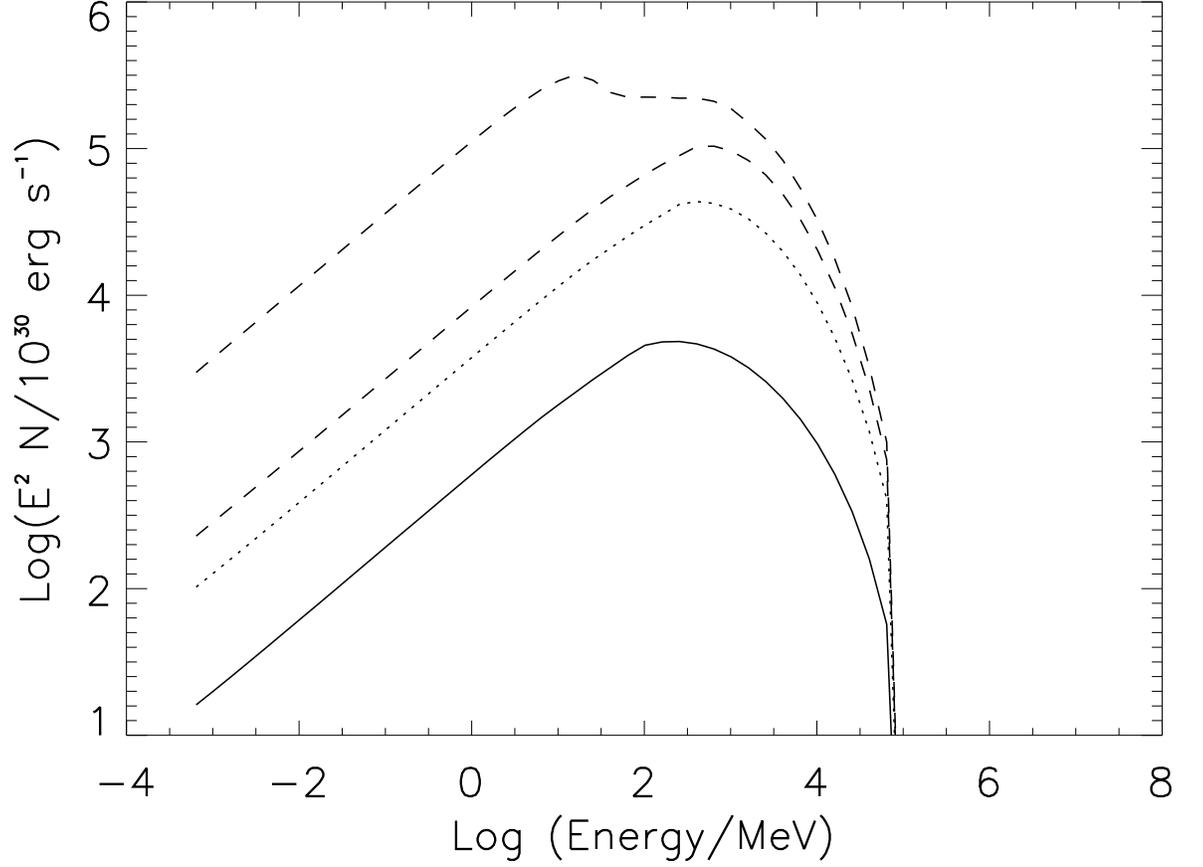}
\caption{
IC spectra from the acceleration region for $D=10^{14}$cm, inclination angle $i=45\degr$ and $\Phi_B=0\degr$ (solid line), 
$90\degr$ (dotted line),
and $180\degr$ (dashed line). For $\Phi_{B}=180\degr$ the total volume-integrated 
(i.e. acceleration plus convection zone) IC spectrum
is also shown. We define $\Phi_B=0$ for the WR-star being in front of the OB-star. 
All other parameters are the same as used in Fig.~\ref{ICincline}.
}
\label{ICphase}
\end{figure}

Inverse Compton (IC) scattering in the dense UV stellar radiation field of the massive
main sequence star often dominates photon production rate at high energies.
The computation of the photon emission is based on the full Klein-Nishina cross section
while for the IC losses of the electrons either the cross section in the Thomson limit or the
Klein-Nishina approximations Eq.~(\ref{KN1}, \ref{KN2}, \ref{KN3}) are applied assuming the losses
to be continuous.
We have further neglected triplet pair production, since the value of $\epsilon_TE/m_e c^2$
that we consider does not exceed $10^3$ \citep{masti91}.
We restrict this work to long-period binary systems, for which the wind momentum from
the WR-star clearly dominates, thus placing the WR-star at a large distance from the collision region.
This allows us to neglect
the stellar radiation field of the WR-star as a target photon field
for IC scattering. For the stellar radiation field of the OB-star the
monochromatic approximation
\begin{equation}
n(\epsilon, \mu_{\rm ph}) = n_0\, \delta(\epsilon-\epsilon_T)\, \delta(\mu-\mu_{\rm ph})
\end{equation}
with $\epsilon_T=2.7\,k_BT_{\rm eff}$ is employed.
All seed photons are approaching the emission region from
the same direction. In this case the full angular dependence of the IC scattering rate has to be taken into
account, since
the scattered power per volume element depends on the scattering angle $\theta_{\rm ph}$. In appendix~C we calculate
the IC photon production rate, $\dot n(\epsilon_s,\Omega_s)$, for an arbitrary target photon field 
$n(\epsilon,\mu_{\rm ph})$ that scatters off an isotropic electron
distribution. We find a declining scattering rate with decreasing scattering angle $\theta_{\rm ph}$, in agreement with
earlier works \citep[][; see Fig.~C.1]{Reynolds1,Brunetti}.
The volume-integrated emitted photon power is calculated by
\begin{equation}
\dot N_{\rm IC} (\epsilon_s) = \int dV\ \dot n(\epsilon_s,\Omega_s) = d\times \int dr r \int d\phi \dot n(\epsilon_s,\Omega_s)
\end{equation}
where the integrals have been solved numerically.
The IC flux variations directly translate into a change of IC power and maximum energy with 
viewing angle $\theta_{\rm ph}$ (see Fig.~\ref{ICincline}).
For a given system inclination the total emitted power and maximal scattered energy 
therefore varies with orbital phase (Fig.~\ref{ICphase}).
These anisotropy effects may be detectable with near future $\gamma$-ray experiments like GLAST (see Sect.~6).
For a non-negligible size of the convection zone the volume integration may lead to photon spectra that show
a kink. This feature occurs at energies above which the convection zone is lacking high energy particles.
The combined effect of both, a deficit of high energy particles in the convection zone and a visible increase
of the total flux from the convection zone, results in the kink at 1-10~MeV in Figs.~\ref{ICincline},\ref{ICphase}.

\subsection{Relativistic bremsstrahlung}

\begin{figure}[t]
\includegraphics[height=13cm]{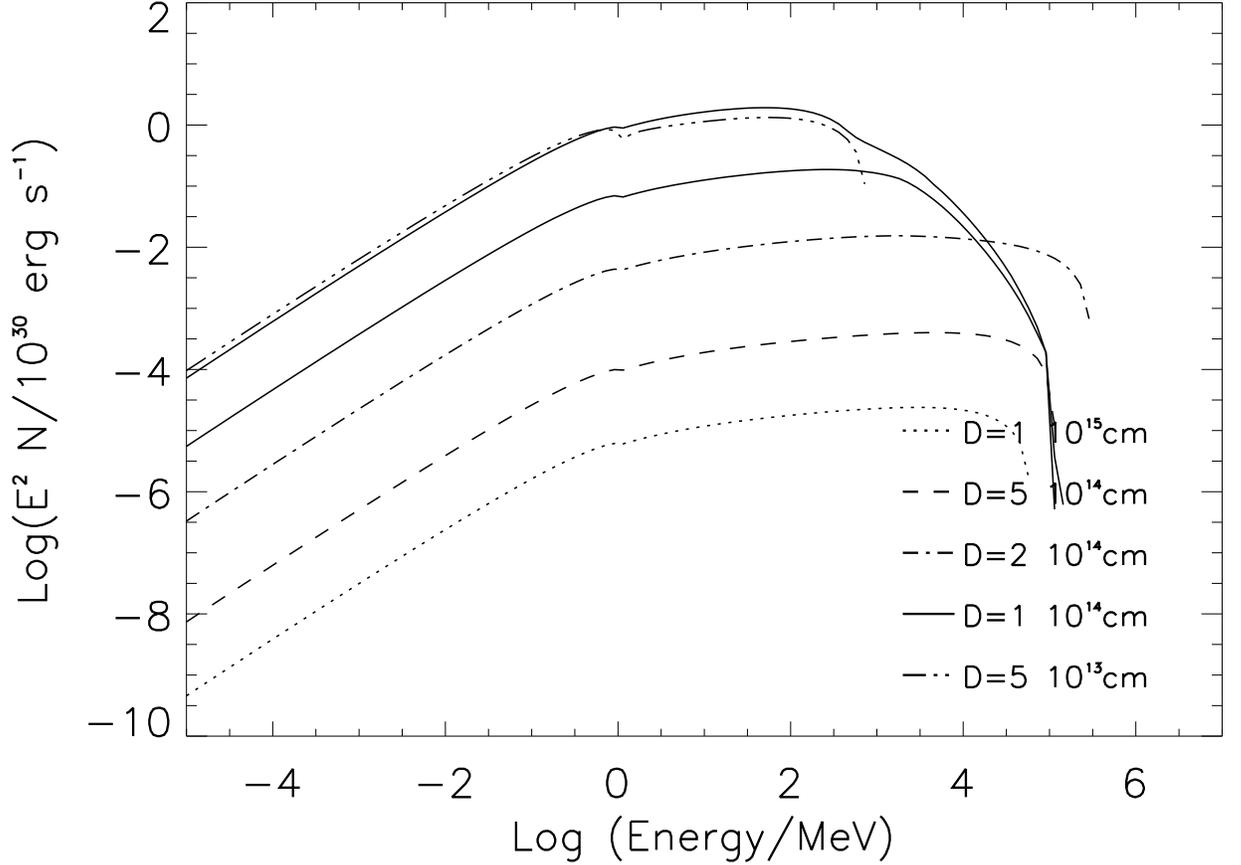}
\caption{Relativistic bremsstrahlung spectra from underlying electron distributions 
as presented in Fig.~\ref{elec_spec} for various binary separations
$D=5\cdot 10^{13}$cm (dashed-triple-dotted line),
$10^{14}$cm (solid line), $2\cdot 10^{14}$cm (dashed-dotted line), $5\cdot 10^{14}$cm (dashed line) 
and $10^{15}$cm (dotted line),
emitted from the acceleration region. 
The normalization corresponds to a injected power of particles with energy $E_0=1$~MeV of 
$\sim$0.14\%, 0.031\%, 0.007\%, 0.001\%, and 0.0003\%, respectively, of the total OB-wind luminosity. 
The emitting volume varies between $3\cdot 10^{37}-5\cdot 10^{38}$cm$^3$.
The upper solid line corresponds to the sum of
bremsstrahlung emission from acceleration and convection region (up to $\Delta r=10^{13}$cm)
for a binary separation of $D=10^{14}$cm.
}
\label{brems}
\end{figure}

Non-thermal relativistic bremsstrahlung losses are non-negligible in the 
stagnation point of the colliding winds where the (compressed) gas density
may reach values of $10^{5\ldots 9}$cm$^{-3}$ whereas in the convection region
the decreasing gas density makes its contribution minor.
The nonthermal bremsstrahlung photon flux using 
a typical ISM metallicity (90\% H, 10\% He, neglecting contributions from higher atomic number particles)
\begin{equation}
\dot N_{\rm ph,br}(E_\gamma) = \frac{1.3 N_H c}{4\pi d_L^2} 
\int dV \int_{\max(E_\gamma m_e c, E_{\rm min})}^{E_{\rm max}} dE\ N(E)\, 
\frac{d\sigma}{dE_\gamma}
\label{brems_phot}
\end{equation}
is calculated numerically in the relativistic limit
(valid for $E/m_ec^2\geq 15/Z$, $Z$ is the atomic number)
using the differential cross section from \cite{Blumenthal1970}. 
Examples of bremsstrahlung spectra are shown in Fig.~\ref{brems} for the
electron spectra in Fig.~\ref{elec_spec}. The shape of the bremsstrahlung
photon spectrum in the relativistic regime reflects the shape of the electron
spectrum. The larger the binary separation is, the larger is the difference between
the turnover energy from the Thomson to KN-loss regime and the cutoff energy (see Fig.~\ref{elec_spec}).
This leads to a decline of the bremsstrahlung spectrum at an energy that increases with
the binary separation. An estimate of the thickness $d$ of the emission region was provided by
\cite{Eichler93} who showed that $d\approx x=x_{\rm OB}$. With $r_0\approx 10^{12}$cm the
total emitting volume in the acceleration region is $\pi r_0^2 d\approx 10^{38\ldots 39}$cm$^3$.
Fig.~\ref{brems} shows the resulting relativistic bremsstrahlung power spectra
for various binary separations. Despite a somewhat larger emitting volume
for colliding wind systems with a large binary separation, the total bremsstrahlung
emission declines with binary separation due to a rapidly decreasing target
gas density in the collision region. 
For $D=10^{14}$cm the compound acceleration and convection region spectrum is also presented
in Fig.~\ref{brems}. Due to the volume effect the dominant contribution to the total emission
spectrum comes from the convection region at large $r$, thus increasing the overall bremsstrahlung
intensity. At the same time the deficit of high energy particles in the convection region at
large distances from the acceleration region (see Fig.~\ref{elec_specconv1}) leads
to a deficit of high energy photons. This causes the feature at $\sim 100$~MeV in the
total bremsstrahlung spectrum shown in Fig.~\ref{brems} (plotted for a $D=10^{14}$cm
binary separation).

\subsection{$\pi^0$-decay $\gamma$-ray emission}

Collisions between cosmic ray protons and nucleons in massive star winds are rather rare
and occur on average a few times per year for gas densities in the wind collision region 
typical for long-period binaries like WR~140.

\begin{figure}[t]
\includegraphics[height=13cm]{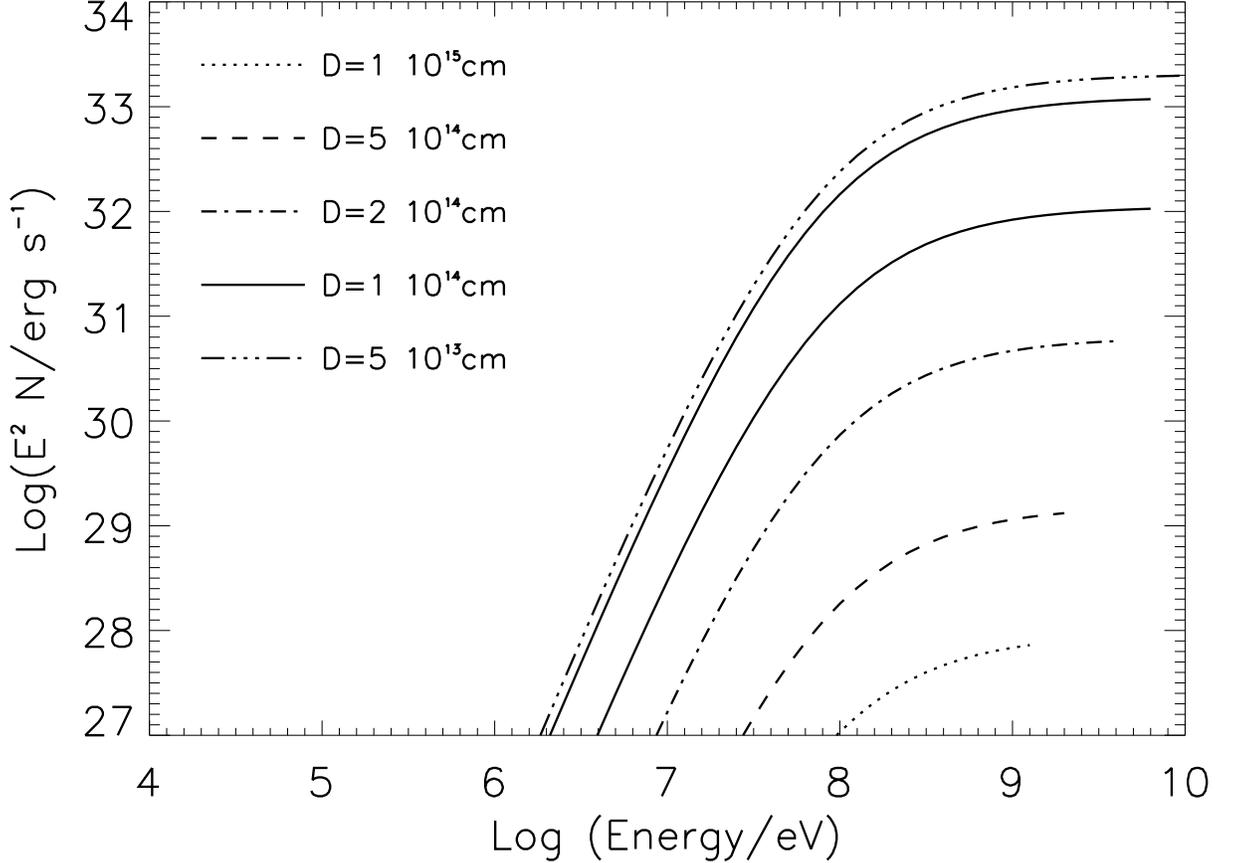}
\caption{
$\pi^0$-decay $\gamma$-ray spectra from underlying proton distributions 
as presented in Fig.~\ref{nucl_specacc} for various binary separations
$D=5\cdot 10^{13}$cm (dashed-triple-dotted line),
$10^{14}$cm (solid line), $2\cdot 10^{14}$cm (dashed-dotted line), $5\cdot 10^{14}$cm (dashed line) 
and $10^{15}$cm (dotted line),
emitted from the acceleration region. 
The normalization corresponds to a injection power of $\sim$0.14\%, 0.031\%, 0.007\%, 0.001\%, and 0.0003\%, 
respectively, of the total OB-wind luminosity in form of thermal particles of energy $E_{0,\rm kin}=1$~MeV. 
The emitting volume varies between $3\cdot 10^{37}-5\cdot 10^{38}$cm$^3$.
The upper solid line corresponds to the sum of
$\pi^0$-decay photon emission from acceleration and convection region (up to $\Delta r=10^{13}$cm)
for a binary separation of $D=10^{14}$cm.
}
\label{pi0}
\end{figure}

The stationary proton spectra as shown in Fig.~(\ref{nucl_specacc}, \ref{nucl_specconv}) are used 
to calculate the $\pi^0$-decay photon spectrum.
Since the proton spectra above the threshold for $pp$-interactions in general reflects
the shape of the acceleration spectrum, one expects pure power law particle spectra in this energy 
range. The formalism developed by \cite{Pfrommer} for the 
$\pi^0$-decay $\gamma$-ray production
of pure power law particle spectra seems therefore appropriate to use.
The resulting $\pi^0$-decay $\gamma$-ray spectra (calculated for a $^4$He mass fraction
of 0.3 for the wind metallicity; see Sect.~4.2) from the acceleration region are shown in 
Fig.~\ref{pi0} 
for various binary separations.
The uppermost curve corresponds to the combined acceleration and convection region
$\pi^0$-decay spectrum for a binary separation of $D=10^{14}$~cm.

For typical parameters of colliding massive wind systems and maximal allowed injection power
the radiative luminosity from
$\pi^0$-decay lies therefore typically on a $\leq 10^{33}$erg/s flux level, 
leading to a $\pi^0$-decay luminosity from wide binary systems 
that can in general be neglected when compared to the expected IC-luminosity $>$1~GeV,
provided the emitting electron spectrum extends to $\geq 10^4$~MeV.

\subsection{$\gamma-\gamma$ opacity in the stellar radiation fields}

Above $\sim 100$~GeV the optical depth $\tau_{\gamma\gamma}$ due to
photon-photon pair production ($\gamma\gamma\rightarrow e^+e^-$) in the intense stellar 
radiation field of the main sequence star
may reach non-negligible values depending on its spectral type.
This may modify the $\gamma$-ray spectrum escaping from the source
by a factor $\exp(-{\tau_{\gamma\gamma}})$. 

The $\gamma-\gamma$ opacity $d\tau_{\gamma\gamma}$ along a path element $dy$ in a radiation field with 
differential photon number density $n(\epsilon,\Omega)$
is given by \citep{sigmagg_ref}
\begin{equation}
d\tau_{\gamma\gamma}(E_\gamma,\epsilon,\Omega) = dy\,n(\epsilon,\Omega)\, \sigma_{\gamma\gamma}(\epsilon,E_\gamma,\mu_{\gamma\gamma}) 
\,(1-\mu_{\gamma\gamma}) .
\end{equation}
With the target photon density 
$$n(\epsilon) = n_0\,{{ r_s^2}\over {\tilde x+y^2+2y\mu_{\rm ph}\sqrt{\tilde x}}}\, 
\delta(\epsilon-\epsilon_T)\,\delta(\mu-\mu_{\gamma\gamma})$$ 
above the stellar radius $r_s$, and $\tilde x=x^2+r^2$, this can be re-written as
\begin{equation}
\tau_{\gamma\gamma}(E_\gamma,\mu_{\rm ph},x,r) = \frac{n_0(\epsilon=\epsilon_T) r_s^2}{8\pi E_\gamma\epsilon_T} \int d\phi
\int_{0}^\infty dy \frac{s\sigma_{\gamma\gamma}(s)}{\tilde x+y^2+2\,\sqrt{\tilde x}\,y\,\mu_{\rm ph}(\theta_L, \phi_L, \cos{\theta}=x/\sqrt{\tilde x},\phi)}\,
H(s-s_{\rm thr})
\end{equation}
for photons of energy $E_\gamma$ escaping from the emission region at location $r$ 
along a path $y$ with angle $\theta_{\rm ph} =\cos{\theta_L}\,\cos{\theta} + \sin{\theta_L}\, \sin{\theta}\, \cos(\phi-\phi_L)$.
Here, $s=2\, E_\gamma\,\epsilon(1-\mu_{\gamma\gamma}(y))$ is the square of the centre of momentum (CM) energy,
$\sigma_{\gamma\gamma}(s)$ the cross section for photon-photon pair production \citep{sigmagg_ref},
$s_{\rm thr} = (2 m_e c^2)^2$ is the squared minimum CM energy required to overcome the process threshold, 
$\mu_{\gamma\gamma}(y) = \cos{\theta_{\gamma\gamma}(y)}$ is the cosine of the angle between the two 
interacting photons.
and $\cos{\theta}=x/\sqrt{x^2+r^2}$.

\begin{figure}[t]
\includegraphics[height=13cm]{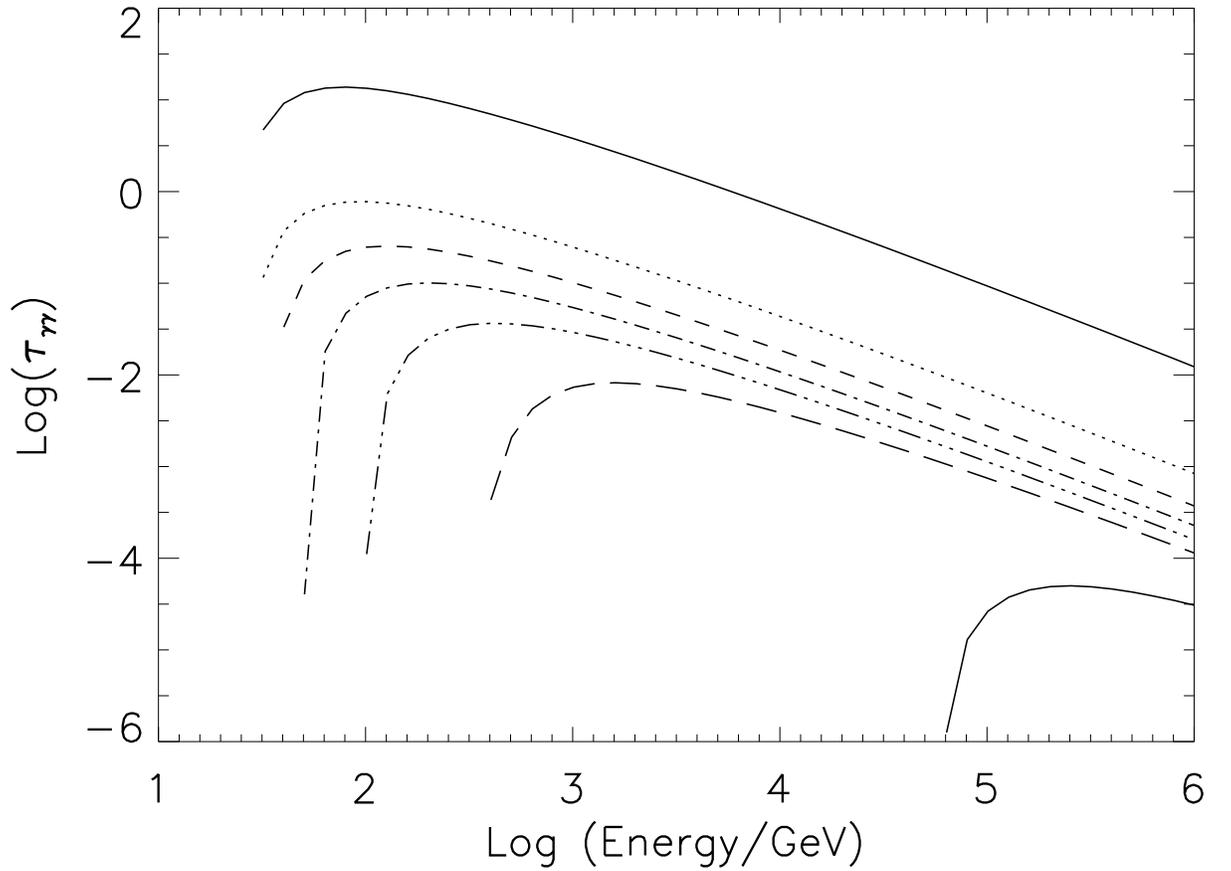}
\caption{$\gamma\gamma$ pair production opacity $\tau_{\gamma\gamma}$ for a binary separation of $D=10^{14}$cm and for angles
$\theta_{\rm L}=0\degr$ (lower solid line), $30\degr$ (long-dashed line), $60\degr$ (dashed-triple-dotted line), $90\degr$ (dashed-dotted line),
$120\degr$ (dashed line), $150\degr$ (dotted line), $180\degr$ (upper solid line),
and fixing $r$ at $10^{12}$cm. 
}
\label{tau_view}
\end{figure}

\begin{figure}[t]
\includegraphics[height=13cm]{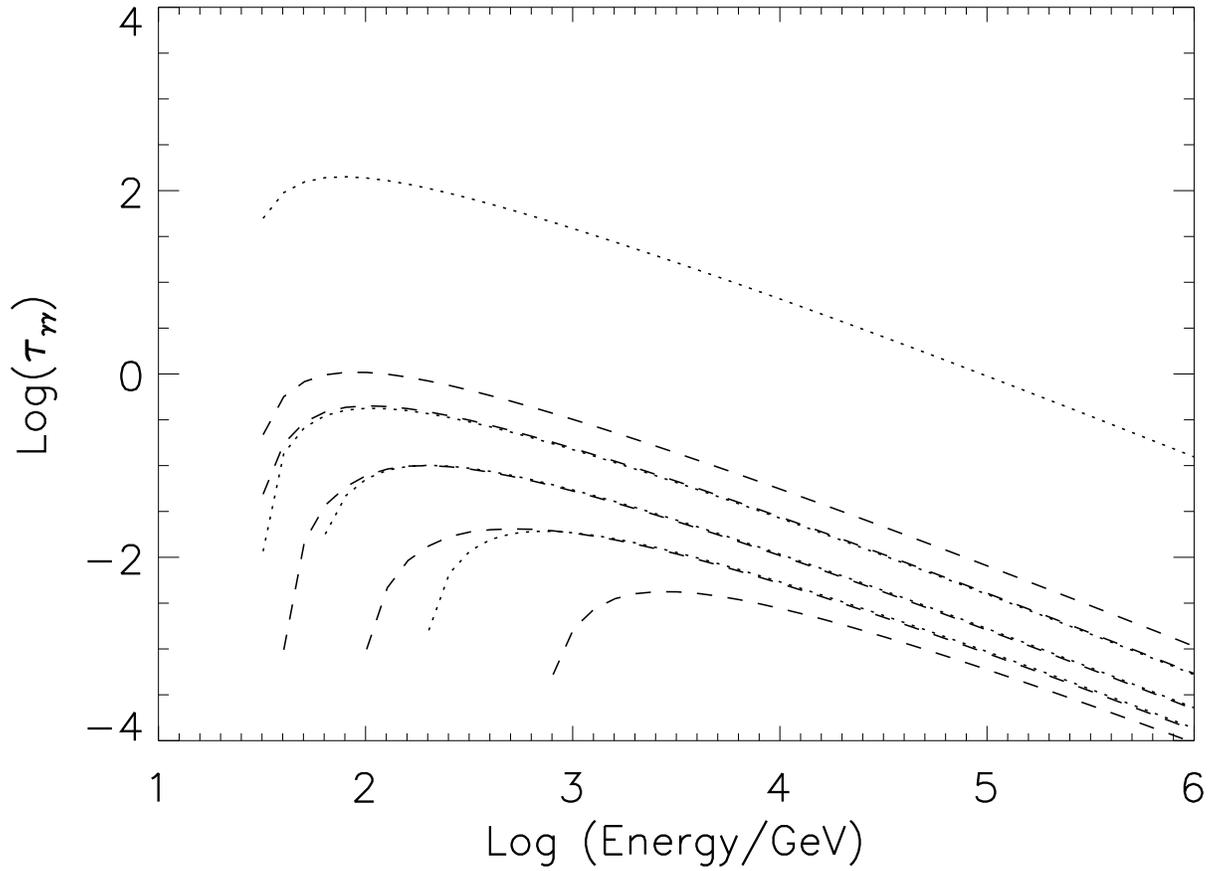}
\caption{$\gamma\gamma$ pair production opacity $\tau_{\gamma\gamma}$ for a binary separation of $D=10^{14}$cm
and varying $r=10^{11}$cm (dotted lines), $r=10^{13}$cm (dashed lines), and for angles
$\theta_{\rm L}=0\degr$, $45\degr$, $90\degr$, $135\degr$, $180\degr$ (from lower to upper curves). 
}
\label{tau_rvar}
\end{figure}

\begin{figure}[t]
\includegraphics[height=13cm]{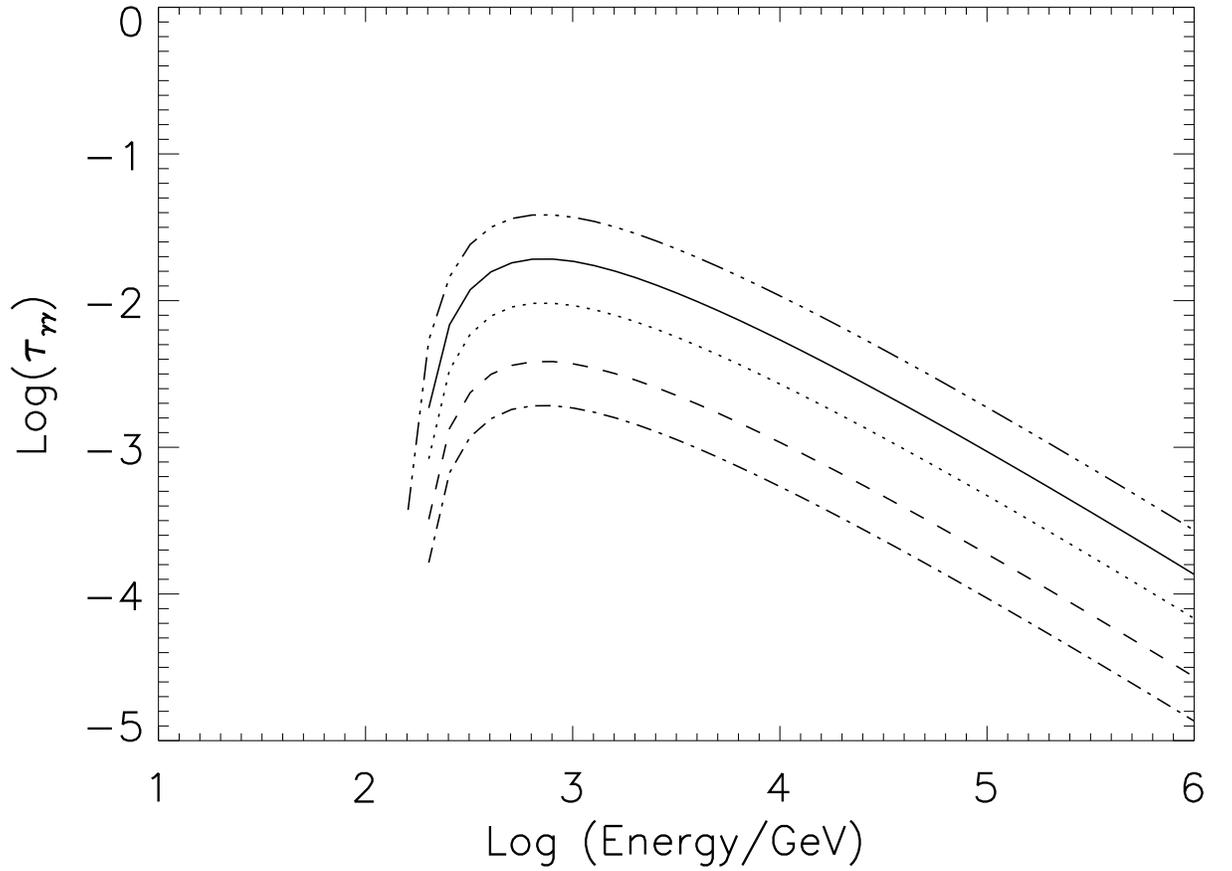}
\caption{$\gamma\gamma$ pair production opacity $\tau_{\gamma\gamma}$ for an angle of $\theta_L=45\degr$ and 
for a separation between the
collision region and the OB-star surface of $r_{\rm OB}=2.4\cdot 10^{14}$cm (dashed-dotted line), 
$1.2\cdot 10^{14}$cm (dashed line), $4.8\cdot 10^{13}$cm (dotted line), $2.4\cdot 10^{13}$cm (solid line)
and $1.2\cdot 10^{13}$cm (dashed-triple-dotted line) and fixing $r$ at $10^{12}$cm.
}
\label{tau_dist}
\end{figure}

In Fig.~(\ref{tau_view}-\ref{tau_dist}) we show the absorption optical depth
due to photon-photon collisions in a $T=43000$~K radiation field (i.e. $\epsilon_T=10$~eV) for
different $r_{\rm OB}=1.2-24\cdot 10^{13}$cm being the separation of the emission region
from the stellar photosphere, and for various angles $\theta_L=0\degr-180\degr$, respectively.
Large viewing angles decrease the process' threshold energy in the observer frame, and increase the 
opacity at the same time due to a longer path way and the rise of the CM energy. 
Fig.~\ref{tau_view} shows that variations of the optical depth with angle $\theta_L$ by several
orders of magnitude are possible, and absorption can be quite severe 
(up to $\tau_{\gamma\gamma}\sim 100$ for $D=10^{14}$cm). Except at large $\theta_L$
the effect of varying $r$ mainly impacts the absorption depth near threshold (see Fig.~\ref{tau_rvar}).
The opacity is also strongly dependent on the radiation field density
which is dependent on the massive star's luminosity as well as on $r_{\rm OB}$. This can be
observed in Fig.~\ref{tau_dist} where $\tau_{\gamma\gamma}$ ranges between 0.002 to 0.04 at the peak of
the cross section and along the $\theta_L=45\degr$ sight line.

This essay shows that $\gamma$-ray absorption due to pair production
can not be neglected in general for massive star systems, but must be treated individually for each system
and is strongly dependent on the sight line.

\section{Applications}

\subsection{WR 140}

\object{WR 140} is the archetype of a spectroscopic long-period massive binary system of colliding winds which shows 
periodic dust formation around periastron passage as well as nonthermal radio emission at phases where
the colliding wind region is outside the photosphere for free-free absorption. The system has been well
monitored at radio, IR, optical, UV and X-rays \citep[e.g.][]{Williams90,Becker95,Panov2000,setiaIUV}. 
It consists of a WC~7pd and a O4-5~V companian (of bolometric luminosity $L_{\rm bol}=10^{5.8} L_{\sun}$
and effective temperature $T_{\rm eff}=43000$~K) with an orbital period of $2899\pm 10$~days.
Located at a distance of 1.85~kpc \citep{Dough05} the large inclination system 
\citep[$i=122{\degr}\pm 5{\degr}$;][]{Dough05} possesses an excentricity of $0.881\pm 0.04$. 
With periastron passage being defined by phase $\Phi=0$ the argument of periastron is 
$\omega=47\degr$ \citep{Marchenko2003}. In this work we use the basic parameters that has been recently re-determined
by \cite{Dough05}.
The mass loss rate of $8.7\cdot 10^{-6}M_{\sun}$/yr and the wind velocity of the O-star $v_{\rm OB}=3100$~km/s
as compared
to the mass loss rate of $4.3\cdot 10^{-5}M_{\sun}$/yr of the WR-star at velocity $v_{\rm WR}=2860$~km/s
places the collision region at a distance of $x\approx 0.32 D \approx 9.6\cdot 10^{12}\ldots 1.5\cdot 10^{14}$cm
from the O-star due to momentum balance and taking also into account the excentricity of the system.
This is a factor $\sim 2.5$ smaller than used in earlier works \citep[e.g.][]{Eichler93,bene2003}.  
At this distance the wind gas density lies between $\sim 4\cdot 10^6\ldots 9\cdot 10^8$cm$^{-3}$.
For the strength of the shock we assume a compression factor $c_r=4$. 
The radio emission reaches its maximum roughly at phase 0.83.
The nonthermal radio spectrum at phases $\Phi=0.671\ldots 0.955$ shows typically a $\alpha=-0.5$ spectrum, indicating
that the underlying particle distribution obeys a power law with spectral index $p=2$
\citep{Williams90,Dough05}.
The spectral requirements are in agreement with a convection velocity
of $\sim 1410\ldots 1540$~km/s in the framework of our simplified geometrical picture.
For predicting WR~140's high energy emission from primary electrons and protons we have selected 
four orbital phases approximately equally separated where its non-thermal radio emission has been determined \citep{Dough05}, 
namely
$\Phi=0.995$, 0.2, 0.671 and 0.8. In order to reproduce the flux level of the cm radio data for
those phases we require a energy injection rate of thermal particles that varies with orbital phase and lies between
$10^{32.5\ldots 33.5}$~erg/s or $10^{-3}\ldots 10^{-2}\%$ of the OB-wind kinetic energy.
The total emission volume $V=r^2_{\rm max}\pi d$ with $d\sim x$ is also assumed to vary with orbital phase
(we used $r_{\rm max}=4\cdot 10^{14}$cm for $\Phi=0.2$, 0.671 and 0.8, and $r_{\rm max}=10^{13}$cm for $\Phi=0.995$)
as suggested by the cm radio observations \citep{Dough05}.
The magnetic field at the O-star surface is assumed of order 100~G, which leads to field values
of 0.2 \ldots 3.5 G in the wind collision zone  
\citep[in agreement with estimates for the equipartition magnetic field:][]{bene2003}. 
The diffusion coefficient $\kappa_{\rm a}$ determines the acceleration rate and, together with the
energy loss channels, the maximum particle energy.
In order to allow for relativistic electron energies, necessary to explain the observed synchrotron
radiation, $\kappa_{\rm a}$ must be low enough that acceleration gains are able to overcome
the Coulomb losses. 
On the other side, Bohm diffusion is the limit for the diffusion coefficient $\kappa_{\rm d}$, which is
connected to $\kappa_{\rm a}$. 
This leads to a rather narrow range of possible values for the diffusion coefficients if one
requires the production of $>100$~MeV photons via the IC process at least at phases close to apastron.
For our models presented in Fig.~\ref{WR140_IC}, \ref{WR140_ph00}, \ref{WR140_ph50} we used $\kappa_{\rm a}=3.76\cdot 10^{19}$~cm$^2$/s which results
in an acceleration region size of $r_0=2.4\ldots 2.6\cdot 10^{11}$~cm and escape time
of $T_0=1.6\ldots 1.9$ks. 
Detections of the maximum photon energy will shed light on the exact value of $\kappa_a$.

\begin{figure}[t]
\includegraphics[height=13cm]{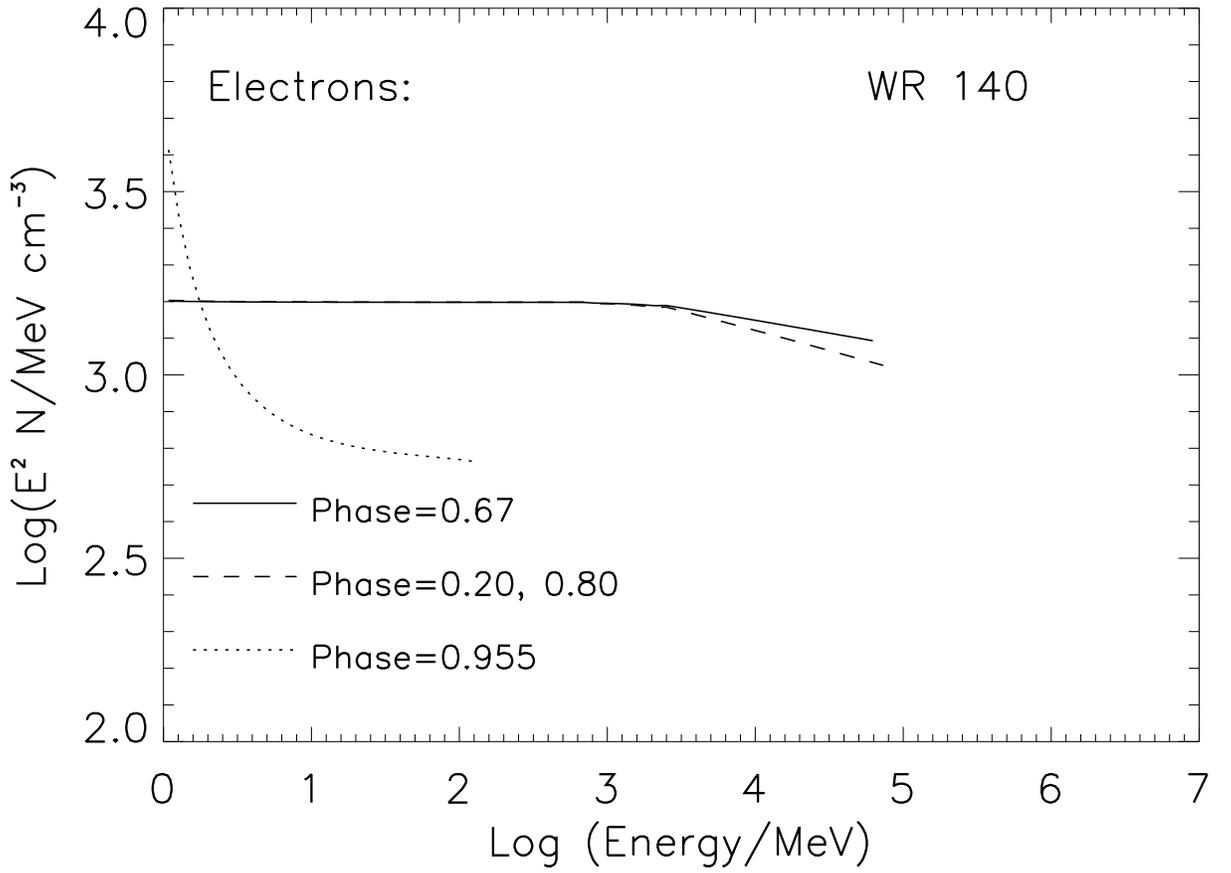}
\caption{Steady-state electron spectra for WR~140 at orbital phases 0.955, 0.2, 0.671 and 0.8, $Q_0=1$.
See text for parameters and discussion.
}
\label{WR140_e}
\end{figure}

\begin{figure}[t]
\includegraphics[height=13cm]{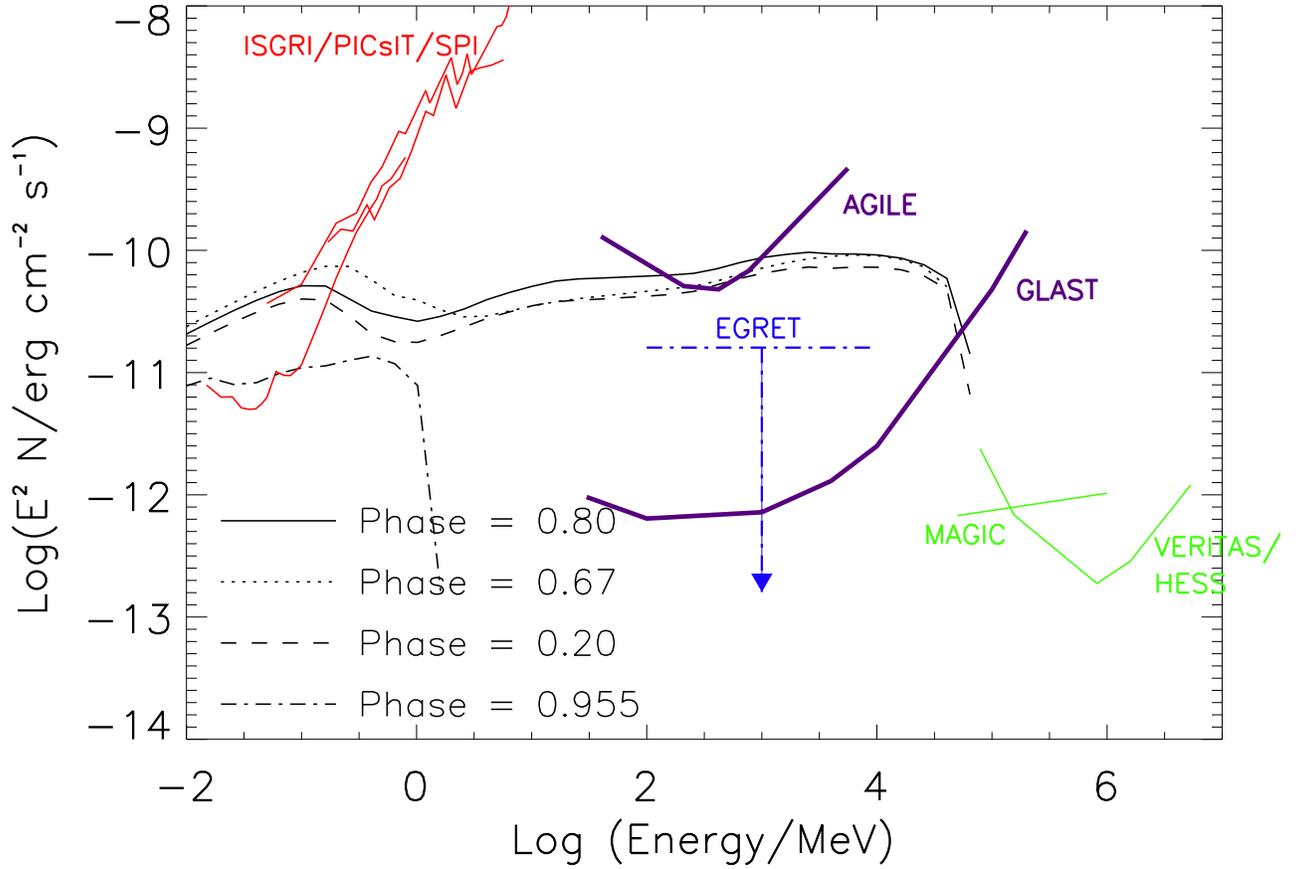}
\caption{IC spectra for WR~140 at phases 0.955, 0.2, 0.671 and 0.8 from electron spectra as shown in
Fig.\ref{WR140_e}. The spectral changes from $\gamma$-ray absorption are not shown here.
The EGRET $2\sigma$ upper limit \citep{Muecke} is based on observations that correspond to a superposition 
of orbital states rather determined by periastron phase, and is therefore compared to phase 0.955 here.
}
\label{WR140_IC}
\end{figure}

\begin{figure}[t]
\includegraphics{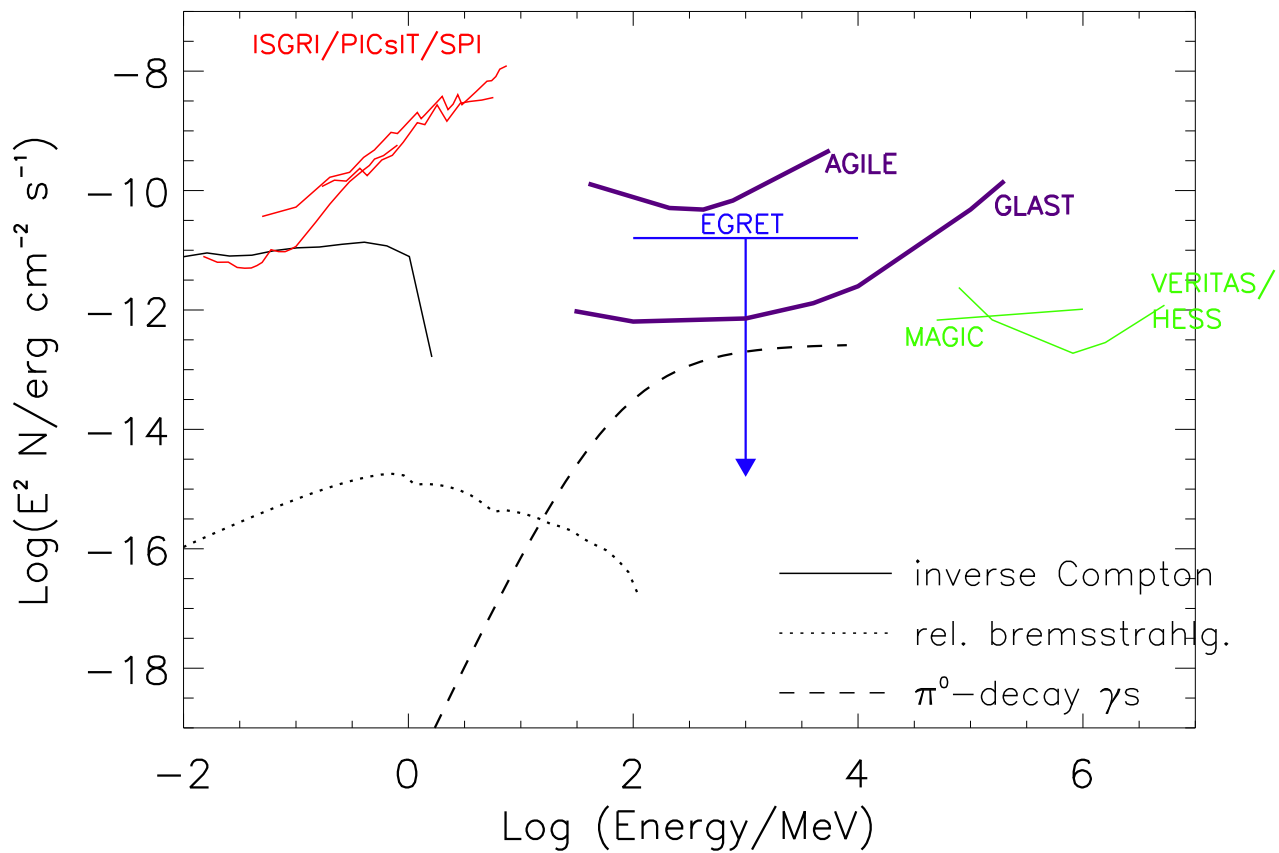}
\caption{Broad band SED of WR~140 at orbital phase 0.955 (close to periastron). 
The injected energy in electrons and protons is equal and $\sim 10^{-5}$ times the kinetic
OB-wind energy.
See text for further parameters and discussion. $\gamma$-ray absorption alters the spectrum significantly 
only above $\sim 60$~GeV.
The escaping photon spectrum remains therefore unchanged at this orbital phase.
}
\label{WR140_ph00}
\end{figure}

\begin{figure}[t]
\includegraphics[height=13cm]{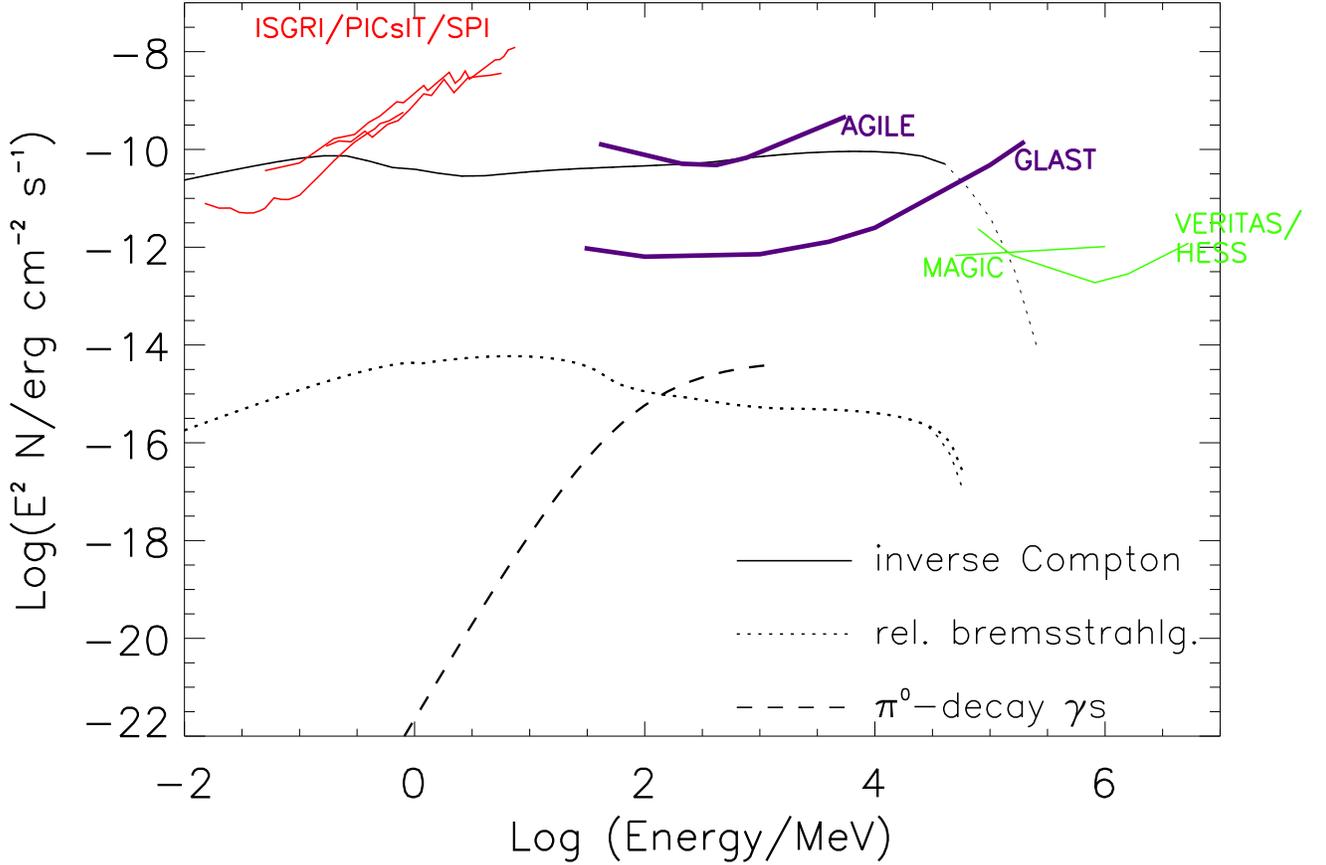}
\caption{Broad band SED of WR~140 at orbital phase 0.671. 
An injected electron energy of $3\cdot 10^{33}$erg/s in total are required to account for the
synchrotron flux at this phase.
For the protons we used $Q_0 E_{0,{\rm kin}} \sim 8\cdot 10^{-7} L_{\rm w}$ ($E_{0,{\rm kin}}=1$~MeV).
See text for further parameters and discussion. $\gamma$-ray absorption due to photon collisions
in the radiation field of the OB-star has been taken into account for all radiation processes: 
IC (solid line) and bremsstrahlung (dotted lines) and $\pi^0$-decay $\gamma$-rays (dashed line).
The corresponding upper and lower curves belong to the unabsorbed and absorbed fluxes, respectively.
The cutoff in the IC spectrum has been extended (dotted line) using an exponential shape to guide
the reader's eye. See text for details.
}
\label{WR140_ph50}
\end{figure}

Fig.~\ref{WR140_e} shows the steady-state electron spectra at orbital phases 0.955 (close to periastron), 0.2, 0.671 
(close to apastron) and 0.8 for these parameter values. Due to the excentricity of the system the wind collision region 
is located closest to the main sequence star at periastron while it is furthest away at apastron. 
All spectra except at periastron extend to very high energies with the cutoff due to the 
finite size of the acceleration region,
and a smooth change in the spectral index where Klein-Nishina effects set in. 
The sharpness of the cutoff is a result of the approximations employed in this work,
and may appear somewhat softened in more sophisticated (however then indispensably non-analytical)
calculations.

At periastron the wind collision region 
is closest to the OB-star, and the electron spectrum is cutoff
already at $\sim$100 MeV due to (Thomson regime) IC losses in the intense photospheric radiation field. 
Fig.~\ref{WR140_ph00} and \ref{WR140_ph50} shows the corresponding broad band SEDs in comparison with the
sensitivity of the INTEGRAL instruments, GLAST, AGILE and the current
generation IACTs like H.E.S.S., MAGIC and VERITAS. The EGRET upper limit has been derived from the summed 
P1-P4 EGRET data at the location of WR~140 \citep[see also][]{Muecke}.
At all orbital phases IC emission dominates over all other radiation processes except close to periastron. Here the
IC spectrum cuts off already at a few MeV, corresponding to the cutoff in the electron spectrum.
The $\gamma$-ray domain is therefore only covered by relativistic bremsstrahlung radiation (up to
$\sim$100 MeV following the electron spectrum) and $\gamma$-rays from the $\pi^0$-decay. 
(Ion-electron) bremsstrahlung emission
turns out to lie always below the IC photon output at a level that is not detectable for current and near future
instruments. 
Assuming the maximal possible injection rate of thermal particles into the acceleration region,
$\pi^0$-decay $\gamma$-rays may possibly be detected with GLAST at orbital phases close to periastron where
the density of the target material is enhanced.
At apastron the losses in the photospheric radiation field as well as due to the wind particle density are 
low enough to allow the photon spectra to extend into the 10-100 GeV regime with a IC flux level detectable,
particularly with low-energy threshold IACTs (e.g. MAGIC) and GLAST. 
This is true despite photon absorption from pair production which leaves its fingerprint
above $\sim$100~GeV (see Fig.~\ref{WR140_ph50}: at $\Phi=0.67$ $\tau_{\gamma\gamma}(r=r_0,
E\approx 100\mbox{GeV})\sim 1$ while at $\Phi=0.95$ $\tau_{\gamma\gamma}(r=r_0,
E\approx 100\mbox{GeV})\sim 3$).

Fig.~\ref{WR140_IC} zooms onto the IC spectra. The flux variations due to the
anisotropy effect at different orbital
phases are blurred by flux variations due to the changes in the radiation field density in the 
strongly excentric system.
The feature at $\leq 1$ MeV stems from the deficit of high energy particles in the convection zone.
For this parameter setting we predict a maximum of the IC flux at phase $\Phi \simeq 0.95$, the
minimum IC flux level would occur at phase $\sim 0.01$. Thus the next maximum $\gamma$-ray flux level
is expected around August 2008, just in time for a clear detection with GLAST.

\subsection{WR 147}

\object{WR 147}, a WN8(h) plus B0.5 V (with bolometric luminosity of $L_{\rm bol} = 5\cdot 10^4 L_{\sun}$
and effective temperature $T_{\rm eff}=28500$~K, i.e. $\epsilon_T\approx 6.6$~eV)
 massive binary system is among the closest and brightest systems that show
non-thermal radio emission in the cm band. Owing to its proximity this system has
been resolved in a northern non-thermal component (WR~147N) and
a southern thermal one (WR~147S) with a separation of $575\pm 15$~mas using the MERLIN instrument 
\citep{Churchwell92,Williams97}. 
The observed radio morphology and spectrum supports a colliding wind scenario
for \object{WR 147}, as first proposed by \cite{Williams97}.
At a distance of 650 pc the implied binary separation is estimated
to 417 AU. The mass loss rates ($\dot M_{\rm WR}=2.5\cdot 10^{-5}M_{\sun}$/yr, 
$\dot M_{\rm OB}=4\cdot 10^{-7}M_{\sun}$/yr)
and wind velocities ($v_{\rm WR}=950$~km/s, $v_{\rm OB}=800$~km/s) place the stagnation point at
$6.6\cdot 10^{14}$cm, in agreement with the MERLIN observations. 
A comprehensive study of the WR~147's radio emission and the geometry of the system has been presented
by \cite{setia}. Any excentricity nor the inclination of the system are known
so far, hence we assume $i=90{\degr}$ and $e=0$ for this application to WR~147. 
The non-thermal flux component can be well fitted by a power law with
spectral index $\alpha=-0.43$ with, however, poor statistical significance. For our presentation
we use again the canonical value of  $\alpha=-0.5$ corresponding to an $E^{-2}$ power law particle spectrum.
Assuming strong shocks, a convection velocity of
399~km/s and a B-star surface magnetic field of 30 G (translating into a 25~mG field strength in the
wind collision region, close to its equipartition value \citep{bene2003}) the observed radio flux level 
can be reproduced if $\sim 0.15\%$ of the
OB-wind kinetic energy is injected as thermal particles into the emission region 
of a total estimated volume of $\sim 2.6\cdot 10^5$AU$^3$.
Again the diffusion coefficient is chosen such to overcome Coulomb losses and allow the production of
$> 100$~MeV through the inverse Compton process without violating the Bohm-limit at the same time. 
With $\kappa_{\rm a}=2\cdot 10^{21}$~cm$^2$/s convection is dominant in $\sim 90\%$
of the total emission region, and the acceleration site covers a size of $r_0=5\cdot 10^{13}$~cm (escape time
is $T_0\approx 10^6$s).

\begin{figure}[t]
\includegraphics[height=13cm]{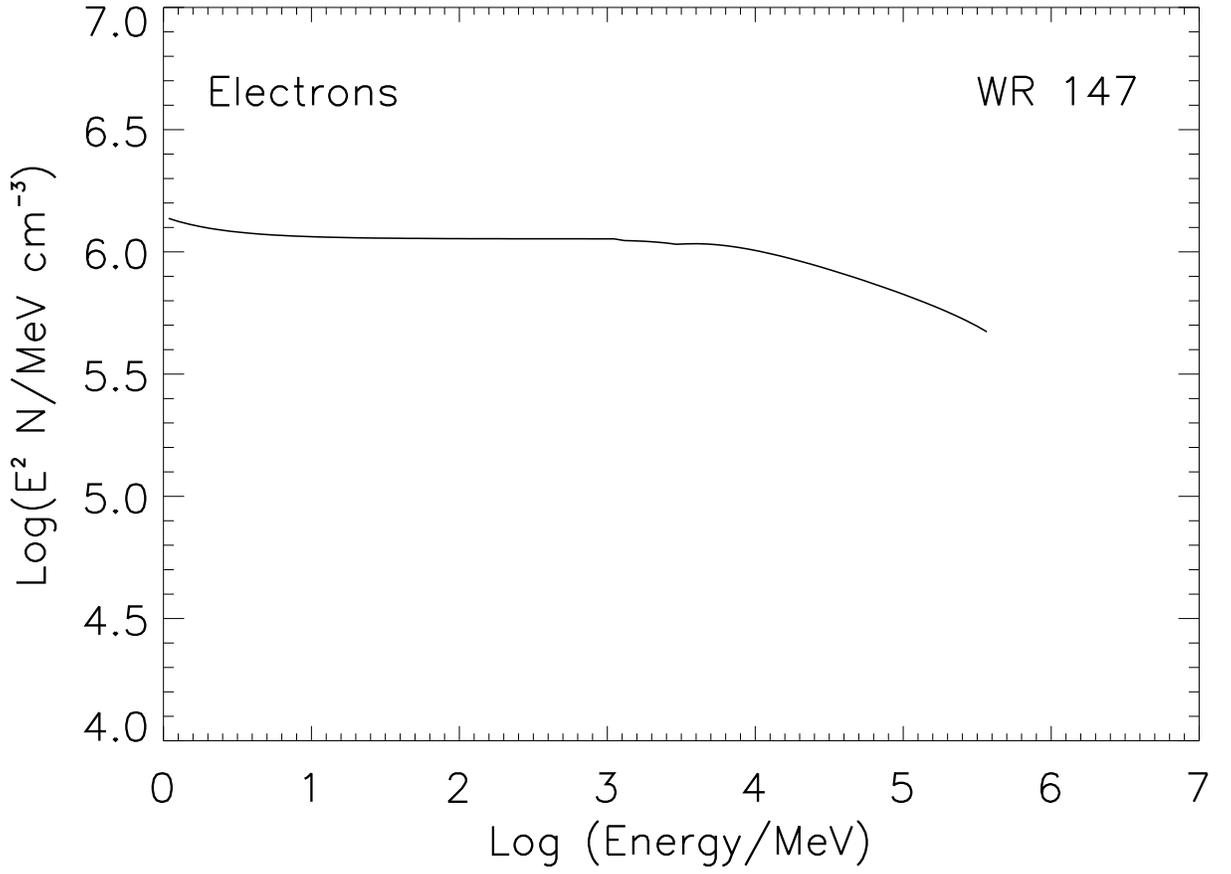}
\caption{Steady-state electron spectrum for WR~147, $Q_0=1$.
See text for parameters and discussion.
}
\label{WR147_e}
\end{figure}

\begin{figure}[t]
\includegraphics[height=13cm]{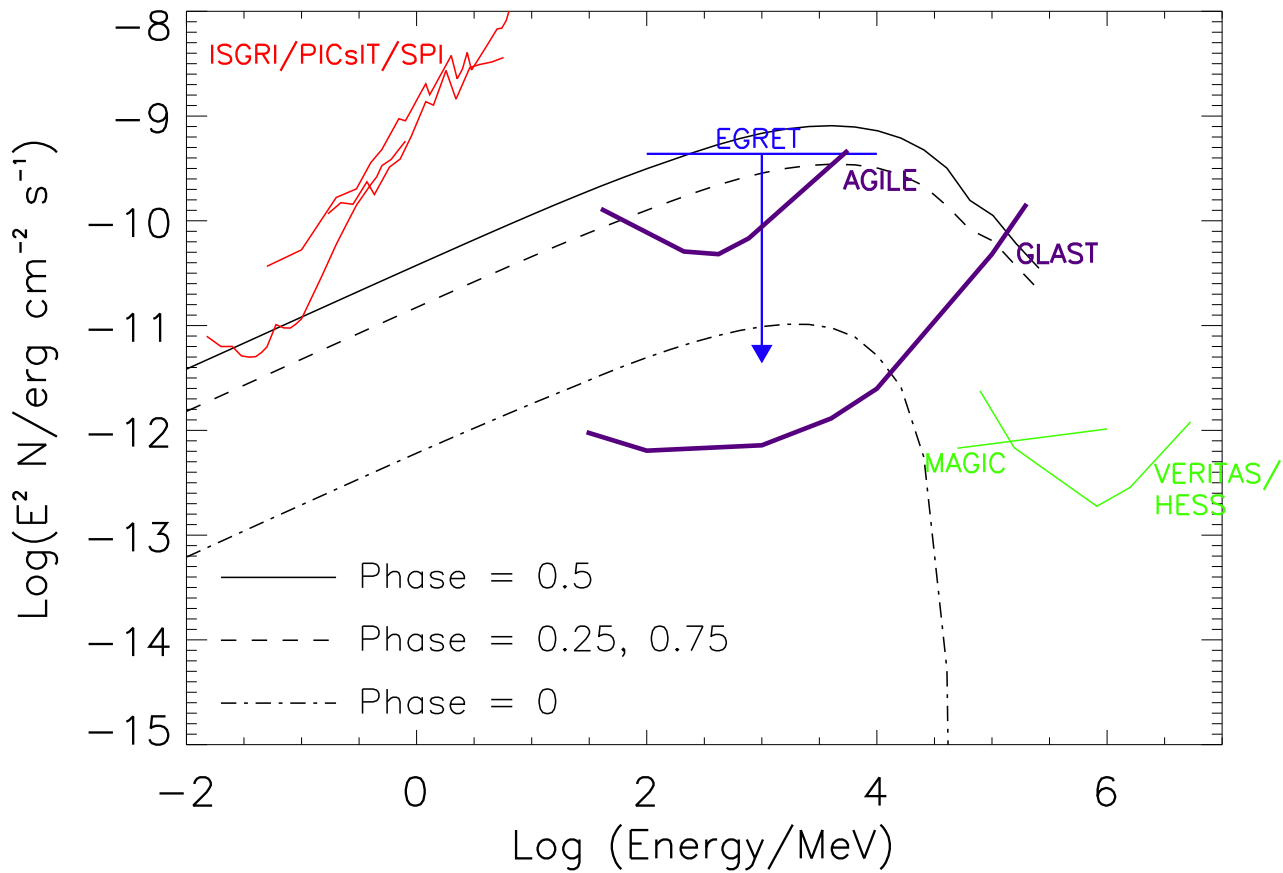}
\caption{IC spectra of WR~147 for orbital phases 0, 0.25, 0.5 and 0.75 for an underlying electron spectrum
as shown in Fig.~\ref{WR147_e}, neglecting any eccentricity of the system and assuming $i=90\degr$. 
See text for further parameters and discussion. $\gamma\gamma$ pair production 
absorbes not more than $\leq$0.3\% ($>50$~GeV) and $\leq$18\% ($>100$~GeV) of the produced flux at
orbital phases 0.25 and 0.5, respectively (not shown in figure). No absorption takes place
at phase 0.
}
\label{WR147_IC}
\end{figure}

In Fig.~\ref{WR147_e} the resulting steady-state electron spectrum is shown. For a negligible system 
eccentricity it is the same
at all orbital phases. The steepening at $10^{4\ldots 5}$MeV is due to synchrotron losses, the slight 
upturn towards lower energies reflects the influence of the Coulomb-losses.
Above $\sim 10^{5.5}$MeV the Bohm-limit cuts off the particle spectrum. 
Fig.~\ref{WR147_IC} demonstrates the orbital IC flux variations due to the anisotropic nature of the 
IC scattering of more 
than one order of magnitude for the chosen
system inclination. The maximum flux level is expected when the WR-star is behind the OB-star along the sight line.
These flux changes will be in principle detectable for GLAST at all orbital phases, and also for low-threshold 
IACTs provided high particle energies are reached, which by itself depends on the acceleration rate.
Thus measurements with IACTs will be able to place important constraints on the acceleration
efficiency in these environments. The EGRET upper limit is taken from \cite{bene2003}.
For an estimated wind gas density of up to $1.4\cdot 10^5$cm$^{-3}$
non-thermal bremsstrahlung emission stays several orders of magnitude below the IC flux level, not detectable
with current instruments to date. Also the contribution from $\pi^0$-decay $\gamma$-rays
remains negligible, even for a maximal possible thermal particle injection rate. 
For WR~147, being a extremely
long-period binary with an assumed small excentricity, the photospheric UV radiation field at the location of the 
collision region is low enough to allow here the neglection of
$\gamma$-ray absorption due to photon-photon collisions.

\section{Detectability from soft to very high energy gamma-rays}

\subsection{Soft gamma-ray instruments}

The regime of soft gamma-rays, here considered to be typical ranging from 511 keV up to several MeV, 
is currently only accessible by the IBIS and SPI instruments aboard INTEGRAL. Previous mission like 
SIGMA/GRANAT and COMPTEL did not reported the detection of soft gamma-ray emission from the WR systems 
studied here. Given the continuum sensitivity of the IBIS and SPI instruments\footnote{INTEGRAL AO-3 documentation, 
SPI and IBIS Observer's Manual, ESA 2004, http://www.rssd.esa.int/Integral/AO3/}, 
accordingly scaled for a common observation time of $10^6$ sec in order to achieve a 3 $\sigma$ detection, 
the detection of WR~140 appears to be possible at soft gamma-rays only if significantly more observation time than 
1 Msec will be dedicated to observations towards such an object. Hard X-ray emission may be seen by ISGRI, most favorably 
in orbital phases where the line of sight towards an observer is parallel to the contact discontinuity of 
the wind collision zone or when both the WR and O star are nearly aligned and the wind collision zone is most 
pronouncedly exposed towards the observer. At soft gamma-rays, even in the most favorable orbital states, several 
Msec may be required to pick up emission from WR~140 (Fig.~\ref{WR140_IC}). This situation is very similar for WR147 
(Fig.~\ref{WR147_IC}).

\subsection{High-energy gamma-ray instruments}

\begin{figure}[t]
\includegraphics[height=13cm]{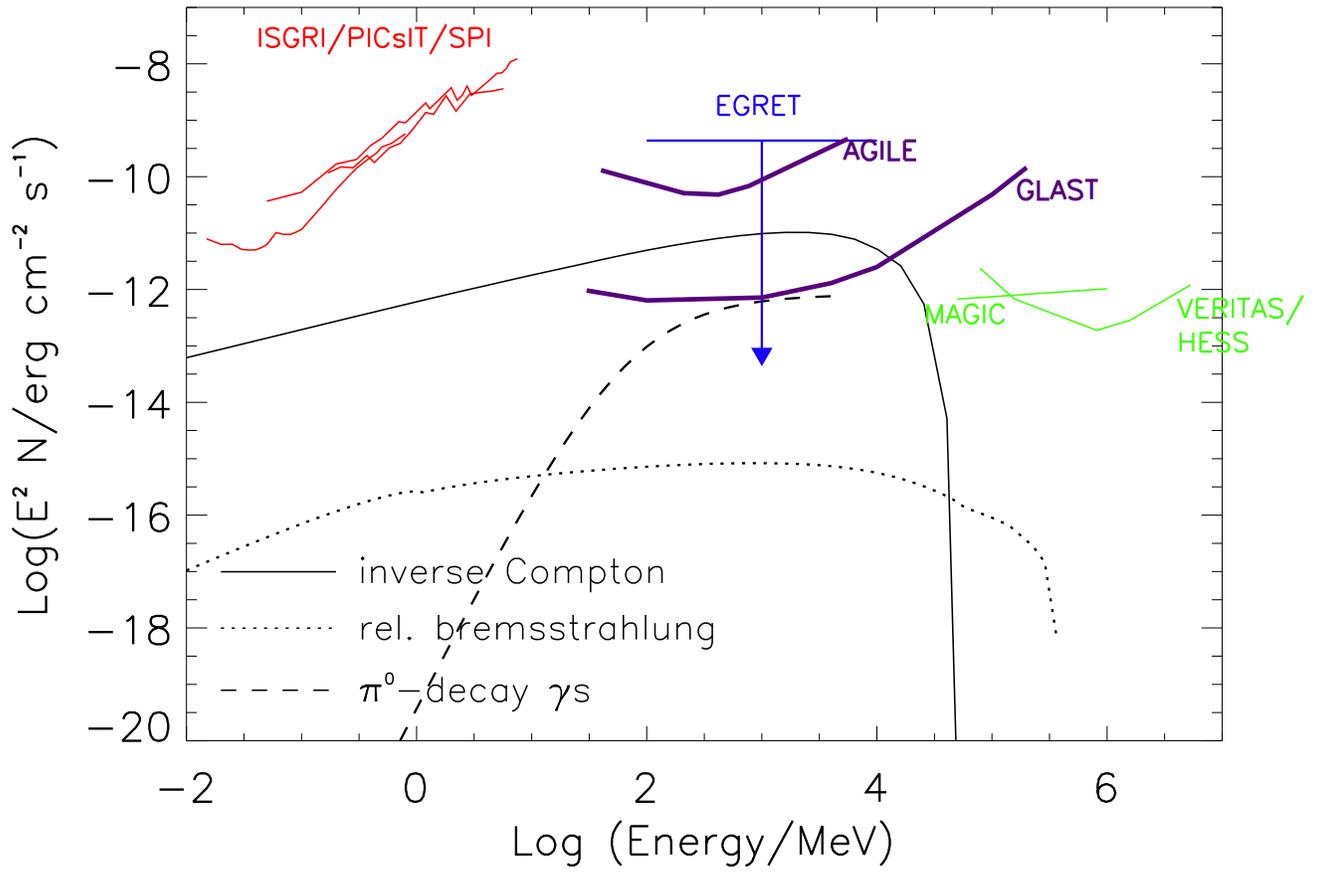}
\caption{Broad band SED of WR~147 at orbital phase 0.
See text for further parameters and discussion.
}
\label{WR147_ph00}
\end{figure}

At energies between 30 MeV and 10 GeV, detection claims of gamma-ray emission from WR binary systems have been made 
already from COS-B observations. More particular, WR~140 (HD193793) was considered to be associated with the variable 
COS-B source 083+03 \citep{Pol87}, a source exhibiting a photon flux at the $5\cdot $10$^{-7} $ph cm$^{-2} $s$^{-1}$ 
level at E $>$ 300 MeV. However, this individual association could not be confirmed by observations of the 
EGRET instrument aboard CGRO, which would have seen a source at the $\sim 1.5\cdot $10$^{-6} $ph cm$^{-2} $s$^{-1}$ 
at E $>$ 100 MeV. Although a number of positional coincidences between unidentified EGRET sources and 
colliding wind binary systems has been noted by \citep{Kau97} and \citep{corr_analysis}, the individual case for an association 
between WR~140 and the unidentified EGRET source 3EG~J2022+4317 appears to be vague on the basis of the 
given observational evidence at $\gamma$-rays. 3EG~J2022+4317 is cataloged \citep{Har99} with a flux of 
$2.5\cdot $10$^{-7} $ph cm$^{-2} $s$^{-1}$, and the spectrum fitted with a power-law with an index of 2.3 $\pm$ 0.2; 
it is further to be characterized by a rather irregular uncertainty contour pointing towards source extension, 
and indication of source confusion above E $>$ 100 MeV. The EGRET observations were taken within the first four 
years of the CGRO mission between 1991 and 1994, with the gross obtained toward the periastron phase of the binary system. 
In the sparse EGRET observations above the detection threshold of the 3EG catalog there is no evidence for 
variable $\gamma$-ray emission \citep{Nol03}.
Since WR~140 is at a distance of 0.67$^o$ to the nominal position of 3EG~J2022+4317, and only barely consistent with
the 99$\%$ source location uncertainty contour, which prevents any conclusive identification between both objects, 
it is appropriate to determine an upper limit at the position of WR~140 under consideration that $\gamma$-ray emission of
3EG~J2022+4317 is fully taken into account but not been associated with WR~140 \citep{Muecke}. This upper limit is consistent 
with our predictions given in Fig.~\ref{WR140_IC}, considering that EGRET observations were performed over a superposition of orbital 
states including the periastron phase, where the cutoff due to IC losses in the Thomson regime will not allow any 
detectable $\gamma$-ray emission at all at energies above a few MeV. 
Since AGILE will exhibit a similar sensitivity characteristics compared to the EGRET instrument, chances for AGILE 
to clarify on 3EG~J2022+4317 are only given if AGILE is operated in orbit over a long period. The instrumental 
sensitivity of AGILE\footnote{http://agile.mi.iasf.cnr.it/Homepage/performances.shtml} is indicated for a 5 $\sigma$ detection on 
the basis of $10^6$ sec observation time. Thus, it'll be the Gamma-ray Large Area Space Telescope (GLAST) to give an 
observational reassessment of the unidentified EGRET source 3EG~J2022+4317 and its association or non-association with WR~140. Clearly, 
GLAST\footnote{http://www-glast.slac.stanford.edu/software/IS/glast\_lat\_performance.htm} 
exhibits a sensitivity characteristics to achieve a detection not only from observations accumulated over various orbital states 
of the colliding wind binary, it has a dedicated chance to provide results from individually selected orbital states. As mentioned above,
during periastron phase there is no high-energy $\gamma$-ray emission predicted (Fig.~\ref{WR140_ph00}), but the changes in the 
$\gamma$-ray flux of WR~140 when going into periastron or coming out of periastron phase 
will be detectable, since the sensitivity of GLAST will be indeed sufficient to detect this system in the more favorable
orbital states when the edges of the contact discontinuity of the wind collision zone lines up in the line-of-sight 
or the wind collision zone is most extensively exposed towards the observer. GLAST will have this detection potential up to energies 
where the the finite acceleration site cuts off the emission, approximately up to $\sim$ 50 GeV. 
The chance to detect high-energy emission in case of WR~147 with future instruments is even somewhat better than for WR~140. 
If AGILE with its better instrumental point spread function than EGRET will be able to distinguish the 
location of WR~147 from the bright $\gamma$-ray source 3EG~J2033+4118, its sensitivity will be sufficient 
to observe $\gamma$-ray emission over the majority of favorable orbital states (Fig.~\ref{WR147_IC}). However, 
the periastron phase will still be out of reach for AGILE, see Fig.~\ref{WR147_ph00}. GLAST, again, will have 
the instrumental capability to both distinguish a test position at the location of WR~147 ($l = 79.85, b = -0.32$)
from the presumably bright $\gamma$-ray source 3EG~J2033+4118. Depending on a possible contribution of further
sub-threshold sources in the vicinity of 3EG~J20233+4118, this might not be too straight forward, though.

\subsection{VHE gamma-ray instruments}

Complementary to the satellite experiments operated at soft and high-energy $\gamma$-rays, the continuous 
development of the Imaging Atmospheric Cherenkov Telescopes (IACTs) towards better sensitivity, even more improved 
angular resolution, and lower energetic thresholds, clearly exhibit the capability to detect $\gamma$-ray emission 
produced in colliding winds of massive stars in binary systems. Especially the chance to accumulate a wealth of photons 
on sub-hour time scales will enable the IACTs to test the predicted orbital flux variations precisely. A unique chance 
to detect emission from WR~140 is given for those experiments able to work at the lowest possible threshold, i.e. MAGIC below 100 GeV. 
Depending on the actual shape of the cutoff due to the finite site of the acceleration site, events may be seen only in extreme 
low-energetic event selections, and the respective sky location is further characterized by an absence of higher energetic photons. 
For an array of IACTs like VERITAS \citep{Weekes} the best sensitivity is achieved at energies above the cutoff due to the finite 
acceleration site in WR~140, subsequently making any detection prospects heavily dependent on the actual shape of the cutoff. 
IACT arrays located in the southern hemisphere like H.E.S.S., or CANGAROO may not have the chance detect this system at all due  
to the higher energetic threshold when observing under low zenith angle conditions. In case of WR~147, where the orbital parameters 
are more promising to detect high-energy $\gamma$-ray emission up to several hundreds of GeVs, low threshold IACTs located in the 
northern hemisphere will have a distinct chance to detect WR147 in non-periastron orbital phases (Fig.~\ref{WR147_IC}), provided 
that the $\gamma$-ray analysis will not introduce hard cuts for preference of a higher energetic event selection. 
For both satellite and ground-based instruments the colliding wind zone will not appear to be spatially resolved,
presenting individual colliding wind binary systems as point-source candidates at the $\gamma$-ray sky.

\section{Conclusions and discussion}

In this work we have calculated the emission from non-thermal steady-state particle spectra built up in the regions of
colliding hypersonic winds (assumed to be homogeneous) of long-period massive binary systems
with the stagnation point defined by balancing the wind momenta and under the assumption of spherical winds.
The shocked high-speed winds are creating a region of hot gas that is separated by a 
contact discontinuity. The gas flow in this region away from the stagnation point
will be some fraction of the wind velocity which we kept constant here.
A simplification of the geometry
from a bow-shaped to a cylinder-shaped collision region
allowed us to solve the relevant diffusion-loss equations analytically.
We considered first order Fermi acceleration out of a 
pool of thermal particles, and
took into account radiative losses (synchrotron, inverse Compton including Klein-Nishina effects, 
bremsstrahlung and Coulomb losses), (energy-independent) diffusion by introducing a constant escape time $T_0$ 
and convection/advection with constant speed.
Above a certain distance from the stagnation point convection dominates over diffusion with the transition point
determined by balancing the diffusion and convection loss time. Correspondingly,
we devided the emission region into a region where acceleration/diffusion dominates,
the ''acceleration zone'', and the outer region where convection/advection dominates, the ''convection zone''.

Electrons may reach relativistic energies, once they overcome the heavy Coulomb losses in the
dense shocked material, through diffusive shock acceleration up to the Bohm diffusion limit. For wide binary systems
this latter constraint is often severe, while for close binaries, radiative losses mostly cause the cutoff.
Taking into account existing upper limits of the stellar surface magnetic field strength of massive stars,
inverse Compton losses in general dominate over synchrotron losses if in the Thomson loss regime. We have shown, however,
that losses may well extend into the transition region leading to the extreme Klein-Nishina regime. The flattening
of the Compton loss rate there may in some cases cause the synchrotron losses dominate eventually.
{\it Thus a rigorous treatment of the Compton losses must include Klein-Nishina effects.} For this purpose we have derived
analytical approximations for Klein-Nishina losses that are suitable for massive binary systems, and allow at the
same time to solve the relevant diffusion-loss equation analytically.
 Despite the high density environment of the emitting collision region, non-thermal bremsstrahlung losses prove
general to be of minor importance. This turns out to be true also for the corresponding radiation.

We have studied inverse Compton radiation, the main emission channel for relativistic electrons in these 
systems, in severe detail. 
The use of the full Klein-Nishina cross section leads to a spectral softening at the high-energy end 
of the emitted radiation. 
Since the stellar target photons for inverse Compton scattering arrive at the
collision region from a prefered direction, the full angular dependence of the scattering process has to be 
considered. Its anisotropic nature leads to variations of the flux level by up to several orders of magnitude
(depending on system inclination and eccentricity) as well as cutoff energy with orbital phase.
The maximum flux and cutoff energy occurs when the WR-star lies behind the OB-star. We consider therefore
massive binary systems as $\gamma$-ray sources that are variable on the time scale of their orbital period
even in the absence of a strong system eccentricity.
The inclusion of convection/advection effects into the calculation of the particle spectra reveals a 
possibly visible
spectral feature, too. Because of a deficit of low-energy particles in the convection zone, a softening of
the volume-integrated radiation spectrum may occur if the convection zone is sufficiently large compared 
to the acceleration zone. A detection of this feature would give valuable information about the particle
propagation properties in the emission region.

Since thermal protons are most likely wind constituents as well, diffusive shock acceleration implies
the presence of relativistic protons in the wind collision region. If they reach energies
of several GeV, their presence may show up as $\pi^0$-decay $\gamma$-rays produced through inelastic proton-proton
collisions. Their detectability, however, depends not only on the relativistic electron-to-proton ratio and the
instrument capabilities, but also on the importance of the competing radiation mechanisms. E.g. in the case of
WR~140 close to periastron, the otherwise dominating inverse Compton radiation most likely cease to reach
sufficient high energies that would allow MeV-GeV emission, which increases the chance of detecting 
$\pi^0$-decay $\gamma$-rays.

Finally, we find that photon-photon pair production can not be neglected if the produced radiation
exceeds energies of $\sim (kT/{\rm eV})^{-1}$ TeV, which lies typically at 50-100 GeV. The absorption optical 
depth thereby depends sensitively on orbital phase and system inclination.

Although many free parameters are involved in the presented model for high energy emission
from the wind collision region of massive binaries, few are those which are unrelated to observations, 
and even fewer those which -- if changing -- may have a significant impact on the predicted $\gamma$-ray 
intensity.
Indeed, since IC emission seems the dominant emission process at high energies in most cases, the
high energy output can directly be deduced from the knowledge of the synchrotron emission.
While the non-thermal radio flux level provides information on the required
injected energy in form of electrons if the magnetic field is known, its radio spectrum constrains 
acceleration and propagation parameters. Equipartition arguments together with 
lower magnetic field limits from the Razin-Tsytovich effect \citep[e.g.][]{Chen92,bene2003},
and observational limits on the stellar surface magnetic field in connection with a 
plausible dipole field configuration
can be used to estimate the magnetic field strength in the collision region.
The fact that relativistic electrons exist, supplies a lower limit on the
acceleration rate, while an upper limit is given by the Bohm diffusion regime
for the likely case of diffusive shock acceleration operating in these objects.

We applied our model to two archetypical WR-systems: WR~140 is arguable the most popular
among these sources. We predict WR~140 to be detectable with GLAST and MAGIC, except at phases
close to periastron due to an early cutoff of the electron spetrum already at $\sim 100$~MeV.
This may lead to the dominance of bremsstrahlung and hadronically produced $\gamma$-rays above $\sim 1$MeV,
at this phase, while at phases far from periastron 
inverse Compton radiation is predicted to dominate at all energies.
Orbital flux variations at high energies far from periastron are expected with amplitudes that vary by a factor $\sim 2$.

The $\sim$factor 10 wider binary system WR~147 is notable for being the brightest (because closest) system
at radio frequencies, and for being one of the few systems where the thermal and non-thermal radio emission
are observed to arise from spatially different resolved regions. Due to a lack of knowledge of the system
parameters we model WR~147 face-on with no significant eccentricity. The low target photon density
at the collision location makes photon absorption negligible here, and at the same time allows
the electron spectrum to extend up to sub-TeV-energies if the acceleration efficiency is favorable. 
This would lead to radiation up to the 100 GeV-region on a flux level possibly detectable
even with VERITAS at some orbital phases, while GLAST has good chances to trace this system
at all phases. INTEGRAL's sensitivity at $\gamma$-ray energies will most likely be insufficient to 
discover these sources as $\gamma$-ray emitter given the finite amount of observation time in 
individual instrumental pointings.

In this work we concentrated on the emission from a steady-state particle spectrum.
Being time-dependent systems in general, 
a time-dependent diffusion-loss equation shall give a more realistic description of the
emitted intensity. In this case we expect that, similar to supernova remnants, the electron spectrum 
will slowly be built up
with the maximum particle energy increasing with time. Typically the electron spectrum is fully developped
after a few tens of hours. The uncertainty of a given phase corresponding to the here calculated
steady-state emission therefore lies typically in this time range.
A more comprehensive discussion of the behaviour of massive colliding wind systems in the framework of
a time-dependent diffusion-loss equation will be considered in a forthcoming paper.

In conclusion, we consider colliding wind regions of massive binary systems that are wide enough to
avoid radiative braking, as promising sources of high energy emission that may extend far beyond
the X-ray band. 
High energy observations of these systems by sensitive, low-threshold IACTs and satellite instruments 
can be used not only to derive geometrical details but also to explore the efficiency of diffusive shock 
acceleration at densities much higher than in other astronomical objects with high Mach number shocks, 
e.g. supernovae.

\acknowledgments
This work was partially supported by DESY-HS, project 05CH1PCA/6.

\newpage

\appendix

\section{Solutions for steady-state electron distributions in the acceleration region}
\label{app_e}

Here we derive the analytical solutions of Eq.~\ref{PDE1} for electrons suffering radiative losses following 
Eq.~\ref{elec_radlosses}.
 
If $\Delta>0$ then
\begin{eqnarray}
N(E) & = & \frac{Q_0}{(a-b_{\rm br})E-b_{\rm syn\& IC}E^2-b_{\rm coul}} \\
& &\left[\frac{(H(E)-\sqrt{\Delta})(H(E_0)+\sqrt{\Delta})}
{(H(E)+\sqrt{\Delta})(H(E_0)-\sqrt{\Delta})}\right]^{-\frac{1}{\sqrt{|\Delta|}T_0}}
\label{elec_solution1}
\end{eqnarray}
for $E\leq E_c = \min\left[(a-b_{\rm br}-\sqrt{\Delta})/2b_{\rm syn\& IC},(a-b_{\rm br}+\sqrt{\Delta})
/2b_{\rm syn\& IC}\right]$
and with $b_{\rm syn\& IC} = b_{\rm syn} + b_{\rm IC,TL}$, $H(E) = a-b_{\rm br}-2b_{\rm syn\& IC}E$
and $\Delta=(a-b_{\rm br})^2-4b_{\rm syn\& IC}b_{\rm coul}$,
$N(E) = 0$ for $E>E_c$.
If $\Delta\leq 0$ then
\begin{eqnarray}
N(E) & = & \frac{Q_0}{(a-b_{\rm br})E-b_{\rm syn\& IC}E^2-b_{\rm coul}} \\
& &\exp\left[-\frac{2}{T_0\sqrt{|\Delta|}}\arctan\left(\frac{2b_{\rm syn\& IC}(E_0-E)\sqrt{|\Delta|}}
{((a-b_{\rm br})-2b_{\rm syn\& IC}E)((a-b_{\rm br})-2b_{\rm syn\& IC}E_0)-\Delta}\right)\right]
\label{elec_solution2}
\end{eqnarray}
It is an exact solution for $E<E_{\rm TL}$.
With the approximation Eq.~\ref{KN1},\ref{KN2} for the inverse Compton losses Eq.~(\ref{elec_solution1}, 
\ref{elec_solution2}) describes the electron spectrum
for $E>E_{\rm s}$ if one substitutes
$b_{\rm syn\& IC}\rightarrow b_{\rm syn}$, $b_{\rm br}\rightarrow b_{\rm br}+ q_a$ and
$b_{\rm coul}\rightarrow b_{\rm coul}+ q_b$.
In practice, however, relativistic bremsstrahlung and Coulomb losses above $E>E_{\rm TL}$ can usually be neglected
in the systems considered here.
For $E_{\rm TL}\leq E\leq E_{\rm s}$ the solution of
Eq.~\ref{PDE1} reads:
\begin{eqnarray}
N(E) & = & \frac{Q_0}{E[(a-b_{\rm br})-b_{\rm syn\& IC}E+b_{\rm IC,TL}E^2/E_g]}\\
& & \left(\frac{E^2((a-b_{\rm br})-b_{\rm syn\& IC}E_0+b_{\rm IC,TL}E_0^2/E_g)}
{E_0^2((a-b_{\rm br})-b_{\rm syn\& IC}E+b_{\rm IC,TL}/E_g E^2)}\right)^{-1/[2(a-b_{\rm br})T_0]}\\
& & \left(\frac{(2b_{\rm IC,TL}E/E_g-b_{\rm syn\& IC}-\sqrt{\Delta}) 
(2b_{\rm IC,TL}E_0/E_g-b_{\rm syn\& IC}+\sqrt{\Delta})}
{(2b_{\rm IC,TL}E/E_g-b_{\rm syn\& IC}+\sqrt{\Delta}) 
(2b_{\rm IC,TL}E_0/E_g-b_{\rm syn\& IC}-\sqrt{\Delta})}\right)^{-b_{\rm syn\& IC}/[2T_0(a-b_{\rm br})\sqrt{\Delta}]}
\label{elec_solution3}
\end{eqnarray}
for $\Delta=b_{\rm syn\& IC}^2-4(a-b_{\rm br})b_{\rm IC,TL}/E_g >0$ and $N(E>E_c) = 0$ with 
$E<E_c=min[(b_{\rm syn\& IC}+\sqrt{\Delta})E_g/2b_{\rm IC,TL},
(b_{\rm syn\& IC}-\sqrt{\Delta})E_g/2b_{\rm IC,TL}]$. 
For $\Delta<0$ the solution is:
\begin{eqnarray}
N(E) & = & \frac{Q_0}{E[(a-b_{\rm br})-b_{\rm syn\& IC}E+b_{\rm IC,TL}E^2/E_g]}\\
& & \left(\frac{E^2((a-b_{\rm br})-b_{\rm syn\& IC}E_0+b_{\rm IC,TL}E_0^2/E_g)}
{E_0^2((a-b_{\rm br})-b_{\rm syn\& IC}E+b_{\rm IC,TL}E^2/E_g)}\right)^{-1/[2(a-b_{\rm br})T_0]}\\
& & \exp\left[-\frac{b_{\rm syn\& IC}}{(a-b_{\rm br})T_0\sqrt{|\Delta|}} 
\arctan\left({\frac{2b_{\rm IC,TL}/E_g\sqrt{|\Delta|}(E-E_0)}
{(2b_{\rm IC,TL}E/E_g-b_{\rm syn\& IC})(2b_{\rm IC,TL}E_0/E_g-b_{\rm syn\& IC})-\Delta}}\right)\pm \pi\right] .
\label{elec_solution4}
\end{eqnarray}


\section{Solutions for steady-state nucleon distributions in the acceleration region}
\label{app_p}

Here we derive analytical solutions for steady-state nucleon spectra solving Eq.~\ref{PDE1}. 

At non-relativistic energies and below the Coulomb-barrier the solution is:
\begin{equation}
N(E)  =  \frac{Q_0}{(a-b_{\rm bel})E}\left(\frac{E}{E_0}\right)^{-\frac{1}{(a-b_{\rm bel})T_0}}
\end{equation}
for $E<E_c$ and $N(E>E_c)=0$, 
while above the Coulomb-barrier $E>E_m$ we find:
\begin{equation}
N(E)  =  \frac{Q_0}{aE-b_{\rm ab}E^{-0.5}} \left(\frac{E_c^{1.5}-E^{1.5}}{E_c^{1.5}-E_0^{1.5}}\right)^{-\frac{2}{3aT_0}}
\end{equation}
for $E_0>E_m$, $E_c = (b_{\rm ab}/a)^{2/3}$,
\begin{equation}
N(E)  =  \frac{Q_0}{aE-b_{\rm ab}E^{-0.5}} \left(\frac{E_c^{1.5}-E^{1.5}}{E_c^{1.5}-E_0^{1.5}}\right)^{-\frac{2}{3aT_0}}
\left(\frac{E_m}{E_0}\right)^{-\frac{1}{(a-b_{\rm bel})T_0}}
\end{equation}
for $E_0<E_m$.

At relativistic energies $> m_p c^2=E_{\rm rel}$ and above $E>E_{\rm thr}$, 
the $pp$-interaction threshold, one finds:
\begin{eqnarray}
N(E) & = & \frac{Q_0}{(a-b_{\rm pp})E-b_{\rm rel}}\left(\frac{(a-b_{\rm pp})E-
b_{\rm rel}}{(a-b_{\rm pp})E_0-b_{\rm rel}}\right)^{-\frac{1}{(a-b_{\rm pp})T_0}}
\left(\frac{E_m}{E_0}\right)^{-\frac{1}{(a-b_{\rm bel})T_0}} \\
& & \times \left(\frac{E_c^{1.5}-E_{\rm rel}^{1.5}}{E_c^{1.5}-E_m^{1.5}}\right)^{-\frac{2}{3aT_0}}
\left(\frac{a E_{\rm thr}-b_{\rm rel}}{a E_{\rm rel} - b_{\rm rel}}\right)^{-\frac{1}{aT_0}}
\end{eqnarray}
for $E_0\leq E_m$,
\begin{equation}
N(E)  =  \frac{Q_0}{(a-b_{\rm pp})E-b_{\rm rel}}\left(\frac{(a-b_{\rm pp})E-b_{\rm rel}}{(a-b_{\rm pp})
E_0-b_{\rm rel}}\right)^{-\frac{1}{(a-b_{\rm pp})T_0}}
\left(\frac{E_c^{1.5}-E_{\rm rel}^{1.5}}{E_c^{1.5}-E_0^{1.5}}\right)^{-\frac{2}{3aT_0}}
\left(\frac{a E_{\rm thr}-b_{\rm rel}}{a E_{\rm rel} - b_{\rm rel}}\right)^{-\frac{1}{aT_0}}
\end{equation}
for $E_m<E_0\leq E_{\rm rel}$, and
\begin{equation}
N(E)  =  \frac{Q_0}{(a-b_{\rm pp})E-b_{\rm rel}}\left(\frac{(a-b_{\rm pp})E-b_{\rm rel}}{(a-b_{\rm pp})
E_0-b_{\rm rel}}\right)^{-\frac{1}{(a-b_{\rm pp})T_0}}
\left(\frac{a E_{\rm thr}-b_{\rm rel}}{a E_0 - b_{\rm rel}}\right)^{-\frac{1}{aT_0}}
\end{equation}
for $E_{\rm rel}<E_0\leq E_{\rm thr}$.

Finally, at relativistic energies but below $E<E_{\rm thr}$ the steady-state spectrum turns out to follow:
\begin{equation}
N(E)  =  \frac{Q_0}{aE-b_{\rm rel}}\left(\frac{aE-b_{\rm rel}}{aE_m-b_{\rm rel}}\right)^
{-\frac{1}{aT_0}}
\left(\frac{E_m}{E_0}\right)^{-\frac{1}{(a-b_{\rm bel})T_0}}
\left(\frac{E_c^{1.5}-E_{\rm rel}^{1.5}}{E_c^{1.5}-E_m^{1.5}}\right)^{-\frac{2}{3aT_0}}
\end{equation}
for $E_0\leq E_m$, and
\begin{equation}
N(E)  =  \frac{Q_0}{aE-b_{\rm rel}}\left(\frac{aE-b_{\rm rel}}{aE_m-b_{\rm rel}}\right)^
{-\frac{1}{aT_0}}
\left(\frac{E_c^{1.5}-E_{\rm rel}^{1.5}}{E_c^{1.5}-E_0^{1.5}}\right)^{-\frac{2}{3aT_0}}
\end{equation}
for $E_m<E_0\leq E_{\rm rel}$.


\section{Analytical calculation of the inverse Compton scattering rate}
\label{app_IC}

In the lab frame we choose a polar coordinate system such that
the line-of-sight marks the z-axis.
A single incident electron is then fully described by its Lorentz
factor, $\gamma$, and the polar angle, $\mu_e= \cos\theta_e$,
and azimuthal angle, $\phi_e$, to mark its direction of flight.

The photon field is described by the differential photon spectrum
\begin{equation}
\nph = {{dn}\over {dV\,d\epsilon\,d\Omega_{ph}}}
\end{equation}
where $\epsilon=E_{ph} /m_e c^2$ is the dimensionless photon energy.

The scattering rate (quantities of the scattered photon are
indexed with $s$) for a given
differential number density of electrons, $\nel$,
is
\begin{displaymath}
\nsca =\oint\int d\gamma d\Omega_e\ \nel 
\end{displaymath}
\begin{equation}
\left[c\oint\int 
(1-\beta \mu_{ph,e})\, \nph\, 
{{d\sigma}\over  {d\Omega_s\,d\epsilon_s}}\  d\epsilon d\Omega_{ph} 
\right] 
\end{equation}
where the part in brackets is the scattering rate for a single electron
and $\mu_{ph,e}$ is the cosine of the angle between the photon
and electron flight directions, i.e.
\begin{equation}
\mu_{ph,e} = \mu_e\mu_{ph} + \sqrt{1-\mu_e^2}\sqrt{1-\mu_{ph}^2} \cos \phi_e
\label{A6}
\end{equation}
The differential cross section is well known in the electron rest frame.
It is therefore attractive to 
calculate the scattering rate per single electron in its rest frame (ERF,
indicated by an asterisk) and then to transform the result back
into the lab frame. Since then only photon spectra have to be transformed
we can use the invariants
\begin{equation}
{\nph\over {\epsilon^2}}\qquad \qquad {{\dot \nph}\over \epsilon} \label{inv}
\end{equation}

\subsection{The scattering rate for a single electron}

In the following we will assume that the electrons have an isotropic
distribution, in which case the scattering rate can not depend on the azimuthal 
angle of the incoming photons. Therefore we can set $\phi_{ph}=0$.

The Lorentz transformation relating lab frame quantities to those in the ERF
are
\begin{equation}
\mu_{ph,e}^\ast = {{\mu_{ph,e} -\beta}\over {1-\beta \mu_{ph,e}}}
\end{equation}
\begin{equation}
\epsilon^\ast = \gamma \,\epsilon\,(1-\beta \mu_{ph,e})
={\epsilon\over {\gamma\,(1+\beta \mu_{ph,e}^\ast)}}
\end{equation}
We also need to know the photons azimuthal angle with respect to the electron.
Let the azimuthal angle of the line-of-sight be $\phi_{los,e}=0$.
Then
\begin{equation}
\cos \phi_{ph,e} ={{\mu_{ph}\sqrt{1-\mu_e^2} 
-\cos\phi_e\,\mu_e \sqrt{1-\mu_{ph}^2}}\over \sqrt{1-\mu_{ph,e}^2}}
\end{equation}
\begin{equation}
\sin \phi_{ph,e} = {{\sin\phi_e\,\sqrt{1-\mu_{ph}^2}}\over
\sqrt{1-\mu_{ph,e}^2}}
\end{equation}
Note that $\phi_{ph,e}=\phi_{ph,e}^\ast$.

The differential cross section in the ERF is
\begin{displaymath} 
{{d\sigma}\over 
{d\Omega_s^\ast\,d\epsilon_s^\ast}} = {{r_e^2\,(1+\cos^2\xi)
\delta\left[\epsilon_s^\ast - {{\epsilon^\ast}\over 
{1+\epsilon^\ast (1-\cos\xi)}}\right]}\over 
{2\,(1+\epsilon^\ast
(1-\cos\xi))^2}} 
\end{displaymath}\begin{equation}
\times\ \left[1+{{{\epsilon^\ast}^2 (1-\cos \xi)^2}\over
{(1+\cos^2\xi)(1+\epsilon^\ast (1-\cos\xi))}}\right]\ 
 \label{cross}
\end{equation}
where $\xi$ is the scattering angle.

Given the photon angles after scattering, $\mu_{s,e}^\ast$ 
and $\phi_{s,e}^\ast= \phi_{s,e}$, the scattering angle can be calculated as 
\begin{displaymath}
\cos\xi = \mu_{ph,e}^\ast\mu_{s,e}^\ast
\end{displaymath}
\begin{equation}
+ \sqrt{1-{\mu_{ph,e}^\ast}^2}
\sqrt{1-{\mu_{s,e}^\ast}^2} \cos (\phi_{ph,e} - \phi_{s,e})
\end{equation}
where
\begin{equation}
\cos (\phi_{ph,e} - \phi_{s,e})= \cos \phi_{ph,e}\,\cos \phi_{s,e}+
\sin \phi_{ph,e}\,\sin \phi_{s,e}
\end{equation}
Then the scattering rate in the ERF is
\begin{equation}
\dot n^\ast (\epsilon_s^\ast, \Omega_s^\ast)
= c\int\oint {{\epsilon^\ast}\over {\gamma\,\epsilon}}\,\nph\,
{{d\sigma}\over {d\epsilon_s^\ast d\Omega_s^\ast}} d\Omega d\epsilon
\end{equation}
which can be transformed back into the lab system using Eq.~\ref{inv}
and
\begin{equation}
\mu_{s,e} = {{\mu_{s,e}^\ast +\beta}\over {1+\beta\mu_{s,e}^\ast}}
\end{equation}
\begin{equation}
\epsilon_s = \epsilon_s^\ast \gamma (1+\beta\mu_{s,e}^\ast)=
{{\epsilon_s^\ast}\over {\gamma (1-\beta\mu_{s,e})}}
\end{equation}
Finally we obtain the scattering rate in the lab frame for a population 
of electrons
\begin{displaymath}
\nsca = \int \oint \int \oint
d\epsilon d\Omega_{ph} d\gamma d\Omega_e 
\end{displaymath}
\begin{equation}
{{c\,(1-\beta \mu_{ph,e})}\over {\gamma (1-\beta \mu_{s,e})}}\,
\nph\,\nel\,{{d\sigma}\over {d\epsilon_s^\ast d\Omega_s^\ast}} 
\end{equation}

\subsection{The scattering rate for an arbitrary isotropic electron spectrum}

Note that the line-of-sight is defined by
$\phi_{s,e}=0$ and $\mu_{s,e}=\mu_e$.
Therefore $\cos (\phi_{ph,e} - \phi_{s,e})= \cos \phi_{ph,e}$. Also 
$\left[1-{\mu^\ast}^2\right]= {{1-\mu^2}\over {\gamma^2 \,(1-\beta\mu)^2}}$,
hence
\begin{displaymath}
\cos\xi = {{(\mu_e -\beta)(\mu_{ph,e}-\beta) + \gamma^{-2}
\sqrt{1-\mu_e^2}\sqrt{1-\mu_{ph,e}^2}\, \cos \phi_{ph,e}}\over
{(1-\beta\mu_e)(1-\beta\mu_{ph,e})}}
\end{displaymath}
\begin{equation}
\hphantom{\cos\xi}
=1- {{1-\mu_{ph}}\over
{\gamma^2\,(1-\beta\mu_e)(1-\beta\mu_{ph,e})}}
\end{equation}
\begin{equation}
1-\cos\xi = {{1-\mu_e\mu_{ph,e} -\sqrt{1-\mu_e^2}\sqrt{1-\mu_{ph,e}^2}\, 
\cos \phi_{ph,e}}\over{\gamma^2\,(1-\beta\mu_e)(1-\beta\mu_{ph,e})}}
\end{equation}
\begin{equation}
\epsilon^\ast (1-\cos\xi) =\epsilon\,
{{1-\mu_e\mu_{ph,e} -\sqrt{1-\mu_e^2}\sqrt{1-\mu_{ph,e}^2}\, 
\cos \phi_{ph,e}}\over{\gamma\,(1-\beta\mu_e)}}
= \epsilon {{1-\mu_{ph}}\over {\gamma (1-\beta \mu_e)}}
\end{equation}
The scattering rate does not depend on $\phi_{ph}$, this integral thus being
trivial. For the angles $\phi_e$ and $\mu_e$ one is trivially performed
by using the delta-functional in the cross section and the other one
has to be done explicitely.

The delta functional  in Eq.~\ref{cross} can be rewritten as
\begin{equation}
\delta\left[\epsilon_s - {{\epsilon\,\gamma
(1-\beta\mu_{ph,e})}\over {\gamma (1-\beta\mu_e) +\epsilon
\left(1-\mu_{ph}\right)}}\right]= 
\delta\left[f(\mu_e, \mu_{ph}, \phi_e)\right]
\end{equation}
With $n_{ph}(\mu_{ph})= \int d\phi_{ph}\ n_{ph}(\Omega_{ph})$
the scattering rate is
\begin{equation}
\nsca = \int d\epsilon\ \int d\mu_{ph}\ \int d\gamma\ \oint d\Omega_e\ 
{{ r_e^2 c\, \nel\,\nphm\,(1-\beta\mu_{ph,e})}\over 
{2\left[\gamma (1-\beta\mu_e) + \epsilon (1-\mu_{ph})\right]^2}} 
\delta\left[f(\mu_e, \mu_{ph}, \phi_e)\right]
\end{equation}
\begin{displaymath}
\hphantom{\nsca}\times\  
\left[1+\left(1- {{1-\mu_{ph}}\over
{\gamma^2\,(1-\beta\mu_e)(1-\beta\mu_{ph,e})}}\right)^2 +
{{\epsilon^2 (1-\mu_{ph})^2}\over {\gamma (1-\beta\mu_e)
\left[\gamma (1-\beta\mu_e) + \epsilon (1-\mu_{ph})\right]}}\right] \ 
\end{displaymath}
Given the argument $f$ of the delta-functional
\begin{equation}{{
(1-\beta\mu_{ph,e})}\over {\gamma (1-\beta\mu_e) +\epsilon
\left(1-\mu_{ph}\right)}}= {{\epsilon_s}\over {\gamma\epsilon}}
\end{equation}
and
\begin{equation}
{1\over {(1-\beta\mu_{ph,e})}} = {{{\gamma\epsilon}\over {\epsilon_s}}\over 
{\gamma (1-\beta\mu_e) +\epsilon \left(1-\mu_{ph}\right) }}
\end{equation}
as well as
\begin{equation}
\delta (f) = \left[\gamma (1-\beta\mu_e) +\epsilon
\left(1-\mu_{ph}\right)\right] \,\delta\left[{\bf g}=\epsilon_s\left(
\gamma (1-\beta\mu_e) +\epsilon
\left(1-\mu_{ph}\right)\right)-\epsilon\gamma (1-\beta\mu_{ph,e})\right]
\end{equation}
Using the isotropy of cosmic ray electrons, $\nel =
{1\over {4\pi}} \nelg$ we then obtain for the scattering rate
\begin{equation}
\nsca = {{r_e^2 c\,\epsilon_s}\over {8\pi}} \int d\epsilon\ \int 
d\mu_{ph}\ \int d\gamma\ \oint d\Omega_e\ 
{{ \nelg\,\nphm}\over {\gamma\epsilon}} \ 
\delta\left[g(\mu_e, \mu_{ph}, \phi_e)\right] \label{A11}
\end{equation}
\begin{displaymath}
\times\ 
\left[1+\left(1- {{\epsilon (1-\mu_{ph})}\over
{\gamma\epsilon_s\,(1-\beta\mu_e)\left[\gamma (1-\beta\mu_e) + 
\epsilon (1-\mu_{ph})\right]}}\right)^2 +
{{\epsilon^2 (1-\mu_{ph})^2}\over {\gamma (1-\beta\mu_e)
\left[\gamma (1-\beta\mu_e) + \epsilon (1-\mu_{ph})\right]}}\right] 
\end{displaymath}
Now we need to find the zeros of the delta functional.
Inserting Eq.~\ref{A6} into $g=0$ we obtain
\begin{equation}
\epsilon_s\left[1+{\epsilon\over \gamma} (1-\mu_{ph})\right]-\epsilon
\qquad - \beta (\epsilon_s -\mu_{ph}\epsilon)\,\mu_e = -\beta \epsilon
\cos\phi_e\,\sqrt{1-\mu_{ph}^2}\,\sqrt{1-\mu_e^2}
\end{equation}
which is of the form
\begin{equation}
A-B\mu_e =-C\sqrt{1-\mu_e^2} \label{A10}
\end{equation}
where $C$ can be positive and negative, depending on $\phi_e$.
Also $A > 0$ and $B > 0$ for the interesting case $\epsilon_s > \epsilon$.
Real zeroes $\mu_0$ of the above equation exist, if $B^2 +C^2 -A^2 \ge 0$. 

\subsection{The case $\mu_{ph}=\pm 1$}

If $\mu_{ph}= \pm 1$ we have $C=0$ and then $\mu_0 = {A\over B}$.
The delta-functional transforms as
\begin{equation}
\delta (g) = {1\over {\gamma B}}\,
\delta \left(\mu_e -{A\over B}\right)
\end{equation}
and
\begin{equation}\nsca\bigg\vert_{\mu_{ph}= \pm 1} 
= {{r_e^2 c\,\epsilon_s}\over {4}} 
\int d\epsilon\ \int d\gamma\ 
{{ \nelg\,n_{ph}(\epsilon,\mu_{ph}=\pm 1)}\over {\gamma^2\epsilon B}} \ 
\Theta \left( B-A\right)
\end{equation}
\begin{displaymath}
\times
\ \left[1+\left(1- {{\epsilon (1-\mu_{ph})}\over
{\gamma\epsilon_s\,(1-\beta\mu_0)\left[\gamma (1-\beta\mu_0) + 
\epsilon (1-\mu_{ph})\right]}}\right)^2 +
{{\epsilon^2 (1-\mu_{ph})^2}\over {\gamma (1-\beta\mu_0)
\left[\gamma (1-\beta\mu_0) + \epsilon (1-\mu_{ph})\right]}}\right] 
\end{displaymath}
\subsubsection{$\mu_{ph}=+1$}
Here obviously $B/A=\beta <1$, which implies that the argument of
the delta functional, $g$, has no zero in the range of integration, and
thus the scattering rate in the forward direction is precisely zero.
\subsubsection{$\mu_{ph}=-1$}
Here
\begin{equation}
\ B=\beta (\epsilon_s +\epsilon)
\qquad A=\epsilon_s (1+{{2\epsilon}\over \gamma})-\epsilon
\end{equation}
and then
\begin{equation}
\mu_e = {A\over B} \le 1 \qquad
\Rightarrow \quad \epsilon_s \le \left({{1+\beta}\over 
{{{2\epsilon}\over \gamma} +1-\beta}}\right) \epsilon
\end{equation}
Then also
\begin{equation}
1-\beta\mu_0 = {{2\epsilon}\over \gamma} {{\gamma -\epsilon_s}\over 
{\epsilon_s +\epsilon}}
\end{equation}
and
\begin{equation}
\gamma (1-\beta\mu_0)
\left[\gamma (1-\beta\mu_0) + \epsilon (1-\mu_{ph})\right]=
{{4\epsilon^2}\over {(\epsilon_s + \epsilon)^2}} \left[\gamma^2
- (\epsilon_s -\epsilon)\gamma - \epsilon\epsilon_s\right]
\end{equation}
Therefore
\begin{equation}
\nsca\bigg\vert_{\mu_{ph}= - 1} = {{r_e^2 c\,\epsilon_s}\over {4}} 
\int d\epsilon\ \int d\gamma\ 
{{ \nelg\,n_{ph}(\epsilon,\mu_{ph}=- 1)}\over {\beta\gamma^2\epsilon
(\epsilon_s + \epsilon)}} \ 
\Theta \left[ \left({{1+\beta}\over 
{{{2\epsilon}\over \gamma} +1-\beta}}\right)-\epsilon_s\right] \label{A13}
\end{equation}
\begin{displaymath}
\times
\ \left[1+\left(1- {{(\epsilon_s +\epsilon)^2}\over
{2\epsilon\epsilon_s\left[\gamma^2
- (\epsilon_s -\epsilon)\gamma - \epsilon\epsilon_s\right]}}\right)^2 +
{{(\epsilon+\epsilon_s)^2}\over {\left[\gamma^2
- (\epsilon_s -\epsilon)\gamma - \epsilon\epsilon_s\right]}}\right] 
\end{displaymath}
where $\Theta$ denotes a step function.

\subsection{The case $\mu_{ph}\neq\pm 1$}

We obtain formally by squaring Eq.~\ref{A10}
\begin{displaymath}
\mu_e = {{AB\pm \vert C\vert \sqrt{B^2+C^2-A^2}}\over {B^2+C^2}}
\end{displaymath}
Since we have squared the original equation, not all formal solutions
may apply. The ambiguity in sign is combined with the modulus of $C$, therefore
writing $C$ instead of $\vert C\vert$ does not introduce new solutions.
Then, assuming $A>0$ and $B> 0$, i.e. upscattering of photons in energy,
we find
\begin{equation}\mu_e = {{AB\pm C\sqrt{B^2+C^2-A^2}}\over {B^2+C^2}}
\end{equation}
is a solution to the original equation 
\begin{equation} 
{\rm case\ } \oplus\qquad {\rm if}\quad
C\le {B\over A} \sqrt{B^2+C^2-A^2}\hphantom{-}
\end{equation}
\begin{equation}
{\rm case\ } \ominus\qquad {\rm if\quad }
C\le -{B\over A} \sqrt{B^2+C^2-A^2}
\end{equation}
The delta functional in Eq.~\ref{A11} transforms as
\begin{equation}
\delta (g) = {{B\mu_e -A}\over {\gamma C \sqrt{B^2 +C^2 -A^2}}}\,
\delta (\mu_e -\mu_0)
\end{equation}
Then 
\begin{equation}
\nsca = {{r_e^2 c\,\epsilon_s}\over {4\pi}} \int d\epsilon\ \int 
d\mu_{ph}\ \int d\gamma\ \int_0^\pi d\phi_e\ 
{{ \nelg\,\nphm}\over {\gamma^2\epsilon}} \ 
{{\Theta \left(
B^2+C^2-A^2\right)}\over {C\,\sqrt{B^2 +C^2 -A^2}}}
\end{equation}
\begin{displaymath}
\times\ \sum_\pm\ (B\mu_0 -A)\ 
 \Theta\left(\pm {B\over A} \sqrt{B^2+C^2-A^2}-C\right)
\end{displaymath}
\begin{displaymath}
\times\ \left[1+\left(1- {{\epsilon (1-\mu_{ph})}\over
{\gamma\epsilon_s\,(1-\beta\mu_0)\left[\gamma (1-\beta\mu_0) + 
\epsilon (1-\mu_{ph})\right]}}\right)^2 +
{{\epsilon^2 (1-\mu_{ph})^2}\over {\gamma (1-\beta\mu_0)
\left[\gamma (1-\beta\mu_0) + \epsilon (1-\mu_{ph})\right]}}\right] 
\end{displaymath}
The integration variable $\phi_e$ appears only as the argument 
of a cosine function, which itself appears only in the term $C$. 
We can therefore substitute $C$ for $\phi_e$. With 
$C_1= \beta\epsilon\sqrt{1-\mu_{ph}^2}$ 
\begin{equation}
\int_0^\pi d\phi_e = \int_{-C_1}^{C_1} {{dC}\over \sqrt{C_1^2-C^2}}
\end{equation}
and we obtain
\begin{equation}
\nsca = {{r_e^2 c\,\epsilon_s}\over {4\pi}} \int d\epsilon\ \int 
d\mu_{ph}\ \int d\gamma\ \int_{-C_1}^{C_1} dC\ 
{{ \nelg\,\nphm}\over {\gamma^2\epsilon
C \sqrt{C_1^2-C^2}}} \ 
{{\Theta \left(
B^2+C^2-A^2\right)}\over {\sqrt{B^2 +C^2 -A^2}}}\ 
\end{equation}
\begin{displaymath}  
\times\ \sum_\pm\ (B\mu_0 -A)\ 
 \Theta\left(\pm {B\over A} \sqrt{B^2+C^2-A^2}-C\right)
\end{displaymath}
\begin{displaymath}
\times\ \left[1+\left(1- {{\epsilon (1-\mu_{ph})}\over
{\gamma\epsilon_s\,(1-\beta\mu_0)\left[\gamma (1-\beta\mu_0) + 
\epsilon (1-\mu_{ph})\right]}}\right)^2 +
{{\epsilon^2 (1-\mu_{ph})^2}\over {\gamma (1-\beta\mu_0)
\left[\gamma (1-\beta\mu_0) + \epsilon (1-\mu_{ph})\right]}}\right]
\end{displaymath}
It is now useful to discriminate two cases.
\begin{equation}{B \ge A}
\quad\Rightarrow\quad
\epsilon_s \le \epsilon \,{{1-\beta\mu_{ph}}\over
{1-\beta+{\epsilon\over \gamma} (1-\mu_{ph})}}
= {{\gamma^2\,\epsilon\,(1+\beta)(1-\beta\mu_{ph})}
\over {1+\gamma\epsilon(1+\beta)(1-\mu_{ph})}}={\epsilon_s}_1
\end{equation}
for which only the $\oplus$ solution exists for all C. The second case is
\begin{equation}
{B < A \ {\rm and}\ C_1^2 \ge A^2 -B^2}\quad\Rightarrow\quad
{\epsilon_s}_1 <\epsilon_s\ {\rm and\ }\epsilon_s\le {\epsilon_s}_2
\end{equation}
with
\begin{equation}
{\epsilon_s}_2 =
 {{\epsilon\gamma\left[\gamma (1-\beta\mu_{ph})+\epsilon (1-\mu_{ph})\right]}
 \over {1+2\epsilon\gamma (1-\mu_{ph})+\epsilon^2 (1-\mu_{ph})^2}}\,\left[1
+ \sqrt {1-{{1+2\epsilon\gamma (1-\mu_{ph})+\epsilon^2 (1-\mu_{ph})^2}
\over {\left[\gamma^2 (1-\beta\mu_{ph})+\epsilon\gamma (1-\mu_{ph})\right]^2}}}
\right]
\end{equation}
for which both the $\oplus$ and the $\ominus$ solutions exist, but only for
negative $C\le -C_2=-\sqrt{A^2 -B^2}$. If $\mu_{ph}$ is not very close to 1,
${\epsilon_s}_2$ does in fact exceed ${\epsilon_s}_2$, but only by a very small
margin. The relative difference $({\epsilon_s}_2-{\epsilon_s}_1)/{\epsilon_s}_1
<10^{-2}$ for $\gamma =10$ and it decreases rapidly with increasing $\gamma$.

\subsubsection{The case $B \ge A$}

The integration variable $C$ is an argument of $\mu_0$. We may substitute
$\mu = \mu_0$ for $C$ and use
\begin{equation}
\int_{-C_1}^{C_1} dC\ 
{{B\mu_0 -A}\over {C \sqrt{C_1^2-C^2}\sqrt{B^2 +C^2 -A^2}}} 
= {1\over \sqrt{B^2 +C_1^2}}
\int_{\mu_-}^{\mu_+} d\mu\ {1\over \sqrt{(\mu-\mu_-)(\mu_+-\mu)}}
\end{equation}
with
\begin{equation}
\mu_\pm = {{AB\pm C_1 \sqrt{B^2+C_1^2 -A^2}}\over {B^2 +C_1^2}}
\end{equation}
to obtain the scattering rate as
$$\dot n = {{r_e^2 c\,\epsilon_s}\over {4\pi}} \int d\epsilon\ \int 
d\mu_{ph}\ \int d\gamma\ {{ n_e (\gamma)\,n_{ph}(\epsilon,\mu_{ph})}\over 
{\beta\gamma^2\epsilon\sqrt{\epsilon_s^2-2\mu_{ph} \epsilon\epsilon_s + 
\epsilon^2}}} \ 
\Theta (B-A)\ \int_{\mu_-}^{\mu_+} {{d\mu}\over \sqrt{(\mu-\mu_-)(\mu_+-\mu)}}
$$ $$ \times
\quad \left[1+\left(1- {{\epsilon (1-\mu_{ph})}\over
{\gamma\epsilon_s\,(1-\beta\mu)\left[\gamma (1-\beta\mu) + 
\epsilon (1-\mu_{ph})\right]}}\right)^2 +
{{\epsilon^2 (1-\mu_{ph})^2}\over {\gamma (1-\beta\mu)
\left[\gamma (1-\beta\mu) + \epsilon (1-\mu_{ph})\right]}}\right] $$
or
\begin{displaymath}
\nsca = {{r_e^2 c\,\epsilon_s}\over {4\pi}} \int d\epsilon\ \int 
d\mu_{ph}\ \int d\gamma\ {{ \nelg\,\nphm}\over {\beta\gamma^2\epsilon
\sqrt{\epsilon_s^2-2\mu_{ph} \epsilon\epsilon_s + \epsilon^2}}} \ 
\Theta (B-A)\ \int_{\mu_-}^{\mu_+} {{d\mu}\over \sqrt{(\mu-\mu_-)(\mu_+-\mu)}}
\end{displaymath}
\begin{equation}
\times\ \left[2+ {{\epsilon^2 \epsilon_s(1-\mu_{ph})^2-
2\epsilon (1-\mu_{ph})}\over
{\gamma\epsilon_s\,(1-\beta\mu)\left[\gamma (1-\beta\mu) + 
\epsilon (1-\mu_{ph})\right]}} +
{{\epsilon^2 (1-\mu_{ph})^2}\over {\epsilon_s^2\left(\gamma (1-\beta\mu)
\left[\gamma (1-\beta\mu) + \epsilon (1-\mu_{ph})\right]\right)^2}}\right]
\end{equation}
which is essentially the sum of three integrals
$I_1$, $I_2$, $I_3$, which are defined by the 
terms in brackets.

For the first integral note that
\begin{equation}
\int_{\mu_-}^{\mu_+} {{d\mu}\over \sqrt{(\mu-\mu_-)(\mu_+-\mu)}}=
-\arcsin \left({{\mu_+ +\mu_- -2\mu}\over 
{\mu_+ - \mu_-}}\right)\bigg\vert_{\mu_-}^{\mu_+}=\pi
\qquad\Rightarrow\quad I_1 =2\pi
\end{equation}
The second integral has the structure
\begin{equation}
I_2 =D\int_{\mu_-}^{\mu_+} {{d\mu}\over \sqrt{(\mu-\mu_-)(\mu_+-\mu)}}
{1\over {(\mu+a)(\mu+b)}}
\end{equation}
where
\begin{equation}
D={{\epsilon (1-\mu_{ph})}\over {\gamma^2\beta^2\epsilon_s}}\left[\epsilon
\epsilon_s (1-\mu_{ph}) -2\right]\qquad a=-{1\over \beta}\qquad
b= -{1\over \beta}-{{\epsilon (1-\mu_{ph})}\over {\gamma\beta}}
\end{equation}
We have
\begin{equation}
I_2 =D {{\gamma\beta}\over {\epsilon (1-\mu_{ph})}}\left[
\int_{\mu_-}^{\mu_+} {{d\mu}\over \sqrt{(\mu-\mu_-)(\mu_+-\mu)}}
{1\over {(\mu+b)}} - 
\int_{\mu_-}^{\mu_+} {{d\mu}\over \sqrt{(\mu-\mu_-)(\mu_+-\mu)}}
{1\over {(\mu+a)}}\right]
\end{equation}
\begin{equation}
\Rightarrow\quad I_2=\,-\,D {{\gamma\beta}\over {\epsilon (1-\mu_{ph})}}\left[
\int_{1\over {\mu_-+a}}^{1\over {\mu_+ +a}} {{dt}\over 
\sqrt{A+B_at+C_at^2}} - 
\int_{1\over {\mu_-+b}}^{1\over {\mu_+ +b}} {{dt}\over 
\sqrt{A+B_bt+C_bt^2}}\right] \label{A12}
\end{equation}
where with $(a,b)$ standing for $a$ or $b$
\begin{equation}
A_a=-1\qquad B_{a,b}=\mu_+ +\mu_- + 2(a,b)
\end{equation}
\begin{equation}
C_{a,b}= -\mu_- \mu_+ -(\mu_-
+\mu_+)(a,b) -(a,b)^2
\end{equation}
Because
\begin{equation}
C_{a,b}< 0\qquad {\rm and}\qquad \Delta =4A_a C_{a,b}-B_{a,b}^2 <0
\end{equation}
the solution of the integrals in Eq.~\ref{A12} is
\begin{equation}
\int_{1\over {\mu_-+(a,b)}}^{1\over {\mu_+ +(a,b)}} {{dt}\over 
\sqrt{A_a+B_{a,b}t+C_{a,b}t^2}}=-{\pi\over \sqrt{-C_{a,b}}}
\end{equation}
hence
\begin{equation}
I_2 = \,-\,\pi{{\epsilon \epsilon_s (1-\mu_{ph}) -2}\over
{\gamma\beta \epsilon_s}} \left[
{1\over \sqrt{(\mu_+ +b)(\mu_- +b)}} -
{1\over \sqrt{(\mu_+ +a)(\mu_- +a)}}\right]
\end{equation}
The third integral is of the form
\begin{equation}
I_3 =D\int_{\mu_-}^{\mu_+} {{d\mu}\over \sqrt{(\mu-\mu_-)(\mu_+-\mu)}}
{1\over {(\mu+a)^2(\mu+b)^2}}
\end{equation}
where
\begin{equation}
D={{\epsilon^2 (1-\mu_{ph})^2}\over {\epsilon_s^2 \gamma^4\beta^4}}
\end{equation}
We may write
\begin{displaymath}
{1\over {(\mu+a)^2(\mu+b)^2}} = {1\over {(a-b)^2}}
\left[{1\over {\mu+b}} - {1\over {\mu+a}}\right]^2
\end{displaymath}
\begin{equation}
= {1\over {(a-b)^2}}
\left[{1\over {(\mu+b)^2}}+{1\over {(\mu+a)^2}}-
{2\over {(\mu+a)(\mu+b)}}\right]
\end{equation}
so that
\begin{displaymath}
I_3 = {1\over {\epsilon_s^2 \gamma^2\beta^2}} 
\int_{\mu_-}^{\mu_+} {{d\mu}\over \sqrt{(\mu-\mu_-)(\mu_+-\mu)}}
\end{displaymath}
\begin{equation}
\times\ \left[{1\over {(\mu+b)^2}}+{1\over {(\mu+a)^2}}-
{2\over {(\mu+a)(\mu+b)}}\right]
\end{equation}
\begin{displaymath}
= J_1+J_2+J_3
\end{displaymath}
The integral $J_3 $ is formally identical to $I_2$ 
\begin{equation}
J_3 =  {{2\pi}\over {\gamma\beta\epsilon_s^2 \epsilon (1-\mu_{ph})}}
\left[{1\over \sqrt{(\mu_+ +b)(\mu_- +b)}} -
{1\over \sqrt{(\mu_+ +a)(\mu_- +a)}}\right]
\end{equation}
The other integrals are (for $J_1$ replace $a$ by $b$)
\begin{equation}
\epsilon_s^2\gamma^2\beta^2\ J_2
=\,-\,
{\pi\over 2} {{\mu_+ +\mu_- +2(a,b)}\over {\left[(
\mu_+ +(a,b))(\mu_- + (a,b))\right]^{3/2}}}
\end{equation}
The total scattering rate is then
\begin{equation}
\nsca = {{r_e^2 c\,\epsilon_s}\over {2}} \int d\epsilon\ \int 
d\mu_{ph}\ \int d\gamma\ {{ \nelg\,\nphm}\over 
{\beta\gamma^2\epsilon\sqrt{\epsilon_s^2-2\mu_{ph} \epsilon\epsilon_s + 
\epsilon^2}}} \ \Theta ({\epsilon_s}_1-\epsilon_s)\
\end{equation}
\begin{displaymath}
\times\ \left[1-{1\over {4\epsilon_s^2\gamma^2\beta^2}}
\left(
{{\mu_+ +\mu_- +2b}\over \sqrt{(\mu_+ +b)(\mu_- +b)}^3} +
{{\mu_+ +\mu_- +2a}\over \sqrt{(\mu_+ +a)(\mu_- +a)}^3}\right)\right.
\end{displaymath}
\begin{displaymath}
\left. - \left({{\epsilon (1-\mu_{ph})}\over {2\gamma\beta}}-
{{\epsilon_s \epsilon (1-\mu_{ph}) +1}\over
{\gamma\beta\epsilon_s^2\epsilon (1-\mu_{ph})}}\right)
\left({1\over \sqrt{(\mu_+ +b)(\mu_- +b)}} -
{1\over \sqrt{(\mu_+ +a)(\mu_- +a)}}\right)
\right] 
\end{displaymath}
Note that
$(\mu_+ -\mu_-)/\mu_+ $ is an extremely small number as well as the difference of the. 
parameters $a$ and $b$.
For $\mu_{ph}=-1$ we reproduce the earlier solution (\ref{A13}).


\clearpage

\clearpage

\clearpage

\clearpage

\clearpage

\clearpage

\clearpage

\clearpage

\clearpage

\clearpage

\clearpage

\clearpage

\clearpage

\clearpage

\clearpage

\clearpage

\clearpage

\clearpage

\clearpage

\clearpage

\clearpage

\clearpage

\clearpage

\newpage

\begin{figure}[t]
\includegraphics[height=13cm]{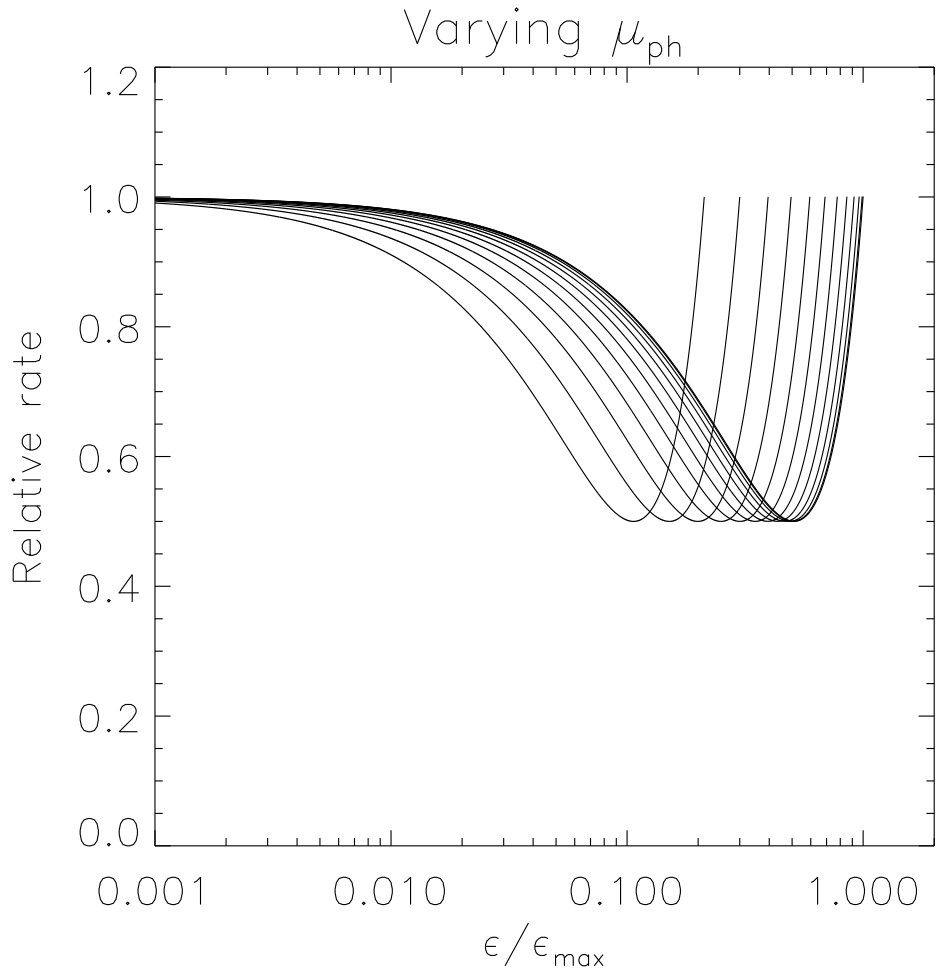}
\caption{Scattering rate for $\epsilon=10^{-7}$, $\gamma=10^5$, and varying
$\mu_{ph}$. For curve $i$ the scattering angle is $\pi -0.2 i$.}
\end{figure}

\end{document}